\documentclass{aa}
\usepackage{graphicx}
\usepackage[colorlinks=true, linkcolor=blue, citecolor=blue, urlcolor=blue]{hyperref}

\usepackage{txfonts}
\usepackage{array}
\usepackage{booktabs}
\usepackage{float}
\usepackage{placeins}
\usepackage{enumitem}
\usepackage{amssymb}
\usepackage{pifont}
\usepackage{multirow}

\usepackage{natbib}
\hypersetup{
    colorlinks=true,
    linkcolor=blue,
    citecolor=blue,
    urlcolor=blue
}

\usepackage{xcolor}
\begin{document} 

\title{\fontsize{16}{18}\selectfont{The kinematic imprinting of environmental quenching in $z<0.2$ galaxies}}

\titlerunning{The kinematic imprinting of environmental quenching}

   \author{Natan de Isídio
          \inst{1}\fnmsep\thanks{\href{www.natanisidio.com}{Natan.Isidio@eso.org}}
          \and
          P.~Popesso \inst{1,2}
          \and
          Y.~Bahé \inst{3,4}
          \and
          B.~Vulcani \inst{5}
          \and
          V.~Toptun \inst{1}
          \and
          I.~Marini \inst{1}
          \and
          B.~Poggianti \inst{5}
          \and
          V.~Biffi \inst{6}
          \and\\
          F.~Belfiore \inst{1,7}
          \and
          C.~Lagos \inst{8,9}
          \and
          K.~Dolag \inst{10,11}
          \and
          D.~Mazengo \inst{1}
}

   \institute{European Southern Observatory, Karl-Schwarzschild-straße-2, 85748 Garching bei München, Germany
    \and Excellence Cluster ORIGINS, Boltzmann-straße-2, 85748 Garching bei München, Germany
    % \and Leiden Observatory, Leiden University, P.O. Box 9513, 2300 RA Leiden, The Netherlands
    \and School of Physics and Astronomy, University of Nottingham, University Park, Nottingham NG7 2RD, UK
    \and Institute of Physics, Ecole Polytechnique F\'{e}d\'{e}rale de Lausanne (EPFL), Observatoire de Sauverny, 1290 Versoix, Switzerland
    \and INAF – Padua Astronomical Observatory, Vicolo Osservatorio 5, I-35122 Padova, Italy
    \and INAF – Astronomical Observatory of Trieste, Via Tiepolo 11, I-34143 Trieste, Italy
    \and INAF – Arcetri Astrophysical Observatory, Largo E. Fermi 5, I-50125 Florence, Italy
    \and International Centre for Radio Astronomy Research (ICRAR), University of Western Australia, Crawley, WA 6009, Australia
    \and ARC Centre of Excellence for All Sky Astrophysics in 3 Dimensions (ASTRO 3D), Sydney, NSW 2000, Australia 
    \and Faculty of Physics, Ludwig-Maximilians University, Scheiner-straße-1, 81679 Munich, Germany
    \and Max Planck Institute for Astrophysics, Karl-Schwarzschild-straße-1, Garching bei München, 85748, Germany
             }
   \date{Received October, 2025; accepted March, 2026}
 
  \abstract

{We present the first systematic census of quenching mechanisms using kinematic asymmetries in a large sample of $\sim$6,700 galaxies from the MaNGA survey, providing a unified view of what halts star formation in the local Universe ($z\lesssim0.2$).
We quantify stellar and nebular gas disturbances through the higher-order terms of a Fourier series expansion, as implemented in the \textsc{Kinemetry} package. These asymmetries serve as powerful diagnostics, as different quenching mechanisms leave distinct kinematic signatures on gas and stars.
Our analysis reveals that the most effective quenching pathways leave minimal kinematic imprints by the time galaxies are fully quenched. This “kinematic regularity” points toward slow-acting processes ($\gtrsim$3~Gyr) such as starvation and maintenance feedback.
A striking finding emerges from our mass-matched analysis: quenched symmetric satellites are significantly more compact than their asymmetric counterparts ($3.4\sigma$), a trend that is even more pronounced for symmetric centrals ($12.3\sigma$).
Our results suggest that environment drives the dominant satellite quenching pathway through rapid gas stripping followed by long-term starvation. These compact, kinematically undisturbed satellites (the most representative case within our sample) have undergone intense gas stripping and central compaction, creating bulge-like structures with old, metal-rich stellar populations. 
Combined with halo gas cut-off and the prevention of cosmological accretion due to starvation, this creates an irreversible quenching path.
Conversely, the larger sizes of disturbed, quenched centrals are consistent with merger-driven growth, where dry and minor mergers account for approximately 30\% of local, massive quenched centrals. Internal processes, likely driven by the AGN cycle over 1-3~Gyr that prevents hot halo gas cooling, sustain quenching maintenance in this population.
The absence of asymmetric satellites in the star-forming regime suggests environmental quenching operates without significant kinematic perturbation.}

{}


\maketitle

\section{Introduction}\label{intro}

Satellite galaxies make up $\sim$30–40\% of the local galaxy population at $z < 0.1$, with most residing in groups \citep{Eke+04, Yang+07, Tempel+15, Tinker+21, Yang+21}. While both centrals and satellites are shaped by their connection to the host dark matter halo, the nature of that connection differs. Central galaxies are mainly regulated by internal mechanisms, particularly feedback between the supermassive black hole and the surrounding circumgalactic medium (CGM). Satellite evolution, by contrast, is more complex, depending on the halo merger history, the thermodynamic and dynamical state of the host halo, and interactions with other galaxies. A satellite may begin as the central of a lower-mass halo before accretion, or remain a satellite throughout, navigating diverse merger histories and environments. These environments primarily quench satellites through gas-removal processes, though AGN feedback–CGM interactions may also contribute.

In the {\it starvation} scenario, first proposed by \citet{Larson+80}, the hot gas reservoir of a satellite is stripped, cutting off the supply of cooling gas needed to replenish the disk and sustain star formation. Simulations show that this can occur as satellites lose their Circumgalactic Medium (CGM) upon infall into a massive host halo \citep[e.g.,][]{Trussler+20}. This process differs fundamentally from rapid gas-removal mechanisms or internal feedback processes, operating on significantly longer timescales (>3~Gyr). 
Cold gas can also be removed through stripping and tidal mechanisms. One of the most effective is {\it ram-pressure stripping}, caused by the interaction between a galaxy and the hot ambient medium in massive halos \citep{Gunn+72,GASP12,GASP33}. This process removes the outer HI disk along the orbit, and its efficiency depends on both the density of the surrounding gas and the galaxy’s velocity—making it particularly strong in clusters. Its most striking observational signatures are “jellyfish galaxies,” which show disturbed morphologies and gaseous tails aligned with their orbital motion \citep{Poggianti+17, Poggianti+19, Vulcani+18}. Alternatively, cold gas may be lost through {\it tidal stripping}, during close gravitational encounters with other satellites or the central galaxy \citep{Kang+08, Pasquali+10, GASP33}. In dense environments, repeated high-speed encounters can enhance this process, a mechanism known as {\it harassment} \citep{Moore+98}.

The combined effect of several—or all—of the aforementioned processes contributes to the higher fraction of passive satellite galaxies observed in massive halos, compared to centrals of similar stellar mass residing in lower-mass halos. In the local Universe, satellites are more likely to fall below the Main Sequence (MS) of star-forming galaxies, where they make up the bulk of the quenched population at stellar masses below $10^{10.8-11}$~M$_{\odot}$ \cite[][]{Popesso+19a}. This phenomenon, commonly referred to as {\it satellite quenching}, plays a central role in shaping the evolution of more than half of the quenched galaxy population observed today \cite[see][for a full review]{Cortese+21}.

Disentangling the relative contributions of different satellite quenching mechanisms remains a complex challenge. Each mechanism is expected to leave distinct kinematic signatures on the gaseous and stellar components of galaxies \citep[e.g.,][]{Bloom+18, Bagge+23}. For example, {\it starvation} is thought to leave the stellar morphology largely intact, gradually quenching star formation as the cold gas reservoir is consumed. In such cases, one would expect a reduced cold gas (HI) fraction while the galaxy morphology remains unchanged. In contrast, tidal processes are expected to disturb both the gas and stellar kinematics. {\it Ram-pressure stripping} alters gas and stellar motions along the satellite’s orbital path, often producing clear kinematic asymmetries \citep[see][]{Poggianti+25}. Meanwhile, {\it tidal disruptions} and {\it harassment} can affect the distribution of gas and stars without a preferred direction, resulting in irregular disk structures or, after multiple high-speed encounters, leaving only a central bulge as the remnant \citep[see][]{Moore+98}.
Although all these processes appear to directly affect the kinematics of galaxies by the time they are fully quenched, we also note that previous studies have shown that quenching and kinematic–morphological transformation are not necessarily closely connected \citep[e.g.,][]{Emsellem+11,Krajnovic+11,Wang+20,Wang+24}.

Identifying the physical drivers of satellite quenching requires a joint analysis of gas and stellar kinematics, typically through IFU spectroscopy, together with cold gas content and host halo properties. Previous studies have used azimuthally averaged profiles of star formation rate (SFR), specific SFR (sSFR), metallicity, and stellar age to reveal gradients tied to galaxy evolution \citep[e.g.,][]{Bluck+20}, but such methods cannot establish direct causal links. Kinematic asymmetries, by contrast, provide a more direct diagnostic, preserving spatial and dynamical information that reflects the underlying environmental interactions. So far, detailed kinematic analyses have focused mainly on cluster environments, most notably the 76 cluster satellites observed with MUSE in the GASP survey \citep{Poggianti+17,Poggianti+25}. Yet clusters contain only 2–3\% of local galaxies, while most satellites reside in groups \citep{Yang+05, Tempel+15, Yang+21}.

The \textit{Mapping Nearby Galaxies at Apache Point Observatory} (MaNGA) survey \citep{MaNGA+15, MaNGA+22}, which observed $\sim$10,000 galaxies up to $z \sim 0.15$, naturally samples the group regime thanks to its volume-limited design. Its spatial and spectral resolution enables velocity maps from both stellar absorption and gaseous emission features, allowing robust measurements of kinematic asymmetries with tools like \textsc{Kinemetry} \citep[e.g.,][]{Kinemetry}. These asymmetries are quantified as deviations in velocity profiles along elliptical annuli, traced by higher-order Fourier coefficients \citep[e.g.,][]{Kinemetry, Barolo, Bloom+17b, DiTeodoro+21, Bagge+24, deIsidio+24}. By combining MaNGA IFU data with SDSS-based group catalogs \citep[e.g.,][]{Yang+05, Tempel+15, Tinker+21}, we directly connect kinematic signatures with environmental processes. Our goal is to disentangle the roles of starvation, gas stripping, and tidal interactions in quenching satellite galaxies by analyzing their distinct imprints on gas and stellar velocity fields, together with cold gas content trends across stellar mass and halo environment. Compared to previous IFU-based studies such as GASP, this work increases the sample size by a factor of $\sim$40 and extends the explored halo mass range by over three orders of magnitude—from galaxy pairs to massive clusters.

\noindent This paper is organized as follows. In Section~\ref{Sec2}, we describe our observational and simulated galaxy samples and the criteria used for their selection.
Section~\ref{Kinemetry} outlines the methodology used to identify and quantify kinematic asymmetries.
In Section~\ref{results}, we present the main results, focusing on the roles of starvation and gas stripping in shaping galaxy evolution, and we compare these with results from the simulated dataset.
Finally, in Section~\ref{sec5}, we summarize our findings and conclusions.

Throughout this paper, we adopt a flat $\Lambda$CDM cosmology with $H_0 = 68.7 \pm 3.1$ km~s$^{-1}$~Mpc$^{-1}$, $\Omega_m = 0.3$, $\Omega_{\Lambda} = 0.7$, and a baryon fraction of $f_b = \Omega_b / \Omega_m = 0.187$ \citep[][]{Planck+18}.

\section{Data and Sample selection}
\label{Sec2}

We describe below the MaNGA survey, our selection criteria and general properties of the sample.

\subsection{The MaNGA survey}
The MaNGA survey is part of the fourth-generation Sloan Digital Sky Survey \citep[SDSS-IV;][]{SDSS-DRIV+17} and is comprised of 10,010 unique galaxies \citep{MaNGA+15,MaNGA+22}.  MaNGA provides spatially resolved spectroscopic data observed with the two identical BOSS spectrographs mounted on the 2.5~m telescope at Apache Point Observatory \citep{Gunn+06,Smee+13,Drory+15}. All MaNGA galaxies are in the nearby Universe at $z\leq0.15$. 
The data covers a spectral range from 3,600 to 10,000~$\AA$, convenient for studying line emissions of ionized gas, such as [OIII] and H$\alpha$, with a resolving power of $R\sim 2000$ at a spatial resolution of $\sim$1.5~kpc. 
The survey is designed to map all galaxies out to at least $1.5~\rm R_{eff}$\footnote{$\rm R_{eff}$ is defined as the radius radius containing half the galaxy's $i$-band luminosity.}.

\subsection{SDSS-related catalogues}\label{MaNGA_catalogs}

\noindent\textbf{\textit{General properties}}. 
The general properties of the sample, such as stellar mass, SFR, and effective radius of MaNGA galaxies are taken from \citet{pyPipe3D}, based on the pyPipe3D pipeline. pyPipe3D processes the IFS data cubes to extract spatially resolved spectroscopic properties of both the stellar population and the ionized gas emission lines. 
The stellar metallicities are taken from the SDSS-based catalog of \citet{Gallazzi+05}, after cross-matching with the MaNGA catalog.

\begin{figure}
  \centering
\includegraphics[width=0.48\textwidth]
{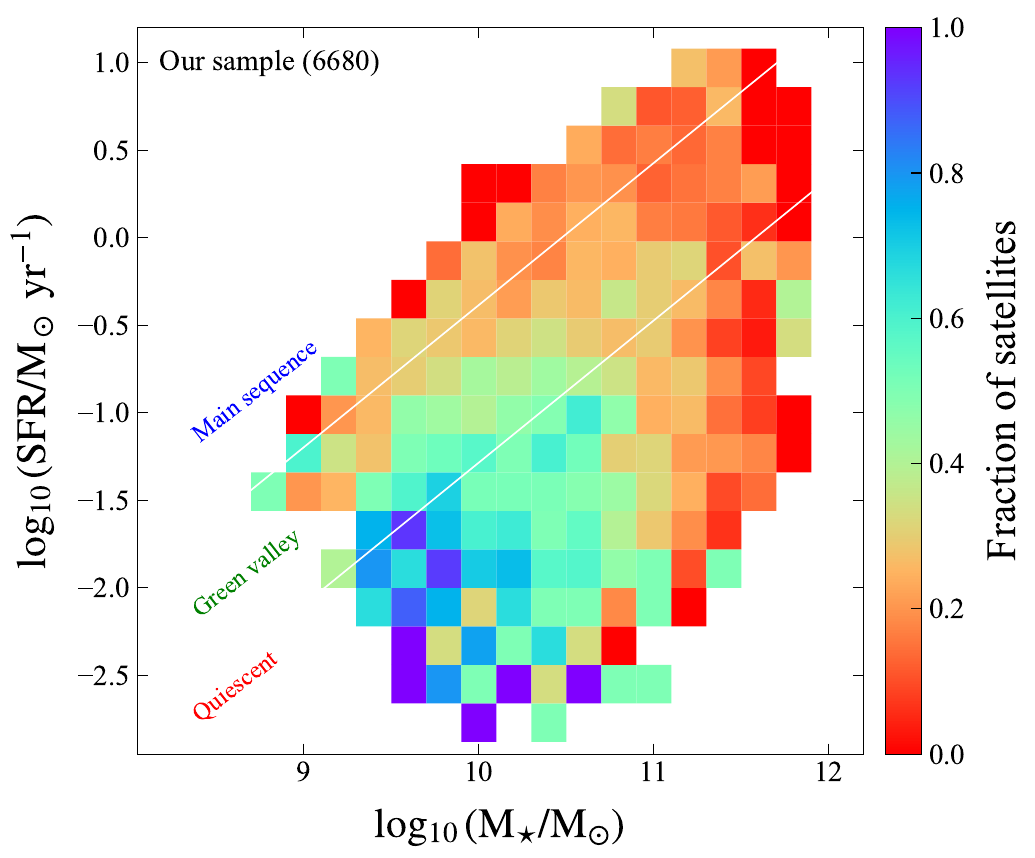}
    \caption{Distribution of our sample of galaxies in the SFR-M$_\star$ plane, colour-coded by the fraction of satellites. As expected, the highest satellite fractions are found among quiescent galaxies with M$_\star \lesssim 10^{10.5}$~M$_\odot$, a region we refer to as the `satellite region'.
    The lines indicate the boundaries separating star forming, green valley, and quiescent galaxies as presented in \cite{Behroozi+19}.}
    \label{fig:satellite_fraction}
\end{figure}

\begin{figure*}[htbp!]
  \centering
\includegraphics[width=1\textwidth]
{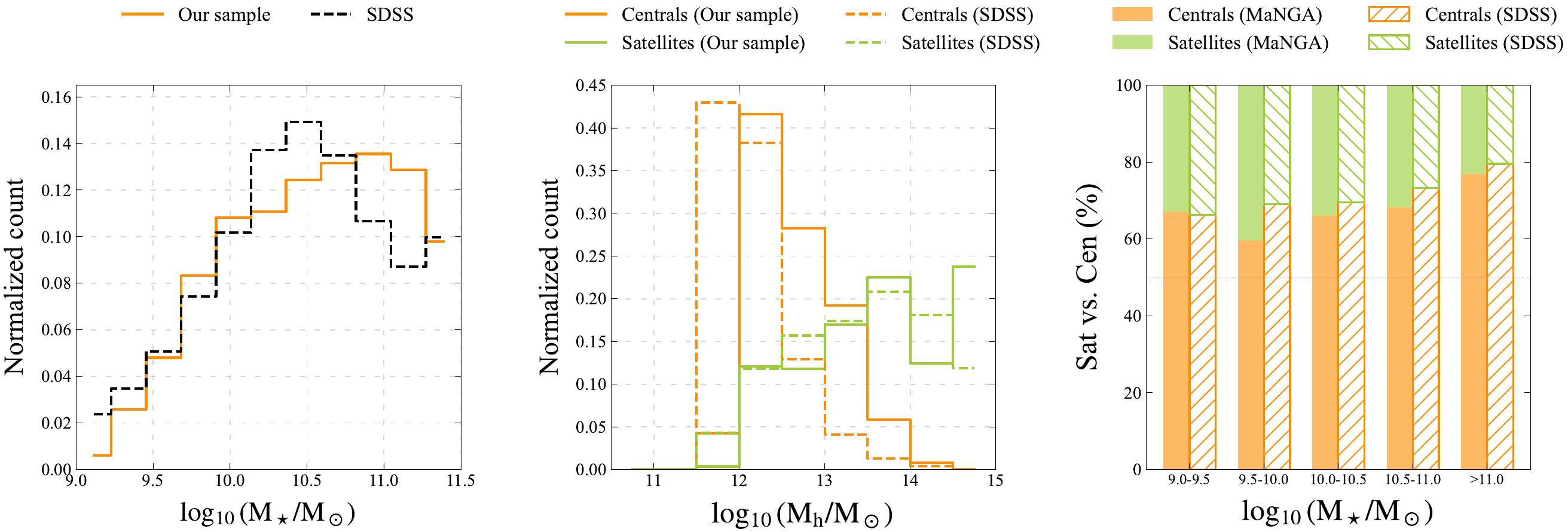}
    \caption{\textit{Left panel:} Stellar mass distribution of our MaNGA sample including both central and satellite galaxies. \textit{Middle panel:} Halo mass distribution of our sample separated by centrals or satellites.
    \textit{Right panel:} Fraction of centrals/satellites per stellar mass bin. In all panels, we show the representativeness of our sample by comparing it to SDSS spectroscopic sample complete in stellar mass down to 10$^{10}~M_\odot$ at $z\sim0.085$.}
\label{fig:Mstar_Mhalo_z_SatCen}
\end{figure*}

\vspace{0.2cm}
\noindent\textbf{\textit{Galaxy environment}}.
Environmental information for MaNGA galaxies comes from an updated version of the \citet{Yang+07} catalog (hereafter Y07), which applies the halo-based group finder of \citet{Yang+05} to SDSS. The catalog contains $\sim$470,000 groups, from clusters to isolated galaxies, and provides central/satellite classifications along with two halo mass proxies: one based on total stellar mass and the other on total $R$-band luminosity. 
The group-finder’s performance has been validated with simulations, showing accuracies of $\sim$80\% for group membership and $\sim$95\% for central identification \citep{Marini24a}. In our sample, only 40\% of centrals have direct halo mass measurements from Y07; the rest are either isolated or fainter than $M_R = -19.5$. For these galaxies, halo masses are inferred from the stellar mass–halo mass relation of \citet{Behroozi+10, Behroozi+13}. In Fig.~\ref{fig:satellite_fraction}, we display the SFR-M$_\star$ plane colour-coded by the fraction of satellites. The halo mass distribution for the measured subset is shown in the middle panel of Fig.~\ref{fig:Mstar_Mhalo_z_SatCen}.
Throughout this paper, we define centrals to be the most massive galaxy within a given dark matter halo and satellites to be any other group/cluster member. Note that an isolated galaxy is taken to be the central of its group of one.

\vspace{0.4cm}
\noindent\textbf{\textit{Morphology}}. 
The morphological classification is provided by the MaNGA Deep Learning Morphological Value Added Catalogue \cite[MDLM-VAC-DR17; ][]{Dominguez-Sanchez+22}. MDLM-VAC-DR17 offers deep-learning-based morphological classifications using convolutional neural networks.
The classification includes Hubble T-Type numerical morphologies, with a refined separation between ellipticals and S0s, and the identification of edge-on and barred galaxies. The T-Type morphologies ranges from -4 to $\sim$9, where the transition between early and late types happens around T-Type$\sim$0. 

\vspace{0.4cm}
\noindent\textbf{\textit{Gas fraction}}. 
HI gas masses and fractions for a subsample of MaNGA galaxies are publicly available in the HI-MaNGA catalog\footnote{Available at: \url{https://greenbankobservatory.org/science/gbt-surveys/hi-manga/}} \citep{Masters+19, Stark+21}, an ongoing HI follow-up program to SDSS-IV MaNGA. The catalog combines Green Bank Telescope (GBT) observations with previously published data. The latest release (DR3.1) includes new GBT measurements for 3,477 MaNGA galaxies and 3,274 additional systems observed in the Arecibo Legacy Fast ALFA survey \citep[ALFALFA;][]{Giovanelli+05}. HI masses are computed following the method described in \citet{Stark+21}.

\subsection{MaNGA gas and stellar velocity maps}
The MaNGA Data Analysis Pipeline (DAP) fits all emission lines with a tied velocity, providing a single velocity measurement per spaxel \citep{Westfall+19}. Gas kinematics are primarily derived from the H$\alpha$ line ($\lambda$6564~$\AA$), which is the strongest and most reliable tracer in both star-forming and low-ionization regions \citep{Belfiore+19}. Although the DAP outputs kinematic maps for all emission lines, we use H$\alpha$ exclusively to maintain consistency and physical interpretability, following the validation of its robustness even at low SNR.

Stellar kinematic maps are obtained from absorption features (e.g., FeI~$\lambda$4383~$\AA$, H$\beta$$\lambda$4861$\AA$) using the penalized pixel-fitting method \textsc{pPXF} \citep{Cappellari+04} as implemented in the DAP. Templates from the MILES-HC library \citep{Sanchez-Blaszquez+06, Falcon-Barroso+11} are fit to Voronoi-binned spectra (SNR > 10), with additive polynomials correcting continuum mismatches. Since the gas kinematics may differ (e.g., due to asymmetric drift or outflows), we adopt H$\alpha$ as our primary gas tracer because of its high SNR and clear mapping of ionized gas dynamics.
In the MaNGA DAP, emission lines are fit using a hybrid binning scheme in which the gas kinematics are extracted from spaxel-level fits, with Balmer and forbidden lines tied kinematically to a common velocity and velocity dispersion. This approach preserves spatial resolution for the ionized gas while ensuring stable kinematic measurements, allowing us to identify non-axisymmetric structures such as bars and spiral arms.

Since MaNGA survey does not provide direct SNR maps of the emission lines, we calculate the SNR per spaxel, here defined as flux divided by the flux error, following the MaNGA Data Analysis Pipeline tutorial\footnote{Available at: \url{https://www.sdss4.org/dr17/manga/manga-tutorials/dap-tutorial/dap-python-tutorial/}} (see Appendix~\ref{Appendix_A}).

\subsection{Sample selection}
\label{selection_criteria}

Since the analysis of the galaxy kinematics requires an accurate measure of the resolved velocity maps, we apply the following selection criteria to ensure high data quality:
   \begin{enumerate}
      \item All spaxels of the stellar continuum and the emission line maps used in the analysis must have SNR~>~5 to be considered “valid”;
      \item Galaxies must have at least 90\% of valid spaxels within 1~$R_{\rm eff}$;
      \item All galaxies must have environment measures in either Y07 or \cite{Argudo-Fernández+15};
      \item Errors in the velocity must be lower than 40\% of the velocity value itself. This ensures that only spaxels with reliable velocity values are considered.
   \end{enumerate}

\begin{figure}[htbp!]
  \centering
\includegraphics[width=0.45\textwidth]
{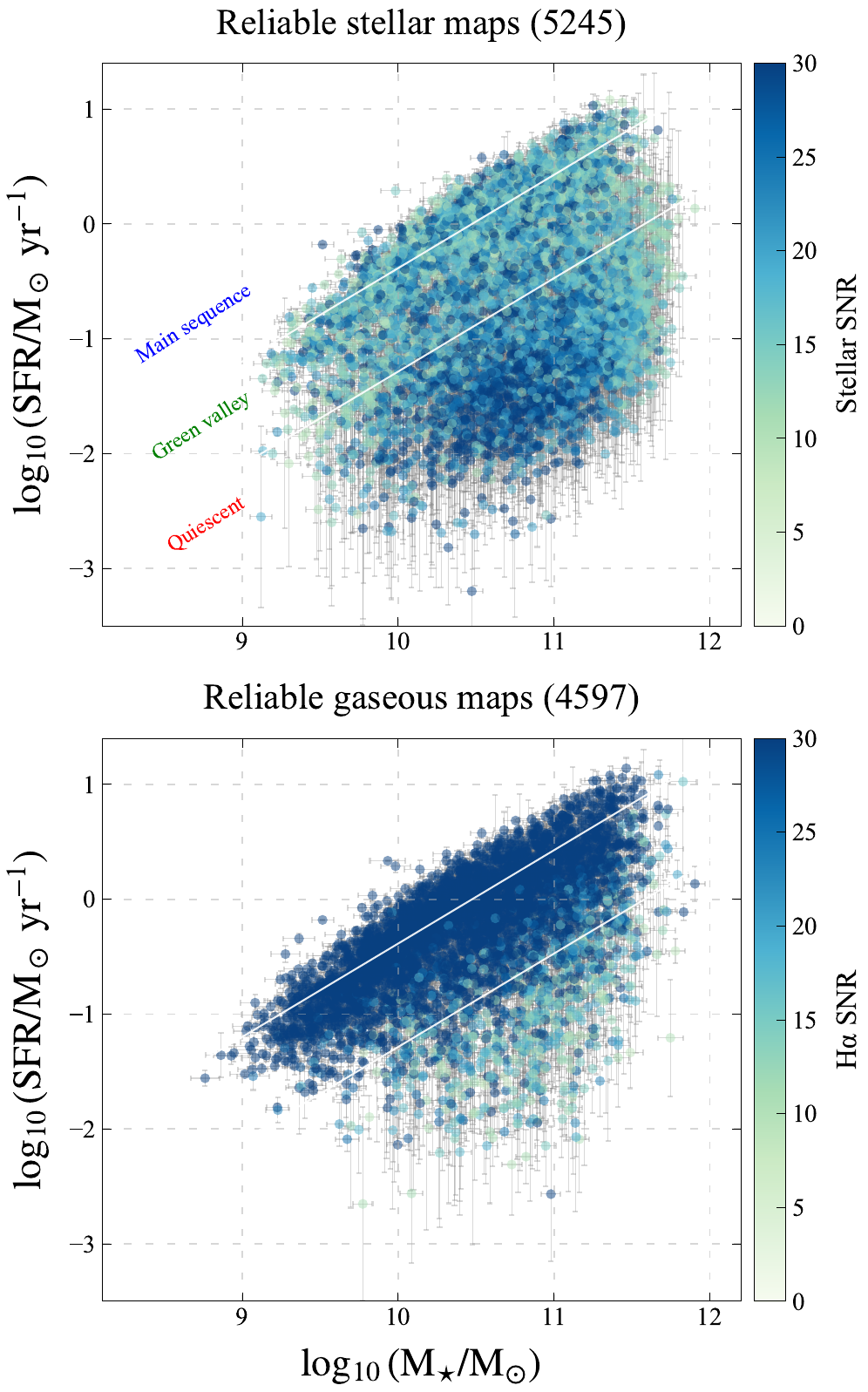}
    \caption{
    SFR–$M_\star$ plane of our galaxy sample, colour-coded by the signal-to-noise ratio of each individual object. The top panel shows our subsample with reliable stellar velocity maps (i.e., maps with $\geq$90\% valid spaxels within 1$R_{\rm eff}$). The bottom panel shows galaxies with reliable gaseous velocity maps.
    The lines indicate the boundaries separating star-forming, green valley, and quiescent galaxies as defined in \cite{Behroozi+19}.}
    \label{fig:3plots}
\end{figure}

The selection criteria outlined above yield a final sample of 6,680 galaxies. This represents approximately 80\% of the original 8,471 galaxies with available stellar velocity and velocity dispersion maps derived from the continuum. 
About $\sim$70\% of our final sample have valid gas kinematics from emission lines following our selection criteria outlined above, while $\sim$80\% of them have valid stellar kinematics.
To check how representative the selected sample is of the underlying local galaxy population,  Fig.~\ref{fig:Mstar_Mhalo_z_SatCen} shows the stellar mass, halo mass and satellite and central fractions for the final sample compared to the SDSS spectroscopic sample complete in stellar mass down to $10^{10}$ $M_{\odot}$ at $z<0.085$. Our final sample is representative of the bulk of the galaxy population above $10^{10}$~M$_{\odot}$ and provides a fair representation of the local galaxy population in terms of halo mass distribution.
Figure~\ref{fig:3plots} shows the distribution of these samples in the SFR–$M_\star$ plane. 
The \textit{top~panel} displays all galaxies with reliable stellar kinematics, colour-coded by the SNR of the stellar continuum. It illustrates that both high- and low-mass galaxies across the star-forming MS, green valley, and quiescent region maintain a consistently high SNR ($\gtrsim$15). 
The \textit{bottom~panel} of Fig.~\ref{fig:3plots} presents the distribution of galaxies with reliable gas velocity and dispersion maps, also in the SFR–$M_\star$ plane, colour-coded by the average SNR of the H$\alpha$ line. This distribution diverges from the stellar sample, as galaxies with weak or no emission—especially those below the MS—do not satisfy the selection thresholds for gas kinematics. The overlapping subsample of 3,162 galaxies, with both stellar and gas kinematic maps simultaneously, primarily occupies the MS and green valley regions. 
It is sparsely populated in the starburst region, where strong emission lines are present but the stellar continuum typically has too low SNR, and in the quiescent region, where gas emission is often negligible. Satellite galaxies comprise approximately 30\% of each sample and subsample, representing a ten- to twenty-fold increase in satellite statistics compared to the GASP survey. 
The 30\% of the population corresponds to satellite galaxies in a precise region at stellar masses lower than $\sim$10$^{10.5}$ M$_{\odot}$ and below the MS consistent with the trend identified in the overall local SDSS galaxy population in \citet{Popesso+19a}. The sizes of all samples and sub-samples are summarized in Table~\ref{tab:kinemetry}. 

\section{Tracing kinematic asymmetries}
\label{Kinemetry}
\subsection{The \textsc{Kinemetry} package}
To identify kinematically asymmetric galaxies, we use the \textsc{Kinemetry} package\footnote{The package is available at: \url{https://www.aip.de/en/members/davor-krajnovic/kinemetry/}}, developed by \cite{Kinemetry}.
This code analyzes 2D maps of the moments of the line-of-sight velocity distribution (intensity, velocity, velocity dispersion) by performing harmonic expansions along best-fitting ellipses, providing a robust quantification of kinematic structures and subcomponents \citep{Kinemetry}. It models velocity profiles as a Fourier series,
\begin{equation}\label{eq:kinemetry}
    K(a,\psi)=A_0(a)+\sum\limits_{n=1}^{N} A_n(a)~\sin(n\psi)+B_n(a)\cos(n\psi),
\end{equation}
where $\psi$ is the eccentric anomaly, $a$ the semi-major axis length, and $A_0$ the systemic velocity. Outputs include the position angle, ellipticity, and asymmetry coefficients. 
The final ellipse parameters obtained by minimization are then used to describe an elliptical ring from which a kinematic profile is extracted and expanded on to the harmonic series of equation (\ref{eq:kinemetry}), where the coefficients ($A_n$, $B_n$) are determined by a least-squares fit with a basis $\{1, \cos(\psi), \sin(\psi), ... , \cos(N\psi), \sin(N\psi)\}$ \cite[][]{Kinemetry}.
\textsc{Kinemetry} has been widely validated as a tracer of kinematic asymmetries, successfully identifying features linked to mergers, gas stripping, bars, and feedback-driven turbulence \citep[e.g.,][]{Shapiro+08, Liu+13, Holmes+15, Bloom+18, Simons+19, Slater+19, Feng+22}.
We quantify asymmetries using the parameter
\begin{equation}\label{eq:asym}
    \overline{I_{asym}}=\overline{\left(\frac{k_2+k_3+k_4+k_5}{4\,k_1}\right)},
\end{equation}
where $k_1$ is the amplitude of the rotating component and higher-order terms ($k_2$–$k_5$) represent non-rotating contributions \citep[][]{Kinemetry, Shapiro+08}. 
While some studies consider only odd modes \cite[e.g., $k_3$ and $k_5$;][]{Bloom+17, Bloom+18}, we include all terms to also capture signatures of major mergers. Equal contributions from rotating and non-rotating components yield $I_{asym} \geq 0.25$ ($\log I_{asym} \geq -0.60$). \citet{Bloom+17} adopted a threshold of $I_{asym} \geq 0.065$ ($-1.19$ in log), while \citet{Feng+22} used $I_{asym} \geq 0.039$ ($-1.41$ in log). Here, we adopt an intermediate threshold of $I_{asym} \geq 0.05$ ($-1.30$ in log), corresponding to a $\sim$18\% non-rotating contribution.

We apply \textsc{Kinemetry} to both stellar and gaseous velocity maps in our MaNGA sample. Examples of galaxies with different asymmetry levels are shown in Fig.~\ref{fig:first_fig}, with the last three rows illustrating asymmetric stellar kinematics. Table~\ref{tab:kinemetry} reports the number of asymmetric galaxies identified in stars and gas, including the fraction of asymmetric satellites relative to the valid sample (see $\S\ref{selection_criteria}$).

\begin{table}[htbp!]
\begin{center}
\caption{Modelling results with \textsc{Kinemetry}.}
\label{tab:kinemetry}
    \begin{tabular}{cccc}
    \hline
    & \multicolumn{3}{c}{--------- Number of galaxies ---------} \\
    Kinematic & Parent & \textsc{Kinemetry} & Asymmetric\\
     map & sample & analyzed & $(I_{asym}\geq0.05)$\\
     (1) & (2) & (3) & (4)\\
    \midrule
    Stars            & 8,471  & 5,245 & 1,465 (27.9$\pm$ 0.6\%)\\
    Gas (H$\alpha$)  & 8,471  & 4,597 & 1,322 (28.8 $\pm$ 0.7\%)\\
    Stars \& Gas     & 8,471  & 3,162 & 351 (11.1 $\pm$ 0.6\%)\\
    Stars only     & 8,471  & 2,083 & 1,079 (53.5 $\pm$ 1.1\%)\\
    Gas only     & 8,471  & 1,435 & 867 (67.7 $\pm$ 1.2\%)\\
    \hline
    \multicolumn{4}{c}{Stellar kinematic map}\\
    Centrals & 5,769 & 3,633 & 1,046 (28.8 $\pm$ 0.8\%)\\
    Satellites & 2,579 & 1,612 & 419 (26.0 $\pm$ 1.1\%)\\

    \multicolumn{4}{c}{Gaseous kinematic map}\\
    Centrals & 5,769 & 3,431 & 908 (26.5 $\pm$ 0.8\%)\\
    Satellites & 2,579 & 1,166 & 414 (35.5 $\pm$ 1.4\%)\\
\hline
\end{tabular}\\
\vspace{0.1cm}
\end{center}
    \footnotesize \bf{Notes:} \normalfont As the primary criterion, all galaxies considered in this work have environmental information available from Y07 or \cite{Argudo-Fernández+15}.
    The percentages in column (4) represent the fraction of asymmetric galaxies relative to the valid sample (i.e., galaxies with >90\% of valid spaxels shown in column (3)), along with the corresponding binomial errors.
    The final number of galaxies with more than $90\%$ of valid spaxels (SNR$_{\rm spx}>5$) within 1~R$_{\rm eff}$ is 6,680, of which 4,681 are centrals and 1,999 are satellites.
\end{table}

\subsection{Identification of ram pressure stripped galaxies in MaNGA}\label{jellyfish}

Jellyfish galaxies have been widely studied as probes of gas-stripping processes and their kinematic impact on galaxy evolution \citep[e.g.,][]{Poggianti+17, Jaffe+18, Vulcani+18}. These systems are characterized by extended gaseous tails stripped from the galaxy body. Because MaNGA’s FoV is relatively small, it will not capture the full extent of these features. To test whether \textsc{Kinemetry} can still identify such systems in MaNGA, we evaluated its performance using gas velocity maps from the \textit{GASP} survey, which provides wide-field MUSE observations ideally suited for studying stripping.

Unlike GASP, which reaches several $R_{\rm eff}$, MaNGA typically covers only $\sim$1.5~$R_{\rm eff}$, while stripping tails often extend much farther \citep[e.g.,][]{Poggianti+25}. To assess the impact of this limitation, we randomly selected 23 GASP galaxies\footnote{Data accessed via the ESO archive: \url{www.archive.eso.org/}}, including 13 with confirmed and strong ram pressure stripping signatures (see Appendix~\ref{Appendix_B}; Table~\ref{tab:GASP}; Fig.~\ref{fig:MaNGA_FoV}). After visually inspecting the maps, we (i) ran \textsc{Kinemetry} on the full velocity maps; (ii) cropped them to MaNGA’s average FoV of 24 arcsec (typical range: 12–32 arcsec\footnote{Set by IFU size, from 19 to 127 fibers}); and (iii) re-ran \textsc{Kinemetry} on the cropped maps. The asymmetry parameters measured within MaNGA’s aperture are consistent with those from the full GASP FoV. For 20/23 galaxies, an aperture of $2\times$ MaNGA’s FoV was sufficient to capture all relevant emission (see Fig.~\ref{fig:JO206_FoVs}). Using MaNGA’s average FoV and our adopted asymmetry threshold, \textsc{Kinemetry} correctly identifies as being asymmetric 12/14 (86\%) galaxies with confirmed ram pressure stripping signatures. 
Out of the remaining cases (11 galaxies), 4/11 galaxies are classified as symmetric using either MaNGA or GASP FoV, 2/11 were consistently classified as asymmetric with both apertures, as expected given that they are ongoing mergers, and the remaining are also classified as asymmetric, although not due to ram pressure as we show in Table~\ref{tab:GASP}. Thus, up to $\sim$20\% of ram pressure stripped galaxies with extended tails may go undetected due to MaNGA’s spatial coverage, though this drops to $\sim$7\% with a relaxed threshold of 0.039.
It is important to note that jellyfish galaxies themselves are rare, representing only $\sim$5–15\% of cluster satellites \citep{Poggianti+19,Le2022,Vulcani+22}. Their frequency rises in massive halos, where high ICM densities and orbital velocities enhance stripping, and declines in groups, which are the focus of this work \citep[e.g.,][]{Roberts2021b}. Therefore, even if $\sim$20\% of jellyfish are missed, the overall impact on our group satellite sample is negligible.

Due to the requirement of a signal-to-noise ratio (SNR) $>5$ in over 90\% of the spaxels within R$_{\rm eff}$, the \textsc{Kinemetry} analysis could be applied to 6,680 galaxies, corresponding to 80\% of the initial MaNGA sample. Of these, 5,245 galaxies were suitable for analysis of absorption features in the stellar continuum, while 4,597 galaxies allowed for the analysis of gas asymmetries through emission lines. 
The two subsets overlap for 3,162 galaxies, for which both stellar and gas components could be analyzed.

\section{Results}
\label{results}
Galaxies with \textsc{Kinemetry} results for the stellar velocity maps span the full extent of the SFR–$M_\star$ plane (\textit{top~panel} of Fig.~\ref{fig:3plots}). Systems with the lowest SNR are predominantly extremely gas-rich galaxies, where the spectra are dominated by emission lines and the stellar continuum is weak. The gas kinematic analysis was primarily feasible for galaxies on the star-forming MS, typically disky and gas-rich, as well as for a portion of green valley systems (\textit{bottom~panel} of Fig.~\ref{fig:3plots}). In contrast, the gas kinematic analysis was applicable to only a negligible fraction of quiescent, gas-poor galaxies, where the spectra lack strong emission lines and are dominated by the continuum. Overall, the combined sample with \textsc{Kinemetry} data spans the full SFR–$M_\star$ diagram with high statistics, as shown in Fig.~\ref{fig:3plots}.
Approximately 70\% of the galaxy population exhibits symmetric stellar or gas kinematics, regardless of morphological type. This fraction remains consistent when considering stellar and gas kinematics separately. Only about 5\% of the total sample shows disturbed kinematics in both components.

\begin{figure}[!t]
  \centering
\includegraphics[width=0.48\textwidth]
{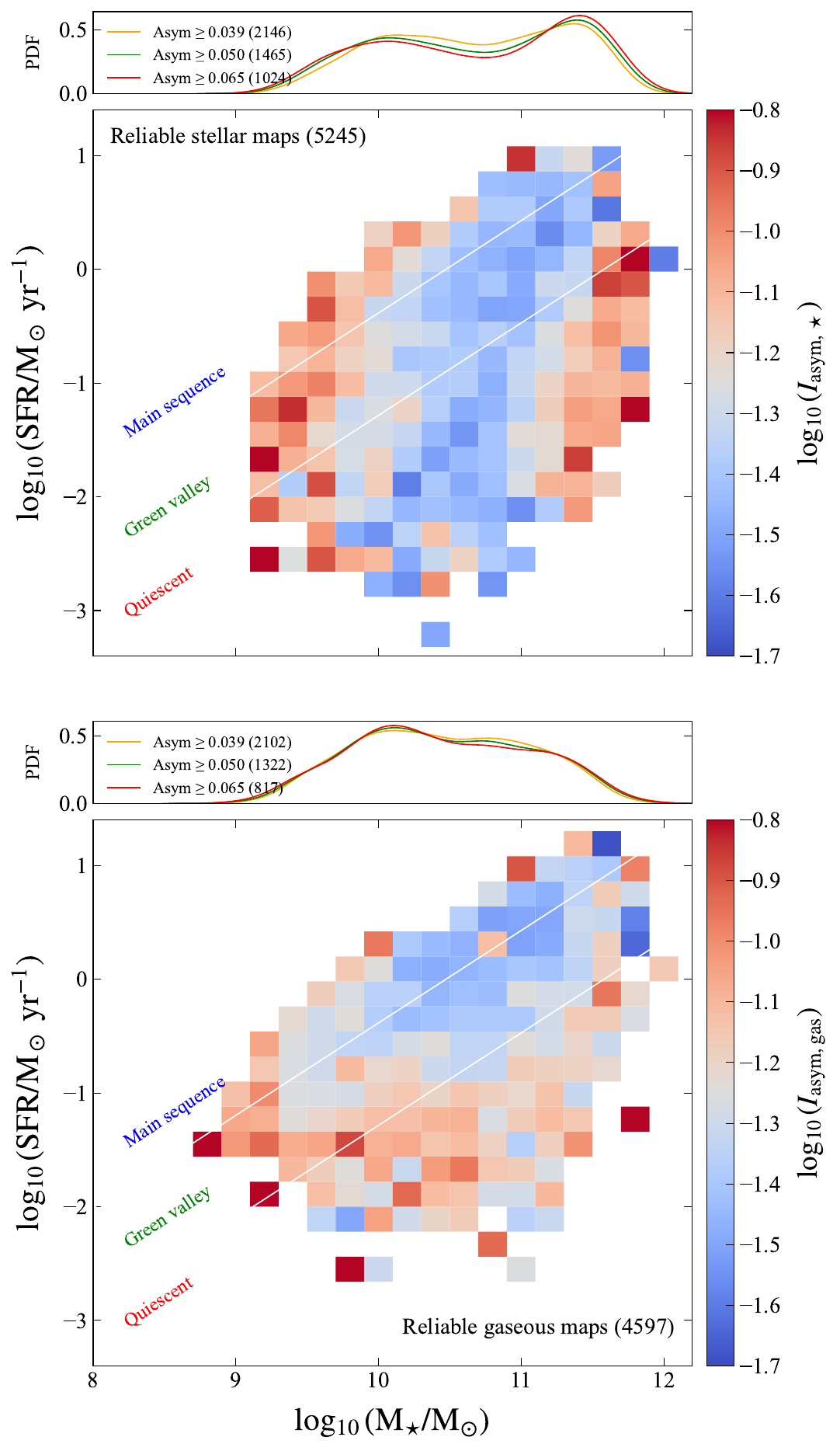}
    \caption{{\it{Top~panel:}} distribution of galaxies with \textsc{Kinemetry} info on the stellar component in the SFR-M$_\star$ plane, colour-coded by the average asymmetry parameter of the stellar kinematic map. As in Fig.~\ref{fig:3plots}, the lines indicate the boundaries separating star forming, green valley, and quiescent galaxies as present in \cite{Behroozi+19}. The upper curves indicate the PDF of galaxies classified as asymmetric according to different \textsc{Kinemetry} thresholds in $I_{\text{asym}}$. {\it{Bottom panel:}} As in the upper panel, now for galaxies with \textsc{Kinemetry} info on the gaseous component.}   \label{fig:satellite_fraction_2}
\end{figure}

Figure~\ref{fig:satellite_fraction_2} shows the SFR–$M_\star$ plane again, this time colour-coded by the average asymmetry index computed from equation~(\ref{eq:asym}), shown separately for the stellar (upper panel) and gaseous (lower panel) components. 
A key feature in both distributions is that galaxies, especially those located along the MS but also in the green valley, with stellar masses above $\sim$$10^{10}~\textrm{M}{_\odot}$, exhibit largely undisturbed stellar and gas kinematics.
This suggests that these systems are not significantly affected by disruptive processes, including mergers. This conclusion is robust, as the signal-to-noise ratio in both components is highest in this mass range, as shown in Fig.~\ref{fig:3plots}. Below $\sim$$10^{10}~\textrm{M}{_\odot}$, the data indicate a trend toward increasingly disturbed kinematics. However, this regime suffers from sample incompleteness: both the MaNGA and SDSS surveys are known to be biased against low-mass galaxies at the redshifts considered. Despite this, the average SNR in both stellar and gas components remains relatively high (around $\sim$10), supporting the idea that low-mass galaxies in this sample are genuinely more kinematically disturbed than their higher-mass counterparts.

While galaxies on the MS generally exhibit symmetric and undisturbed kinematics in both the stellar and gaseous components, a different trend emerges below the MS, from the green valley toward the quiescent region.
In this regime, the most massive galaxies often exhibit clear signs of kinematic disturbance.
This is especially evident in the stellar component, which benefits from high SNR in these systems. Similar, though less pronounced, indications are present in the gas kinematics. However, this pattern pertains to only a negligible fraction of the overall sample. The quiescent region is dominated by gas-poor galaxies that lack emission lines and typically have a relatively low SNR (below 10), particularly in the gaseous component, which limits the detectability of the gas kinematic analysis in this population.

\subsection{Central vs. satellite galaxies}

Surprisingly, the fraction of disturbed galaxies remains essentially unchanged when central and satellite galaxies are separated. Table~\ref{tab:kinemetry} shows that the fraction of symmetric and disturbed galaxies is the same in both the satellite and central galaxy population, $\sim$30\% as in the global population. 
In principle, one might expect a higher incidence of disturbed kinematics among satellite galaxies, as they are more susceptible to environmental processes that are less effective on central galaxies residing at the bottom of their halo's potential well. However, this expected difference is not observed in the data.

We emphasize that this result is not due to misclassification of centrals and satellites by the \citet{Yang+05} algorithm. A thorough test of the algorithm, applied to a mock catalog designed to mimic a GAMA-like survey, demonstrated high accuracy—approximately 95\%—in identifying central galaxies across halos spanning nearly three orders of magnitude in mass, from Milky Way–like groups to massive clusters \citep[see][]{Marini24a}. Furthermore, the relative fraction of disturbed galaxies remains consistent between centrals and satellites even when applying more relaxed or more stringent thresholds for kinematic disturbance in the \textsc{Kinemetry} analysis. While the absolute fraction of disturbed systems changes with threshold choice, the trend of similarity across galaxy types holds. This indicates that the observed parity in disturbed kinematics between centrals and satellites is not driven by classification errors or threshold definitions. This conclusion is reinforced by the consistent distribution of disturbed galaxies across stellar mass, shown in the upper subpanel of Fig.~\ref{fig:satellite_fraction_2}.

\begin{figure*}[htbp!]
  \centering
\includegraphics[width=1\textwidth]
{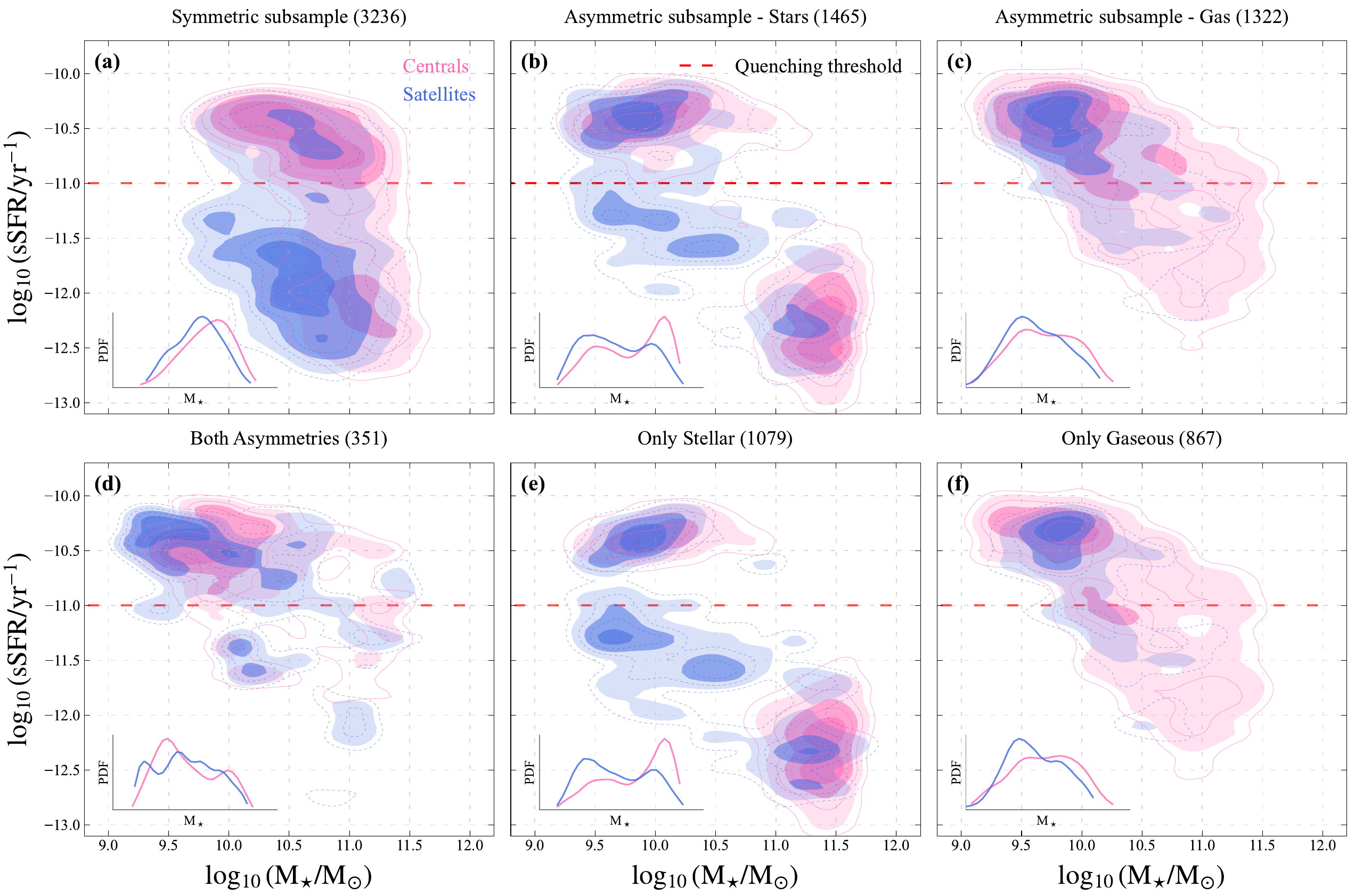}
    \caption{We show in panel a the sSFR-M$_\star$ plane displaying all symmetric galaxies ($I_{asym}=0.05$) in our sample, including both centrals and satellites.
    Subsequent panels highlight galaxies with kinematic asymmetries in the stellar component (panel b), and gaseous component (panel c).
    Panel d displays galaxies with asymmetries in both maps, followed by galaxies with asymmetries solely in stars (panel e) and gas (panel f).
    To guide the eye, we show the corresponding PDFs in the lower left corner of each panel.
    We also show a dashed red line separating quiescent and star-forming galaxies.
    Following our selection criteria, only galaxies with at least 90\% of valid spaxels (spaxels with SNR$_{\star,\rm gas}>5$) are included.
    The absence of asymmetric galaxies in the so-called `satellite region' of the (s)SFR–$M_\star$ plane (see also Fig.~\ref{fig:satellite_fraction}) strongly suggests that the mechanism suppressing star formation in our sample does not significantly impact the kinematic structure of the galaxy.}
    \label{fig:Mstar_SFR}
\end{figure*}

Some differences between centrals and satellites become more apparent when examining the distribution of disturbed galaxies in the sSFR–$M_\star$ plane. Figure~\ref{fig:Mstar_SFR} shows the distribution of symmetric systems (panel a), galaxies with disturbed stellar kinematics (panel b), and galaxies with disturbed gas kinematics (panel c), with centrals and satellites distinguished in each. Galaxies with symmetric stellar or gas kinematics—comprising the majority of the population—follow the global distribution, as confirmed by a Kolmogorov–Smirnov (KS) test. In the diagram, the satellite region is dominated by symmetric systems.
In contrast, symmetric centrals are concentrated along the MS and populate the high-mass end of the green valley and quiescent regions.

When considering only the stellar component, we find that disturbed central galaxies occupy two distinct regions of the sSFR–$M_\star$ plane. Among star-forming systems, they lie along the upper envelope of the MS, well above the quiescent threshold defined at sSFR $= 10^{-11}\ \text{yr}^{-1}$ \citep[e.g.,][]{Wetzel+12,Behroozi+19}. In contrast, quiescent disturbed centrals are found at the high-mass end, with stellar masses $>10^{11}\ \rm M{_\odot}$. 
Satellites with disturbed stellar kinematics, by comparison, are more uniformly distributed across the plane. For both centrals and satellites, the fraction of disturbed galaxies remains similar—ranging from 25\% to 30\%—above and below the quiescence threshold.

The distribution of disturbed galaxies differs markedly in the gaseous component. For both centrals and satellites, disturbed gas kinematics are predominantly found in galaxies on the MS. This is largely a result of selection effects: galaxies below the quiescence threshold typically lack detectable gas or do not exhibit emission lines with sufficient SNR for reliable analysis. Unlike the stellar component, disturbed gas kinematics in the quiescent region are more uniformly distributed at stellar masses below $10^{11}\ \rm{M}_{\odot}$. Notably, among the small subset of galaxies in the quiescent region with detectable H$\alpha$ emission at sufficient SNR, a significant fraction—44\% of centrals and 47\% of satellites—exhibit disturbed gas kinematics. 

\begin{table}[htbp!]
\begin{center}
\caption{Results for symmetric and asymmetric satellites and centrals.}
\label{tab:results}
    \begin{tabular}{cccc}
    \multicolumn{3}{c}{Stellar kinematics}\\
    \hline \hline
     & Centrals & Asymmetric Centrals \\
    \midrule
    Parent sample & 3,634 & 1,046 (28.8\%)\\
    Non-quenched & 1,604 & 434 (27.1\%)\\
    Quenched & 2,030 &  612 (30.1\%)\\
\hline
     & Satellites & Asymmetric Satellites\\
    \midrule
    Parent sample & 1,612 & 419 (26.0\%)\\
    Non-quenched & 485 & 163 (33.6\%)\\
    Quenched & 1,127 & 256 (22.7\%)\\
    
    \multicolumn{3}{c}{Gaseous kinematics}\\
    \hline \hline
     & Centrals & Asymmetric Centrals \\
    \midrule
    Parent sample & 3,432 & 908 (26.5\%)\\
    Non-quenched & 2,465 & 496 (20.1\%)\\
    Quenched & 967 & 412 (42.6\%)\\
\hline
     & Satellites & Asymmetric Satellites\\
    \midrule
    Parent sample & 1,166 & 414 (35.5\%)\\
    Non-quenched & 792 & 244 (30.8\%)\\
    Quenched & 374 & 170 (45.5\%)\\
\hline
\end{tabular}\\
\vspace{0.1cm}
\end{center}
    \footnotesize \bf{Note:} \normalfont All numbers are relative to galaxies with >90\% of valid spaxels (SNR$_{\rm spx}>5$).
\end{table}

\subsection{Identifying environmental processes}\label{obs_results}

To create a causal link between the kinematic signatures in the star and gaseous velocity maps and the possible environmental processes, we define six classes of galaxies depending on their SFR, gas fraction, and asymmetries. We present each of them below.

\underline{\bf{Case 1.}}\\
{\bf{Star-forming galaxies with no sign of asymmetry in the gas and stellar velocity maps}} are considered evolving through secular evolution without environmental effects both in the central and satellite populations. We define the star-forming galaxies as those above the quiescent threshold (sSFR>10$^{-11}$~yr$^{-1}$).
Within the star-forming population with valid gaseous and stellar kinematic maps simultaneously (2,039), Case 1 galaxies account for the majority (1,305; 64\%). Symmetric central galaxies dominate this population, accounting for 79.6\%, while the remaining are satellites. Central galaxies are distributed along the whole MS, while satellites tend to be mainly at low masses and in the green valley as indicated in Fig. \ref{fig:satellite_fraction}.

\underline{\bf{Case 2.}}\\
{\bf{Galaxies with both gas and stellar velocity asymmetry}} might be related to {\it{i)}} ram pressure stripping, if the asymmetry direction of the velocity maps is along a preferential direction (possibly aligned with the galaxy orbit in the halo), {\it{ii)}} mergers or tidal interaction coincident with the presence of double source or close pair, {\it{iii)}} ongoing harassment if the asymmetry has an irregular distribution and there is no clear double source or close pair.
Galaxies with disturbed kinematics in both stellar and gaseous components are spread across the full sSFR–$M_\star$ plane (panel d of Fig.~\ref{fig:Mstar_SFR}). They represent 5.1\% of centrals and 5.5\% of satellites, computed relative to the total number of each population, independent of whether both kinematic maps are available. Despite their broad distribution, these galaxies cluster toward the low-mass end of the MS.

For satellites, systems with asymmetric kinematics while still on the star-forming sequence are strong candidates for environmental disturbance via ram-pressure stripping or tidal interactions \citep[e.g.,][]{Smith+10, Poggianti+17}. To distinguish between these mechanisms, we identify close companions—defined as galaxies within 200~kpc and a line-of-sight velocity difference of less than 300 km/s. These thresholds are designed to capture recent flybys consistent with the velocity dispersion of group environments, with typical halo masses of $10^{12.5}\ \text{M}{_\odot}$ \citep{Yang+05}. Robustness checks using thresholds relaxed or tightened by $\pm$25 kpc and $\pm$50 km/s yield the error bars reported from now on. We find that 58$_{-9}^{+3}$\% of satellites with disturbed kinematics (gas and stars) have a close companion and are thus likely undergoing tidal interactions, while the remaining $\sim$40\% are candidate ram-pressure stripped systems.

Centrals, by contrast, are unlikely to experience ram-pressure stripping since they typically reside at the bottom of their halo potential, with minimal relative velocity to the intra-group medium. Applying the same companion search, we find only 31$_{-5}^{+4}$\% of centrals with disturbed kinematics host a close companion, suggesting tidal interactions or minor mergers. %Visual inspection further reveals merger signatures—such as double-peaked light distributions—in roughly 40\% of these galaxies. 
Visual inspection further reveals merger signatures—such as double-peaked light distributions—in roughly 40\% of these galaxies, supporting minor mergers as an important, though not dominant, mechanism for disturbed kinematics in centrals.

Finally, we test for AGN-driven feedback using the BPT diagnostic diagram (Fig.~\ref{fig:BPT}), with emission-line ratios derived via \texttt{pyPipe3D} ($\S$\ref{MaNGA_catalogs}). Only 3\% of the sample shows clear AGN outflow signatures, indicating that AGN feedback plays a negligible role in producing the observed disturbances.\\

\underline{\bf{Case 3.}}\\
{\bf{Galaxies with symmetric stellar velocity maps and asymmetric gas velocity maps}} are interpreted as in the initial stage or mild case of {\it{i)}} ram pressure stripping if without companion, {\it{ii)}} initial stage of tidal stripping if in a close pair or {\it{iii)}} with ongoing outflows if the integrated H$\alpha$ emission shows a double Gaussian component.
Galaxies showing asymmetries only in their gaseous kinematic maps represent 13.0\% of the full valid sample (6,680), spanning the entire SFR–$M_\star$ plane (panel f of Fig.~\ref{fig:Mstar_SFR}). BPT diagnostics indicate that their ionized gas is primarily excited by star formation. Since their stellar kinematics remain symmetric, the gas disturbances likely trace recent events, consistent with the shorter response timescale of gas compared to stars.

Among satellites, this class makes up 12.9\% of the subsample. Using the companion selection described above, we find that 51.0$_{-7}^{+5}$\% have a clear companion, while the remaining do not, suggesting that the majority are undergoing (or recently underwent) tidal interactions.
For both centrals and satellites, we analyze the contributions from AGN outflows (see Appendix~\ref{Appendix_C}). 
Stacked spectra reveal a possible broad H$\alpha$ component in AGN hosts, but non-AGN satellites show emission lines well fit by a single Gaussian (Fig.~\ref{fig:stacked}), ruling out significant outflows. 
Thus, gaseous asymmetries are attributed to tidal stripping in systems with companions and to ram-pressure stripping otherwise. 
For star-forming satellites (sSFR $>10^{-11}\mathrm{yr}^{-1}$), 53$_{-7}^{+5}$\% have close companions and are linked to tidal interactions. The remaining $\sim$47\% are strong candidates for “jellyfish” galaxies undergoing ram-pressure stripping, though MaNGA’s FoV limits detection of extended gas tails.
For quenched satellites (sSFR $<10^{-11}\mathrm{yr}^{-1}$), AGN-driven outflows account for 6.2\%, tidal interactions dominate (49.1$_{-5}^{+5}$\%), and ram pressure has a minor contribution of less than 20\%.

Among central galaxies, 13.0\% show kinematic disturbances in the gaseous component only. Since ram-pressure stripping is ineffective at halo centers, these asymmetries are unlikely to result from this process.
We find that 22.4$_{-5}^{+4}$\% have a close companion, indicating that ongoing interactions are not the dominant drivers, but most likely remaining effects of past minor mergers. 
Only 5.3\% of these centrals show clear AGN activity.

\underline{\bf{Case 4.}}\\
{\bf{Gas-rich galaxies with stellar velocity asymmetry}} are interpreted as an indication of minor merger.
We define gas-rich galaxies as those with a cold gas fraction\footnote{Here, the cold gas fraction is defined as the ratio between the HI gas mass and the stellar mass ($M_{\rm{HI}}/M_\star$).} $>10$\% and with more than 70\% of valid spaxels in the gaseous component. To trace their distribution in the sSFR–$M_\star$ plane, we use HI masses from the HI-MaNGA catalog ($\S$\ref{MaNGA_catalogs}), matching $\sim$65\% of the full sample and $\sim$71\% of satellites. As shown previously \citep[e.g.,][]{Brown+17}, the cold gas mass fraction declines with sSFR. Consistently, roughly all gas-rich galaxies in our sample lie above the quiescent threshold and cluster above the MS, extending to M$_\star \sim 10^{11}~$M$_\odot$. 
Centrals dominate, comprising 73\% of this population.

Within this group, only 27.0$_{-4}^{+3}$\% host a close companion, while 58\% show no gas asymmetry despite high SNR. This suggests that any disruptive event occurred in the past, with gas kinematics relaxing faster than the stellar component \citep[see also][]{Lotz2008}. We interpret these cases (i.e., Case 4 galaxies) as signatures of past minor mergers (no companion) or tidal disruption (close companion).

\underline{\bf{Case 5.}}\\
{\bf{Gas-poor galaxies with stellar velocity asymmetry}} are interpreted as an indication of dry merger.
We define gas-poor galaxies as those with a cold gas fraction $<10$\% and fewer than 70\% of valid spaxels within 1R$_{\rm eff}$ in the gaseous component. The HI gas mass fraction is measured as in Case 4. These galaxies all lie below the quiescent threshold, represent $26.4$\% of the overall sample, and are concentrated at M$_\star \gtrsim 10^{11}$~M$_\odot$ (Fig.~\ref{fig:satellite_fraction_2}). Centrals dominate, accounting for 58\% of this group. 
Our companion search indicates that 31.1$_{-7}^{+6}$\% of the galaxies host a close neighbor. However, only in 6.3$_{-1}^{+3}$\% of cases is the companion sufficiently close ($\leq$50~kpc) and has a low enough line-of-sight velocity difference ($\leq$250~km/s) to suggest an ongoing merger.
For the remaining galaxies with companions ($\sim$25\%), the observed disturbances are likely caused by tidal interactions. 
The kinematic disturbances in galaxies without a detected companion may result from recent interactions or dry mergers within the last 2 Gyr. 
Together, this suggests that both recent and ongoing interactions—ranging from tidal encounters to minor or major mergers—are the primary drivers of the stellar kinematic disturbances in quenched, gas-poor galaxies as we further discuss in $\S$\ref{discussion:results}.

\underline{\bf{Case 6.}}\\
{\bf{Galaxies with symmetric stellar velocity maps, no ionized gas and low HI gas mass fraction}} are interpreted as {\it{i)}} the end product of starvation/strangulation if they are satellites and {\it{ii)}} feedback quenching maintenance if they are centrals.
Galaxies with symmetric stellar velocity maps, no ionized gas (<20\% of valid spaxels), and low HI content are the most represented class in the quiescent region, making up 49\% of its population. This group includes 58\% of quiescent satellites and 35\% of all satellites (Fig.~\ref{fig:satellite_fraction_2}). We note that the absence of stellar asymmetries is not due to poor data quality, as this region shows the highest continuum SNR (Fig.~\ref{fig:3plots}), confirming the robustness of this result.

A comparable fraction (43\%) is found among quiescent centrals, implying either that environment has little impact on stellar kinematics or that different quenching mechanisms converge to similar end states. In the latter case, quenching could arise from processes that leave stellar structures largely intact—such as starvation, strangulation, or repeated AGN feedback cycles \citep[e.g.,][]{Dekel+06, Baxter+25} as we discuss in $\S$\ref{discussion:results}.

We summarize the main properties of all analyzed cases in Table~\ref{tab:summary}.
\begin{table*}[htbp!]
\begin{center}
\caption{Summary of the cases discussed in this work.}
\label{tab:summary}
    \begin{tabular}{cccccccc}
     & Star forming & Stellar & Gaseous & \multirow{2}{*}{Gas rich} & \multirow{2}{*}{Gas poor}  & Possible & Percentage\\
     & or quenched? & asymmetry & asymmetry & & & end-product & of the full sample
     \\

\midrule
Case 1 & SF & \ding{55} & \ding{55} & -- & -- & Secular evolution & 26.7\% \\[0.15cm]

\multirow{2}{*}{Case 2}
& \multirow{2}{*}{Both}
& \multirow{2}{*}{\ding{51}}
& \multirow{2}{*}{\ding{51}}
& \multirow{2}{*}{--}
& \multirow{2}{*}{--}
& RPS, tidal interactions,
& \multirow{4}{*}{$
\begin{array}{c}
\mid \\
14.5\% \\
\mid
\end{array}
$}
\\
& & & & & & mergers, harassment \\[0.15cm]

\multirow{2}{*}{Case 3}
& \multirow{2}{*}{Both} & \multirow{2}{*}{\ding{55}} & \multirow{2}{*}{\ding{51}}  & \multirow{2}{*}{--} & \multirow{2}{*}{--} & RPS, tidal interactions,\\
& & & & & & AGN \\[0.15cm]

Case 4 & Both & \ding{51} & --  & \ding{51} & \ding{55} & Minor merger & 11.0\% \\[0.15cm]

Case 5 & Both & \ding{55} & --  & \ding{55} & \ding{51} & Dry merger & 8.7\% \\[0.15cm]

\multirow{2}{*}{Case 6}
& \multirow{2}{*}{Both} & \multirow{2}{*}{\ding{55}} & \multirow{2}{*}{\ding{55}}  & \multirow{2}{*}{\ding{55}} & \multirow{2}{*}{\ding{51}} & Starvation, quenching &\multirow{2}{*}{28.1\%}\\
& & & & & & maintenance \\

\hline
\end{tabular}\\
\vspace{0.1cm}
\end{center}
    \footnotesize \bf{Notes:} \normalfont `--' means that the feature is indifferent for the corresponding case. The total percentage is not equal to 100\% as 11.0\% of the analyzed galaxies did not fit any of the cases discussed. Such cases are displayed as `inconclusive' in Fig.~\ref{fig:Relative_Presence}. 
\end{table*}

\subsection{Occurrence of different quenching processes}

\begin{figure*}[htbp!]
  \centering
\includegraphics[width=0.9\textwidth]
{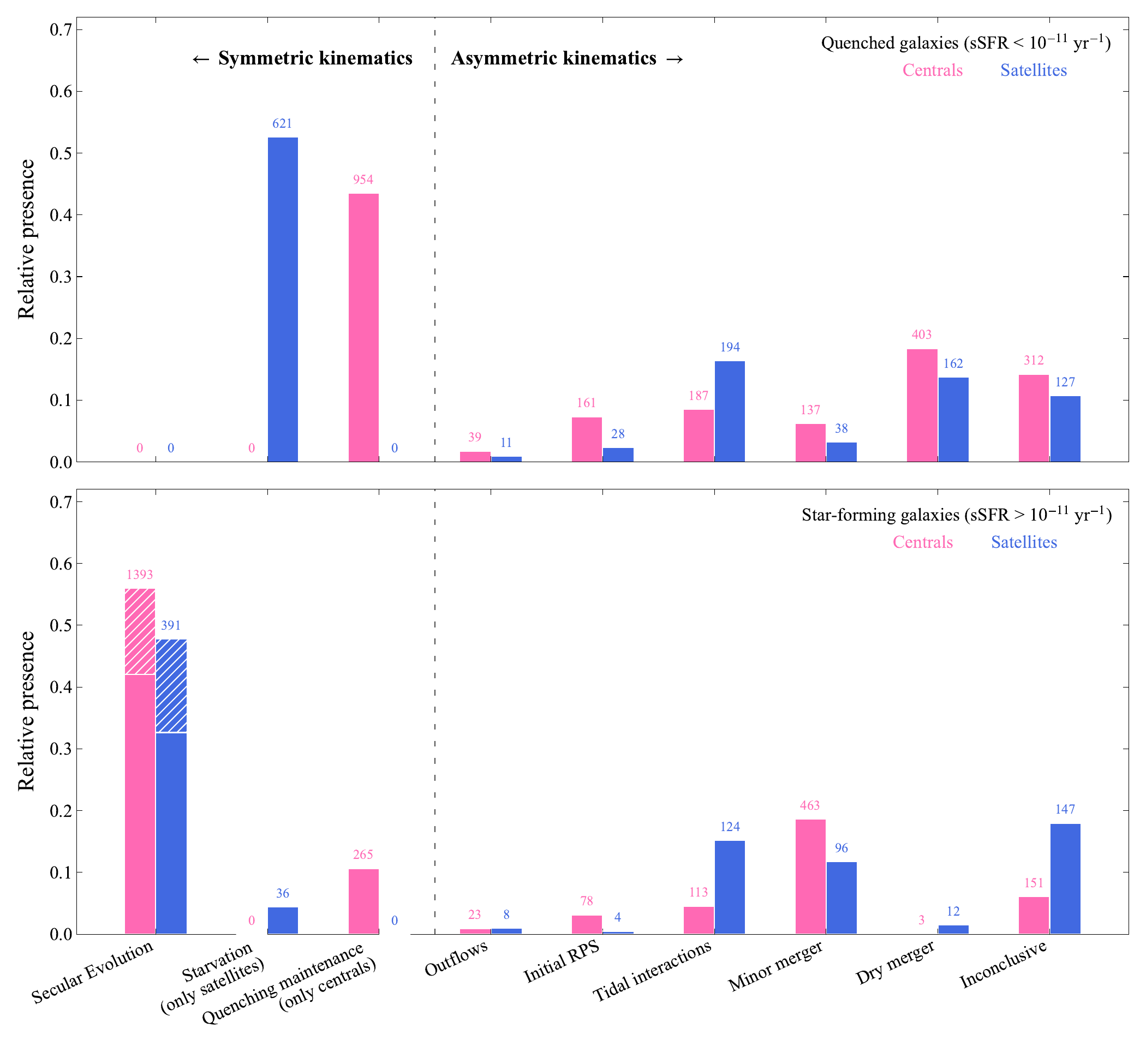}
    \caption{
    The relative presence of the different quenching mechanisms we analyze in this paper.
    In the \textit{top panel}, all values shown are relative to our subset of quenched galaxies (sSFR < $10^{-11}$~yr$^{-1}$) with low gas fractions ($M_{\rm{HI}}/M_\star$<0.2) (\textit{top~panel}). The \textit{bottom~panel} corresponds to the quenching mechanisms hinted in our star-forming subsample.
    Galaxies not meeting our classification criteria are labeled as `inconclusive'.
    We also complement the count of galaxies evolving through secular evolution (hatched bars in the bottom panel) with objects that have symmetric gaseous kinematics but whose stellar maps do not have sufficiently high SNR to be classified as symmetric in that component.}
    \label{fig:Relative_Presence}
\end{figure*}

In Fig.~\ref{fig:Relative_Presence}, we present the relative occurrence of the different quenching features analyzed case-by-case in this work (see $\S$\ref{obs_results}).
The \textit{top~panel} of Fig.~\ref{fig:Relative_Presence} shows the subset of galaxies below our quenching threshold (sSFR < $10^{-11}$~yr$^{-1}$), which accounts for 50.5\% of our sample.
As discussed throughout $\S$\ref{obs_results}, the majority of the galaxies we analyze here show no signs of kinematic asymmetries (in either gaseous or stellar components).
Following our definition of Case 3 in $\S$\ref{obs_results}, galaxies with symmetric stellar velocity maps, no significant ionized gas content, and low HI gas mass fractions are interpreted as the end products of starvation (for satellites) or feedback-maintenance quenching (for centrals).
This is the case for over 50\% of the quenched galaxies analyzed in this work (see Fig.~\ref{fig:Relative_Presence}).

Based on our statistically robust sample, the main result of this work suggests that whatever process drove down the star formation in local quenched satellites either did not significantly disturb their kinematic structure or occurred over 3 Gyr ago, leaving sufficient time for kinematics to relax—particularly the stellar component.
In the first case, the most plausible explanation is that the quenching of satellites is governed by a starvation-driven scenario, which is expected to be much less disruptive to galaxy kinematics compared to scenarios driven by mergers and/or rapid gas stripping \cite[e.g.,][]{Dekel+06, Peng+15, Poggianti+17, Baxter+25}.
Alternatively, this local quenched satellite population could represent the end product of repeated tidal interactions and/or intense gas stripping episodes that occurred more than 3~Gyr ago, providing sufficient time for the stellar kinematics to relax and for the system to re-establish a rotation-supported configuration \cite[e.g.,][]{Smith+22}. 
In this context, gas stripping phenomena (e.g., ram pressure, tidal or viscous stripping) may have played an even more significant role compared to starvation, but their dynamical signatures might no longer be observable \cite[e.g.,][]{Vollmer+01}.
For central galaxies, our results suggest feedback quenching maintenance as a clear mechanism for keeping this population quenched over long time-scales ($>$3 Gyr). 
In this proposed scenario, the galaxy is prevented from cooling its hot gaseous halo and re-igniting star formation, thereby maintaining its quenched state previously triggered by either powerful feedback outbursts, major mergers, or rapid gas consumption \cite[e.g.,][]{Croton+06, Fabian+12}.

As expected based on Fig.~\ref{fig:Mstar_SFR}, dry mergers, particularly in central galaxies, contribute a relevant ($\sim$20\%) portion to the observed kinematics of local quenched galaxies. When dry mergers and minor mergers are considered together, their overall contribution makes up to 27\% of the analyzed cases.
Therefore, merging signatures in the observed kinematics of quenched central galaxies represent the second most common case in our sample.
In satellites, tidal interactions driven by flybys with relatively close companions ($\leq $200~kpc) seem to contribute more significantly ($\sim$20\%) to the overall quenching picture.

Among the star-forming population, passively evolving processes through secular evolution dominate the overall kinematics in both centrals and satellites. This is again consistent with regular kinematics showing no signs of asymmetries.
Since these galaxies are star-forming, high-SNR ($\gtrsim$30) gaseous maps are also available, meaning that any present asymmetry could be easily detected with \textsc{Kinemetry}.
We also complement the count of galaxies evolving through secular evolution with objects that have symmetric gaseous kinematics but whose stellar maps do not have sufficiently high SNR to be classified as symmetric in that component. This is the case for many star-forming galaxies with high SNR in the H$\alpha$ component but very low stellar continuum. For both centrals and satellites, these objects account for $\sim$13\%, as we highlight with hatching in the \textit{bottom panel} of Fig.~\ref{fig:Relative_Presence}.

\section{Discussion and conclusions}
\label{sec5}

\subsection{Linking galaxy quenching, kinematic evolution, and morphological transformation} \label{discussion:results}
As we have shown in Case 5, gas-poor galaxies with asymmetric kinematics are likely the result of both past and ongoing interactions.
To test this hypothesis, we conduct a mass-matched sampling using 1,000 iterations of balanced subsamples of quenched galaxies above and below the asymmetry threshold ($I_{asym,\star}=0.05$).
We find that asymmetric galaxies in the stellar component have consistently more early-type morphologies.
For galaxies with $M_\star \geq 10^{9.5}~M_\odot$, the T-Type difference between symmetric and asymmetric systems reaches $\Delta \mathrm{T\text{-}Type} \approx -1.0$, indicating that asymmetric quenched galaxies are preferentially early-type (>$15\sigma$ significance) compared to their symmetric counterpart.

We find a striking environmental dependence in the size-asymmetry relation among quenched galaxies. Central galaxies exhibit particularly strong enhancement, with asymmetric systems showing $R_{\rm eff}$ values 32.5\% larger than their symmetric counterparts.
In the highest mass bin ($M_\star \geq 10^{11}~M_\odot$), $\overline{R_{\rm eff}} = 6.43 \pm 0.10$~kpc versus $5.46 \pm 0.10$~kpc for symmetric centrals ($9.9\sigma$ significance).
Satellites follow a comparable pattern, though less prominent, with the same $M_\star \sim 10^{10.5}~M_\odot$ threshold distinguishing size regimes. Above this mass bin, asymmetric satellites are systematically larger, reaching $\overline{R_{\rm eff}} = 5.59 \pm 0.22$~kpc compared to $4.98 \pm 0.17$~kpc for their symmetric counterparts ($3.2\sigma$ significance).
This size enhancement in centrals supports the merger-driven growth scenario, consistent with models of merger-induced stellar puffing \citep[e.g.,][]{Lackner+12}.
Stellar mass further modulates this effect. Below $M_\star=10^{10.5}~\rm M_\odot$, symmetric and asymmetric systems have comparable $R_{\rm eff}$ regardless of environment. Above this threshold, the size offset grows with stellar mass, and the most massive asymmetric galaxies are significantly more extended. This trend suggests that size-growth mechanisms—minor mergers or cold gas accretion—are increasingly efficient at high masses, particularly for central galaxies, which dominate the massive end of the population \citep[e.g.,][]{Naab+09}.

To further investigate the morphological transformation of quenched symmetric galaxies, we use the same mass-matched approach described above. Morphologies are similar across both sub-samples (all nearly spheroidal, i.e., T-Type$\sim$0.0), but centrals consistently show larger sizes. 
Symmetric centrals have $\overline{R_{\rm eff}} = 3.92 \pm 0.06$~kpc, whereas symmetric satellites overall display much more compact structures, $2.83 \pm 0.05$~kpc (16.8$\sigma$ significance). 
This offset strongly suggests that centrals and satellites are the products of distinct evolutionary processes as we further discuss in $\S$\ref{discussion:general}.

\begin{figure}[htbp!]
  \centering
\includegraphics[width=0.49\textwidth]
{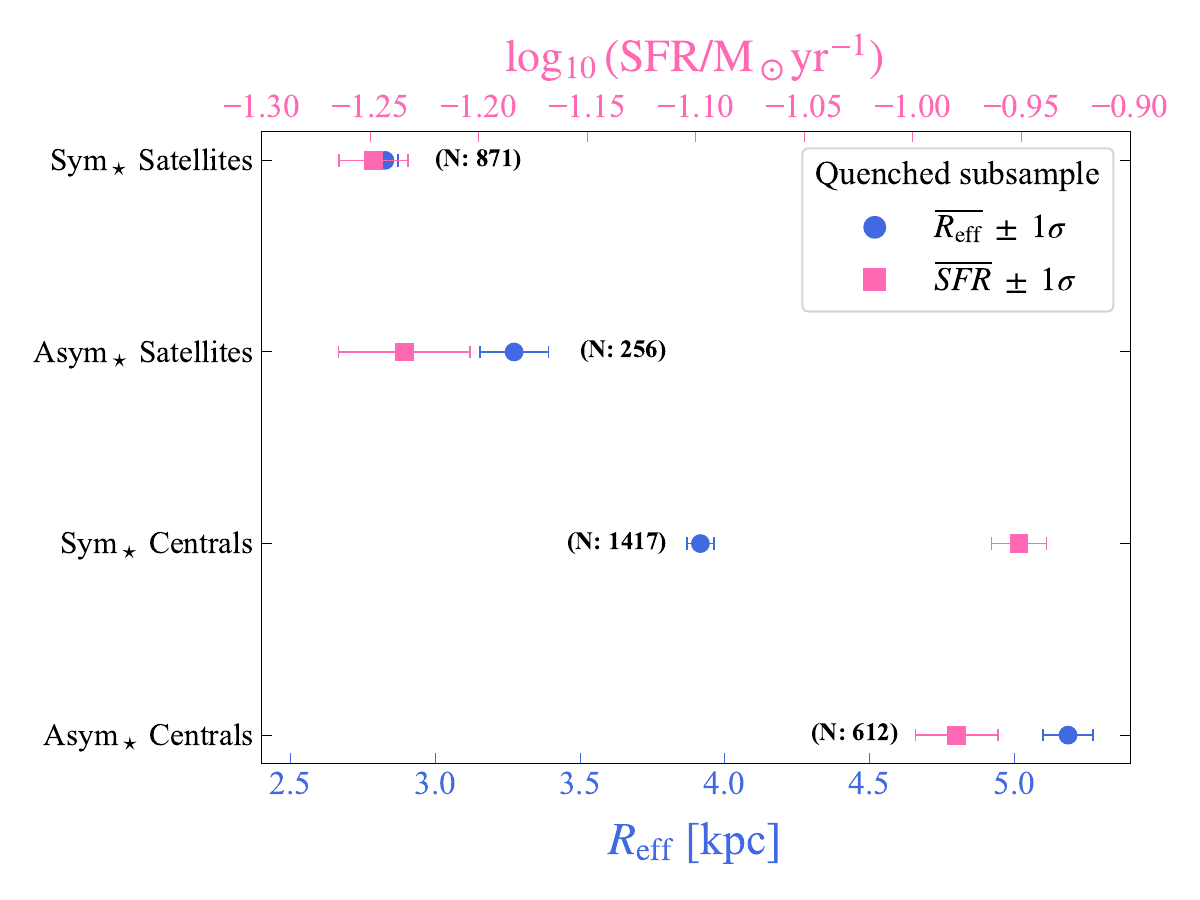}
    \caption{Schematic summary of the SFR- and size-dependent results presented in Cases 5 and 6. Symmetric satellites exhibit the lowest SFRs and the most compact structures among quenched galaxies, suggesting that these objects are the end product of an intense period of mass removal, either via gas striping itself or tidal interactions.}
    \label{fig:Reff_SFR}
\end{figure}

We summarize these results in Fig.~\ref{fig:Reff_SFR}, showing average SFRs and sizes for quenched sub-samples in Cases 5 and 6. Two signatures emerge: (1) gas removal truncates star formation \citep[e.g.,][]{Peng+15}, and (2) tidal forces strip outer disks, leaving compact bulges \citep[e.g.,][]{Lackner+12}. 
The highly significant SFR offset between symmetric centrals and satellites indicates that environmental processes enhance the quenching \citep[e.g.,][]{Poggianti+17, Poggianti+19, Bluck+20, Baxter+25, Lim+25}.
For satellites, the stripping together with the absence of gas replenishment—due to starvation cutting off hot halo gas and preventing cosmological accretion—locks them onto an irreversible quenching pathway.
These compact remnants remain perpetually quiescent, with symmetric stellar kinematics reflecting both disk stripping and the lack of subsequent disturbances.

Our results on size differences between quenched symmetric and asymmetric centrals ($12.3\sigma$) and satellites ($3.4\sigma$) strongly suggest that symmetric quenched galaxies in the local Universe have undergone central-compaction-like quenching mechanisms.
These processes tend to quench galaxies rapidly and drive the size differences shown in Fig.~\ref{fig:Reff_SFR}.
These findings are broadly consistent with recent simulation-based work by \cite{Ni+25} using ASTRID data, which supports that dissipative processes funnel gas into the central regions, triggering starbursts that lead to rapid quenching and more compact morphologies compared to star-forming galaxies. 
In contrast, galaxies that quench over longer timescales typically exhibit larger sizes and older central regions, characteristics we identify here in asymmetric galaxies.

As presented above, the observed offset in the overall size distribution of quenched symmetric centrals and satellites strongly suggests that these two populations represent the end product of distinct evolutionary paths. 
Satellites could result from strong gas stripping or repeated tidal interactions more than 3~Gyr ago, leaving time for the stellar component to relax. Such processes—ram-pressure, tidal, or viscous stripping—may erase their dynamical signatures but still reduce satellite mass and size. Centrals, being more massive, are less prone to stripping and instead deplete gas passively or grow through accretion \citep[e.g.,][]{Cortese+19}.

\subsection{The quenching pathway}\label{discussion:general}

Based on the results we presented in $\S$\ref{results} along with the discussion above, we propose a coherent evolutionary pathway for a typical low- to intermediate-mass quenched satellite:
a satellite galaxy falls into a group or cluster halo. It experiences an episode of gas stripping (ram-pressure and/or tidal interactions), which removes its HI reservoir and may slightly disturb its kinematics.
Repeated episodes of gas stripping and tidal interactions remove the outer stellar disk, truncating the galaxy and increasing its metallicity.
The hot halo gas is stripped upon infall (starvation), severing the supply of future cold gas. 
The satellite consumes its remaining gas and quenches. 
Over the next few Gyr, the kinematic signatures of the initial disturbance dynamically relax, leaving a compact, symmetric, metal-rich, and passive galaxy—the most common type of quenched satellite we observe in this work.
This pathway explains the kinematic regularity, the low SFR, the high metallicity, and the small size simultaneously, and highlights starvation as the crucial maintenance mechanism that makes quenching irreversible in satellites.

Our results are also consistently aligned with recent work by \cite{Leung+25} on post-starburst galaxies in MaNGA. Their findings support mergers or galaxy-galaxy interactions as the primary mechanism that causes quenching, highlighting that large scale feedback from a starburst or a central AGN play a lesser role.
Our findings suggest that for satellite galaxies, a combination of rapid gas stripping followed by long-term starvation can adequately explain the majority of the quenched, yet kinematically undisturbed, satellites observed today.
Furthermore, we highlight that our findings are particularly consistent with \cite{Cortese+19}. Using both observations (SAMI survey) and simulations (EAGLE), the authors suggested that satellites undergo little structural change before and during their quenching phase—the main conclusion of this work. Also, at fixed stellar mass, their satellite sample showed higher passive fractions than centrals, suggesting that environment is directly quenching their star formation, as we also find here.

When analyzing the population of star-forming satellites, Fig.~\ref{fig:Relative_Presence} shows a negligible contribution from starvation. We emphasize that this trend is expected and provides further support for our main conclusion: starvation is ubiquitous and plays a crucial role in quenching satellites across all stellar masses. By definition, one of the most characteristic effects of starvation is low gas content, which represents the key distinction between the starvation and secular evolution categories presented in Fig.~\ref{fig:Relative_Presence}.
The effects of starvation (e.g., low gas content and unperturbed kinematics), however, become measurable only a few Gyr after the galaxy has fallen into its host halo. Prior to this late stage, such galaxies are correctly classified as evolving through secular evolution, as clearly shown in the \textit{bottom panel} of Fig.~\ref{fig:Relative_Presence}.

We also note that previous literature \citep[e.g.,][]{Shapiro+08, Krajnovic+11,Bloom+17,Feng+22} has established that kinematic asymmetry is a reliable proxy not only for galaxy-galaxy interactions but also for the $V/\sigma$ ratio. To verify whether the results presented in $\S$\ref{obs_results} are primarily driven by environmental effects rather than by variations in $V/\sigma$, we investigated the correlation between $V/\sigma$ and the asymmetry parameter in our sample. Overall, we find no significant correlations in either the stellar or gaseous kinematic maps.
To investigate this further, we split our sample into stellar mass bins. The results remain largely unchanged; only in the high-mass end ($M_\star \geq 10^{10.8}~M_\odot$) the Pearson coefficient suggests a correlation between the stellar asymmetry parameter and the stellar $V/\sigma$ within 1$R_{\mathrm{eff}}$. 
In this high-mass population, we note that the observed fraction of dry mergers we show in $\S$\ref{obs_results} may be influenced by variations in $V/\sigma$.
For the gaseous component, we do not find a clear correlation in any mass bin, providing additional support for our main conclusions.
We further investigate the connection between stellar kinematic asymmetry and the galaxy's dynamical state by separating our sample into fast and slow rotators using the spin parameter $\lambda_{R_e}$. Following the established $\lambda_{R_e}$--$\epsilon$ classification\footnote{Fast rotators have $\lambda_{R_e} > 0.31 \cdot \sqrt{\epsilon}$, where $\epsilon$ is the galaxy's ellipticity, following \citet{Emsellem+11}.} \citep[e.g.,][]{Emsellem+11}, we find that within our sample of galaxies with $\lambda_{R_e}$ measurements (2,897) available in \citet{Graham+19}, 241 (8.3\%) are classified as slow rotators and 2,656 (91.7\%) as fast rotators. The mean stellar kinematic asymmetry for slow rotators is significantly higher ($\langle I_{\rm asym} \rangle_{\mathrm{slow}} = 0.147 \pm 0.089$) than for fast rotators ($\langle I_{\rm asym} \rangle_{\mathrm{fast}} = 0.047 \pm 0.044$).
This clear dichotomy supports the interpretation that the kinematic asymmetries reported in this work are more fundamentally tied to interactions. Indeed, slow rotators are thought to form through violent interactions during mergers—particularly dry, intermediate-to-major mergers—which can both build stellar mass and produce the observed kinematic asymmetries \citep[e.g.,][]{Hoffman+10, Bois+11, Naab+14, Lagos+22}. Observationally, slow rotators have been linked to merger histories and signs of kinematic disturbance \citep[e.g.,][]{Ene+18, Krajnovic+20, Loubser+22}. The enhanced asymmetry we find in slow rotators therefore reinforces our conclusion that environmental effects—specifically merger-driven growth—are the dominant driver of the observed kinematic asymmetries in quenched central galaxies.

\subsection{Summary}\label{discussion:general}
This work presents a novel, case-by-case analysis of kinematic asymmetries in a robust sample of 6,680 nearby galaxies ($z < 0.15$) to disentangle the relative prevalence of different quenching mechanisms. 
For the first time, we systematically evaluate a comprehensive suite of processes—including AGN activity, ram pressure stripping, tidal interactions, minor and dry mergers, secular evolution, galaxy harassment, starvation (in satellites), and feedback quenching maintenance (in centrals)—as potential drivers of the observed kinematics, treating them as equally probable a priori to provide an unbiased assessment of their roles in galaxy evolution.
We summarize below the main conclusions of this work.
\begin{enumerate}[label=\roman*.]

    \item 
    The absence of kinematically asymmetric satellites in the so-called “satellite region” of the SFR-M$_\star$ plane strongly suggests that whatever is driving down the star formation in local ($z\sim0$), star-forming satellites does not significantly perturb either the stellar or gaseous structures.\\

    \item Our analysis reveals that the most common quenching mechanisms leave minimal imprints on stellar kinematics by the time a galaxy is fully quenched. This kinematic regularity itself is a crucial diagnostic, pointing towards slow-acting processes ($\gtrsim$3~Gyr) such as starvation and maintenance feedback over violent, recent events.\\

    \item We establish a clear link between quenching history and galaxy size. Quenched satellites are significantly more compact than quenched centrals at fixed mass, a signature consistent with outer disk stripping by rapid environmental effects (e.g., gas stripping, tidal interactions). Conversely, the larger sizes of disturbed, quenched centrals support a merger-driven growth scenario. 
    Furthermore, our results suggest that internal processes, likely mediated by AGN feedback over the past 1-3~Gyr that prevented hot halo gas from cooling, lead to sustained quenching maintenance in this central population, in contrast to satellites.\\

    \item The dominant pathway for quenching satellite galaxies is environmentally driven. A sequence of rapid gas stripping and tidal interactions followed by long-term starvation effectively explains the observed population of compact, kinematically regular, and fully quenched satellites, as revealed by our mass-matching analysis.\\        
 \end{enumerate}

\begin{acknowledgements}
We would like to thank the anonymous referee for a productive report with comments and suggestions that improved this paper. NdI gratefully acknowledges the IMPRS program and ESO for the support and funding of his PhD.
PP, IM and VT acknowledge financial support from the European Research Council (ERC) under the European Union’s Horizon Europe research and innovation programme ERC CoG CLEVeR (Grant agreement No. 101045437). YMB acknowledges support from UK Research and Innovation through a Future Leaders Fellowship (grant agreement MR/X035166/1) and financial support from the Swiss National Science Foundation (SNSF) under project 200021\_213076.
\end{acknowledgements}

\section*{Data availability}
The MaNGA data underlying this article are publicly available and can be accessed from the SDSS data base: \href{https://www.sdss.org/dr15/manga/manga-data/data-access/}{https://www.sdss.org/dr15/manga/manga-data/data-access/}, or through \textsc{Marvin} at \href{https://dr17.sdss.org/marvin/}{https://dr17.sdss.org/marvin/}.

\bibliographystyle{aa} 
\bibliography{lib.bib}

@ARTICLE{Bagge+24,
       author = {{Bagge}, R.~S. and {Foster}, C. and {D'Eugenio}, F. and {Battisti}, A. and {Bellstedt}, S. and {Derkenne}, C. and {Vaughan}, S. and {Mendel}, T. and {Barsanti}, S. and {Harborne}, K.~E. and {Croom}, S.~M. and {Bland-Hawthorn}, J. and {Grasha}, K. and {Lagos}, C.~D.~P. and {Sweet}, S.~M. and {Mailvaganam}, A. and {Mukherjee}, T. and {Valenzuela}, L.~M. and {van de Sande}, J. and {Wisnioski}, E. and {Zafar}, T.},
        title = "{The MAGPI survey: using kinematic asymmetries in stars and gas to dissect drivers of galaxy dynamical evolution}",
      journal = {\mnras},
         year = 2024,
        month = jul,
       volume = {531},
       number = {3},
        pages = {3011-3022},
          doi = {10.1093/mnras/stae1341},
archivePrefix = {arXiv},
       eprint = {2405.11292},
 primaryClass = {astro-ph.GA},
       adsurl = {https://ui.adsabs.harvard.edu/abs/2024MNRAS.531.3011B}
}

@ARTICLE{Graham+19,
       author = {{Graham}, Mark T. and {Cappellari}, Michele and {Bershady}, Matthew A. and {Drory}, Niv},
        title = "{SDSS-IV MaNGA: New benchmark for the connection between stellar angular momentum and environment: a study of about 900 groups/clusters}",
      journal = {arXiv e-prints},
         year = 2019,
        month = oct,
          eid = {arXiv:1910.05139},
        pages = {arXiv:1910.05139},
          doi = {10.48550/arXiv.1910.05139},
archivePrefix = {arXiv},
       eprint = {1910.05139},
 primaryClass = {astro-ph.GA},
       adsurl = {https://ui.adsabs.harvard.edu/abs/2019arXiv191005139G}
}

@ARTICLE{Emsellem+11,
       author = {{Emsellem}, Eric and {Cappellari}, Michele and {Krajnovi{\'c}}, Davor and {Alatalo}, Katherine and {Blitz}, Leo and {Bois}, Maxime and {Bournaud}, Fr{\'e}d{\'e}ric and {Bureau}, Martin and {Davies}, Roger L. and {Davis}, Timothy A. and {de Zeeuw}, P.~T. and {Khochfar}, Sadegh and {Kuntschner}, Harald and {Lablanche}, Pierre-Yves and {McDermid}, Richard M. and {Morganti}, Raffaella and {Naab}, Thorsten and {Oosterloo}, Tom and {Sarzi}, Marc and {Scott}, Nicholas and {Serra}, Paolo and {van de Ven}, Glenn and {Weijmans}, Anne-Marie and {Young}, Lisa M.},
        title = "{The ATLAS$^{3D}$ project - III. A census of the stellar angular momentum within the effective radius of early-type galaxies: unveiling the distribution of fast and slow rotators}",
      journal = {\mnras},
         year = 2011,
        month = jun,
       volume = {414},
       number = {2},
        pages = {888-912},
          doi = {10.1111/j.1365-2966.2011.18496.x},
archivePrefix = {arXiv},
       eprint = {1102.4444},
 primaryClass = {astro-ph.CO},
       adsurl = {https://ui.adsabs.harvard.edu/abs/2011MNRAS.414..888E}
}

@ARTICLE{Poggianti+25,
       author = {{Poggianti}, Bianca M. and {Vulcani}, Benedetta and {Tomicic}, Neven and {Moretti}, Alessia and {Gullieuszik}, Marco and {Bacchini}, Cecilia and {Fritz}, Jacopo and {George}, Koshy and {Gitti}, Myriam and {Ignesti}, Alessandro and {Jaff{\'e}}, Yara and {Lassen}, Augusto and {Marasco}, Antonino and {Radovich}, Mario and {Serra}, Paolo and {Smith}, Rory and {Tonnesen}, Stephanie and {Wolter}, Anna},
        title = "{The MUSE view of ram pressure stripped galaxies in clusters: The GASP sample}",
      journal = {\aap},
         year = 2025,
        month = jul,
       volume = {699},
          eid = {A357},
        pages = {A357},
          doi = {10.1051/0004-6361/202554200},
archivePrefix = {arXiv},
       eprint = {2505.21107},
 primaryClass = {astro-ph.GA},
       adsurl = {https://ui.adsabs.harvard.edu/abs/2025A&A...699A.357P}
}

@ARTICLE{Bloom+18,
       author = {{Bloom}, J.~V. and {Croom}, S.~M. and {Bryant}, J.~J. and {Schaefer}, A.~L. and {Bland-Hawthorn}, J. and {Brough}, S. and {Callingham}, J. and {Cortese}, L. and {Federrath}, C. and {Scott}, N. and {van de Sande}, J. and {D'Eugenio}, F. and {Sweet}, S. and {Tonini}, C. and {Allen}, J.~T. and {Goodwin}, M. and {Green}, A.~W. and {Konstantopoulos}, I.~S. and {Lawrence}, J. and {Lorente}, N. and {Medling}, A.~M. and {Owers}, M.~S. and {Richards}, S.~N. and {Sharp}, R.},
        title = "{The SAMI Galaxy Survey: gas content and interaction as the drivers of kinematic asymmetry}",
      journal = {\mnras},
         year = 2018,
        month = may,
       volume = {476},
       number = {2},
        pages = {2339-2351},
          doi = {10.1093/mnras/sty273},
archivePrefix = {arXiv},
       eprint = {1801.06628},
 primaryClass = {astro-ph.GA},
       adsurl = {https://ui.adsabs.harvard.edu/abs/2018MNRAS.476.2339B}
}

@ARTICLE{Shapiro+08,
       author = {{Shapiro}, Kristen L. and {Genzel}, Reinhard and {F{\"o}rster Schreiber}, Natascha M. and {Tacconi}, Linda J. and {Bouch{\'e}}, Nicolas and {Cresci}, Giovanni and {Davies}, Richard and {Eisenhauer}, Frank and {Johansson}, Peter H. and {Krajnovi{\'c}}, Davor and {Lutz}, Dieter and {Naab}, Thorsten and {Arimoto}, Nobuo and {Arribas}, Santiago and {Cimatti}, Andrea and {Colina}, Luis and {Daddi}, Emanuele and {Daigle}, Olivier and {Erb}, Dawn and {Hernandez}, Olivier and {Kong}, Xu and {Mignoli}, Marco and {Onodera}, Masato and {Renzini}, Alvio and {Shapley}, Alice and {Steidel}, Charles},
        title = "{Kinemetry of SINS High-Redshift Star-Forming Galaxies: Distinguishing Rotating Disks from Major Mergers}",
      journal = {\apj},
         year = 2008,
        month = jul,
       volume = {682},
       number = {1},
        pages = {231-251},
          doi = {10.1086/587133},
archivePrefix = {arXiv},
       eprint = {0802.0879},
 primaryClass = {astro-ph},
       adsurl = {https://ui.adsabs.harvard.edu/abs/2008ApJ...682..231S}
}

@ARTICLE{Smith+10,
       author = {{Smith}, Russell J. and {Lucey}, John R. and {Hammer}, Derek and {Hornschemeier}, Ann E. and {Carter}, David and {Hudson}, Michael J. and {Marzke}, Ronald O. and {Mouhcine}, Mustapha and {Eftekharzadeh}, Sareh and {James}, Phil and {Khosroshahi}, Habib and {Kourkchi}, Ehsan and {Karick}, Arna},
        title = "{Ultraviolet tails and trails in cluster galaxies: a sample of candidate gaseous stripping events in Coma}",
      journal = {\mnras},
         year = 2010,
        month = nov,
       volume = {408},
       number = {3},
        pages = {1417-1432},
          doi = {10.1111/j.1365-2966.2010.17253.x},
archivePrefix = {arXiv},
       eprint = {1006.4867},
 primaryClass = {astro-ph.CO},
       adsurl = {https://ui.adsabs.harvard.edu/abs/2010MNRAS.408.1417S}
}

@ARTICLE{Bloom+17,
       author = {{Bloom}, J.~V. and {Fogarty}, L.~M.~R. and {Croom}, S.~M. and {Schaefer}, A. and {Bryant}, J.~J. and {Cortese}, L. and {Richards}, S. and {Bland-Hawthorn}, J. and {Ho}, I. -T. and {Scott}, N. and {Goldstein}, G. and {Medling}, A. and {Brough}, S. and {Sweet}, S.~M. and {Cecil}, G. and {L{\'o}pez-S{\'a}nchez}, A. and {Glazebrook}, K. and {Parker}, Q. and {Allen}, J.~T. and {Goodwin}, M. and {Green}, A.~W. and {Konstantopoulos}, I.~S. and {Lawrence}, J.~S. and {Lorente}, N. and {Owers}, M.~S. and {Sharp}, R.},
        title = "{The SAMI Galaxy Survey: asymmetry in gas kinematics and its links to stellar mass and star formation}",
      journal = {\mnras},
         year = 2017,
        month = feb,
       volume = {465},
       number = {1},
        pages = {123-148},
          doi = {10.1093/mnras/stw2605},
archivePrefix = {arXiv},
       eprint = {1610.02773},
 primaryClass = {astro-ph.GA},
       adsurl = {https://ui.adsabs.harvard.edu/abs/2017MNRAS.465..123B}
}

@ARTICLE{Vulcani+22,
       author = {{Vulcani}, Benedetta and {Poggianti}, Bianca M. and {Smith}, Rory and {Moretti}, Alessia and {Jaff{\'e}}, Yara L. and {Gullieuszik}, Marco and {Fritz}, Jacopo and {Bellhouse}, Callum},
        title = "{The Relevance of Ram Pressure Stripping for the Evolution of Blue Cluster Galaxies as Seen at Optical Wavelengths}",
      journal = {\apj},
         year = 2022,
        month = mar,
       volume = {927},
       number = {1},
          eid = {91},
        pages = {91},
          doi = {10.3847/1538-4357/ac4809},
archivePrefix = {arXiv},
       eprint = {2201.02644},
 primaryClass = {astro-ph.GA},
       adsurl = {https://ui.adsabs.harvard.edu/abs/2022ApJ...927...91V}
}

@ARTICLE{Poggianti+17,
       author = {{Poggianti}, Bianca M. and {Moretti}, Alessia and {Gullieuszik}, Marco and {Fritz}, Jacopo and {Jaff{\'e}}, Yara and {Bettoni}, Daniela and {Fasano}, Giovanni and {Bellhouse}, Callum and {Hau}, George and {Vulcani}, Benedetta and {Biviano}, Andrea and {Omizzolo}, Alessandro and {Paccagnella}, Angela and {D'Onofrio}, Mauro and {Cava}, Antonio and {Sheen}, Y. -K. and {Couch}, Warrick and {Owers}, Matt},
        title = "{GASP. I. Gas Stripping Phenomena in Galaxies with MUSE}",
      journal = {\apj},
         year = 2017,
        month = jul,
       volume = {844},
       number = {1},
          eid = {48},
        pages = {48},
          doi = {10.3847/1538-4357/aa78ed},
archivePrefix = {arXiv},
       eprint = {1704.05086},
 primaryClass = {astro-ph.GA},
       adsurl = {https://ui.adsabs.harvard.edu/abs/2017ApJ...844...48P}
}

@ARTICLE{Leung+25,
       author = {{Leung}, Ho-Hin and {Wild}, Vivienne and {Papathomas}, Michail and {Carnall}, Adam C. and {Chen}, Yanmei},
        title = "{The diverse quenching pathways of post-starburst galaxies in SDSS-IV MaNGA}",
      journal = {\mnras},
         year = 2025,
        month = sep,
          doi = {10.1093/mnras/staf1493},
archivePrefix = {arXiv},
       eprint = {2509.05172},
 primaryClass = {astro-ph.GA},
       adsurl = {https://ui.adsabs.harvard.edu/abs/2025MNRAS.tmp.1452L}
}

@ARTICLE{Ni+25,
       author = {{Ni}, Yueying and {Chen}, Nianyi and {Zhou}, Yihao and {Park}, Minjung and {Yang}, Yanhui and {Di Matteo}, Tiziana and {Bird}, Simeon and {Croft}, Rupert},
        title = "{The Astrid Simulation: Evolution of Black Holes and Galaxies to z = 0.5 and Different Evolution Pathways for Galaxy Quenching}",
      journal = {\apj},
         year = 2025,
        month = sep,
       volume = {990},
       number = {2},
          eid = {120},
        pages = {120},
          doi = {10.3847/1538-4357/adf3a7},
archivePrefix = {arXiv},
       eprint = {2409.10666},
 primaryClass = {astro-ph.GA},
       adsurl = {https://ui.adsabs.harvard.edu/abs/2025ApJ...990..120N}
}

@ARTICLE{Fabian+12,
       author = {{Fabian}, A.~C.},
        title = "{Observational Evidence of Active Galactic Nuclei Feedback}",
      journal = {\araa},
         year = 2012,
        month = sep,
       volume = {50},
        pages = {455-489},
          doi = {10.1146/annurev-astro-081811-125521},
archivePrefix = {arXiv},
       eprint = {1204.4114},
 primaryClass = {astro-ph.CO},
       adsurl = {https://ui.adsabs.harvard.edu/abs/2012ARA&A..50..455F}
}

@ARTICLE{Vollmer+01,
       author = {{Vollmer}, B. and {Cayatte}, V. and {Balkowski}, C. and {Duschl}, W.~J.},
        title = "{Ram Pressure Stripping and Galaxy Orbits: The Case of the Virgo Cluster}",
      journal = {\apj},
         year = 2001,
        month = nov,
       volume = {561},
       number = {2},
        pages = {708-726},
          doi = {10.1086/323368},
archivePrefix = {arXiv},
       eprint = {astro-ph/0107237},
 primaryClass = {astro-ph},
       adsurl = {https://ui.adsabs.harvard.edu/abs/2001ApJ...561..708V}
}

@ARTICLE{Smith+22,
       author = {{Smith}, Rory and {Calder{\'o}n-Castillo}, Paula and {Shin}, Jihye and {Raouf}, Mojtaba and {Ko}, Jongwan},
        title = "{The First Fall is the Hardest: The Importance of Peculiar Galaxy Dynamics at Infall Time for Tidal Stripping Acting at the Centers of Groups and Clusters}",
      journal = {\aj},
         year = 2022,
        month = sep,
       volume = {164},
       number = {3},
          eid = {95},
        pages = {95},
          doi = {10.3847/1538-3881/ac8053},
archivePrefix = {arXiv},
       eprint = {2207.05099},
 primaryClass = {astro-ph.GA},
       adsurl = {https://ui.adsabs.harvard.edu/abs/2022AJ....164...95S}
}

@ARTICLE{Comerford+24,
       author = {{Comerford}, Julia M. and {Nevin}, Rebecca and {Negus}, James and {Barrows}, R. Scott and {Eracleous}, Michael and {M{\"u}ller-S{\'a}nchez}, Francisco and {Roy}, Namrata and {Stemo}, Aaron and {Storchi-Bergmann}, Thaisa and {Wylezalek}, Dominika},
        title = "{An Excess of Active Galactic Nuclei Triggered by Galaxy Mergers in MaNGA Galaxies of Stellar Mass {\ensuremath{\sim}}{}10$^{11}$ M $_{{\ensuremath{\odot}}}$}",
      journal = {\apj},
         year = 2024,
        month = mar,
       volume = {963},
       number = {1},
          eid = {53},
        pages = {53},
          doi = {10.3847/1538-4357/ad1a15},
archivePrefix = {arXiv},
       eprint = {2404.14490},
 primaryClass = {astro-ph.GA},
       adsurl = {https://ui.adsabs.harvard.edu/abs/2024ApJ...963...53C}
}

@ARTICLE{Falcon-Barroso+11,
       author = {{Falc{\'o}n-Barroso}, J. and {S{\'a}nchez-Bl{\'a}zquez}, P. and {Vazdekis}, A. and {Ricciardelli}, E. and {Cardiel}, N. and {Cenarro}, A.~J. and {Gorgas}, J. and {Peletier}, R.~F.},
        title = "{An updated MILES stellar library and stellar population models}",
      journal = {\aap},
         year = 2011,
        month = aug,
       volume = {532},
          eid = {A95},
        pages = {A95},
          doi = {10.1051/0004-6361/201116842},
archivePrefix = {arXiv},
       eprint = {1107.2303},
 primaryClass = {astro-ph.CO},
       adsurl = {https://ui.adsabs.harvard.edu/abs/2011A&A...532A..95F}
}

@ARTICLE{Sanchez-Blaszquez+06,
       author = {{S{\'a}nchez-Bl{\'a}zquez}, P. and {Peletier}, R.~F. and {Jim{\'e}nez-Vicente}, J. and {Cardiel}, N. and {Cenarro}, A.~J. and {Falc{\'o}n-Barroso}, J. and {Gorgas}, J. and {Selam}, S. and {Vazdekis}, A.},
        title = "{Medium-resolution Isaac Newton Telescope library of empirical spectra}",
      journal = {\mnras},
         year = 2006,
        month = sep,
       volume = {371},
       number = {2},
        pages = {703-718},
          doi = {10.1111/j.1365-2966.2006.10699.x},
archivePrefix = {arXiv},
       eprint = {astro-ph/0607009},
 primaryClass = {astro-ph},
       adsurl = {https://ui.adsabs.harvard.edu/abs/2006MNRAS.371..703S}
}

@ARTICLE{Cappellari+04,
       author = {{Cappellari}, Michele and {Emsellem}, Eric},
        title = "{Parametric Recovery of Line-of-Sight Velocity Distributions from Absorption-Line Spectra of Galaxies via Penalized Likelihood}",
      journal = {\pasp},
         year = 2004,
        month = feb,
       volume = {116},
       number = {816},
        pages = {138-147},
          doi = {10.1086/381875},
archivePrefix = {arXiv},
       eprint = {astro-ph/0312201},
 primaryClass = {astro-ph},
       adsurl = {https://ui.adsabs.harvard.edu/abs/2004PASP..116..138C}
}

@ARTICLE{Kewley+06,
       author = {{Kewley}, Lisa J. and {Groves}, Brent and {Kauffmann}, Guinevere and {Heckman}, Tim},
        title = "{The host galaxies and classification of active galactic nuclei}",
      journal = {\mnras},
         year = 2006,
        month = nov,
       volume = {372},
       number = {3},
        pages = {961-976},
          doi = {10.1111/j.1365-2966.2006.10859.x},
archivePrefix = {arXiv},
       eprint = {astro-ph/0605681},
 primaryClass = {astro-ph},
       adsurl = {https://ui.adsabs.harvard.edu/abs/2006MNRAS.372..961K}
}

@ARTICLE{Kauffmann+03,
       author = {{Kauffmann}, Guinevere and {Heckman}, Timothy M. and {Tremonti}, Christy and {Brinchmann}, Jarle and {Charlot}, St{\'e}phane and {White}, Simon D.~M. and {Ridgway}, Susan E. and {Brinkmann}, Jon and {Fukugita}, Masataka and {Hall}, Patrick B. and {Ivezi{\'c}}, {\v{Z}}eljko and {Richards}, Gordon T. and {Schneider}, Donald P.},
        title = "{The host galaxies of active galactic nuclei}",
      journal = {\mnras},
         year = 2003,
        month = dec,
       volume = {346},
       number = {4},
        pages = {1055-1077},
          doi = {10.1111/j.1365-2966.2003.07154.x},
archivePrefix = {arXiv},
       eprint = {astro-ph/0304239},
 primaryClass = {astro-ph},
       adsurl = {https://ui.adsabs.harvard.edu/abs/2003MNRAS.346.1055K}
}

@ARTICLE{Dominguez-Sanchez+22,
       author = {{Dom{\'\i}nguez S{\'a}nchez}, H. and {Margalef}, B. and {Bernardi}, M. and {Huertas-Company}, M.},
        title = "{SDSS-IV DR17: final release of MaNGA PyMorph photometric and deep-learning morphological catalogues}",
      journal = {\mnras},
         year = 2022,
        month = jan,
       volume = {509},
       number = {3},
        pages = {4024-4036},
          doi = {10.1093/mnras/stab3089},
archivePrefix = {arXiv},
       eprint = {2110.10694},
 primaryClass = {astro-ph.GA},
       adsurl = {https://ui.adsabs.harvard.edu/abs/2022MNRAS.509.4024D}
}

@ARTICLE{Giovanelli+05,
       author = {{Giovanelli}, Riccardo and {Haynes}, Martha P. and {Kent}, Brian R. and {Perillat}, Philip and {Saintonge}, Amelie and {Brosch}, Noah and {Catinella}, Barbara and {Hoffman}, G. Lyle and {Stierwalt}, Sabrina and {Spekkens}, Kristine and {Lerner}, Mikael S. and {Masters}, Karen L. and {Momjian}, Emmanuel and {Rosenberg}, Jessica L. and {Springob}, Christopher M. and {Boselli}, Alessandro and {Charmandaris}, Vassilis and {Darling}, Jeremy K. and {Davies}, Jonathan and {Garcia Lambas}, Diego and {Gavazzi}, Giuseppe and {Giovanardi}, Carlo and {Hardy}, Eduardo and {Hunt}, Leslie K. and {Iovino}, Angela and {Karachentsev}, Igor D. and {Karachentseva}, Valentina E. and {Koopmann}, Rebecca A. and {Marinoni}, Christian and {Minchin}, Robert and {Muller}, Erik and {Putman}, Mary and {Pantoja}, Carmen and {Salzer}, John J. and {Scodeggio}, Marco and {Skillman}, Evan and {Solanes}, Jose M. and {Valotto}, Carlos and {van Driel}, Wim and {van Zee}, Liese},
        title = "{The Arecibo Legacy Fast ALFA Survey. I. Science Goals, Survey Design, and Strategy}",
      journal = {\aj},
         year = 2005,
        month = dec,
       volume = {130},
       number = {6},
        pages = {2598-2612},
          doi = {10.1086/497431},
archivePrefix = {arXiv},
       eprint = {astro-ph/0508301},
 primaryClass = {astro-ph},
       adsurl = {https://ui.adsabs.harvard.edu/abs/2005AJ....130.2598G}
}

@ARTICLE{Drory+15,
       author = {{Drory}, N. and {MacDonald}, N. and {Bershady}, M.~A. and {Bundy}, K. and {Gunn}, J. and {Law}, D.~R. and {Smith}, M. and {Stoll}, R. and {Tremonti}, C.~A. and {Wake}, D.~A. and {Yan}, R. and {Weijmans}, A.~M. and {Byler}, N. and {Cherinka}, B. and {Cope}, F. and {Eigenbrot}, A. and {Harding}, P. and {Holder}, D. and {Huehnerhoff}, J. and {Jaehnig}, K. and {Jansen}, T.~C. and {Klaene}, M. and {Paat}, A.~M. and {Percival}, J. and {Sayres}, C.},
        title = "{The MaNGA Integral Field Unit Fiber Feed System for the Sloan 2.5 m Telescope}",
      journal = {\aj},
         year = 2015,
        month = feb,
       volume = {149},
       number = {2},
          eid = {77},
        pages = {77},
          doi = {10.1088/0004-6256/149/2/77},
archivePrefix = {arXiv},
       eprint = {1412.1535},
 primaryClass = {astro-ph.IM},
       adsurl = {https://ui.adsabs.harvard.edu/abs/2015AJ....149...77D}
}

@ARTICLE{Smee+13,
       author = {{Smee}, Stephen A. and {Gunn}, James E. and {Uomoto}, Alan and {Roe}, Natalie and {Schlegel}, David and {Rockosi}, Constance M. and {Carr}, Michael A. and {Leger}, French and {Dawson}, Kyle S. and {Olmstead}, Matthew D. and {Brinkmann}, Jon and {Owen}, Russell and {Barkhouser}, Robert H. and {Honscheid}, Klaus and {Harding}, Paul and {Long}, Dan and {Lupton}, Robert H. and {Loomis}, Craig and {Anderson}, Lauren and {Annis}, James and {Bernardi}, Mariangela and {Bhardwaj}, Vaishali and {Bizyaev}, Dmitry and {Bolton}, Adam S. and {Brewington}, Howard and {Briggs}, John W. and {Burles}, Scott and {Burns}, James G. and {Castander}, Francisco Javier and {Connolly}, Andrew and {Davenport}, James R.~A. and {Ebelke}, Garrett and {Epps}, Harland and {Feldman}, Paul D. and {Friedman}, Scott D. and {Frieman}, Joshua and {Heckman}, Timothy and {Hull}, Charles L. and {Knapp}, Gillian R. and {Lawrence}, David M. and {Loveday}, Jon and {Mannery}, Edward J. and {Malanushenko}, Elena and {Malanushenko}, Viktor and {Merrelli}, Aronne James and {Muna}, Demitri and {Newman}, Peter R. and {Nichol}, Robert C. and {Oravetz}, Daniel and {Pan}, Kaike and {Pope}, Adrian C. and {Ricketts}, Paul G. and {Shelden}, Alaina and {Sandford}, Dale and {Siegmund}, Walter and {Simmons}, Audrey and {Smith}, D. Shane and {Snedden}, Stephanie and {Schneider}, Donald P. and {SubbaRao}, Mark and {Tremonti}, Christy and {Waddell}, Patrick and {York}, Donald G.},
        title = "{The Multi-object, Fiber-fed Spectrographs for the Sloan Digital Sky Survey and the Baryon Oscillation Spectroscopic Survey}",
      journal = {\aj},
         year = 2013,
        month = aug,
       volume = {146},
       number = {2},
          eid = {32},
        pages = {32},
          doi = {10.1088/0004-6256/146/2/32},
archivePrefix = {arXiv},
       eprint = {1208.2233},
 primaryClass = {astro-ph.IM},
       adsurl = {https://ui.adsabs.harvard.edu/abs/2013AJ....146...32S}
}

@ARTICLE{Gunn+06,
       author = {{Gunn}, James E. and {Siegmund}, Walter A. and {Mannery}, Edward J. and {Owen}, Russell E. and {Hull}, Charles L. and {Leger}, R. French and {Carey}, Larry N. and {Knapp}, Gillian R. and {York}, Donald G. and {Boroski}, William N. and {Kent}, Stephen M. and {Lupton}, Robert H. and {Rockosi}, Constance M. and {Evans}, Michael L. and {Waddell}, Patrick and {Anderson}, John E. and {Annis}, James and {Barentine}, John C. and {Bartoszek}, Larry M. and {Bastian}, Steven and {Bracker}, Stephen B. and {Brewington}, Howard J. and {Briegel}, Charles I. and {Brinkmann}, Jon and {Brown}, Yorke J. and {Carr}, Michael A. and {Czarapata}, Paul C. and {Drennan}, Craig C. and {Dombeck}, Thomas and {Federwitz}, Glenn R. and {Gillespie}, Bruce A. and {Gonzales}, Carlos and {Hansen}, Sten U. and {Harvanek}, Michael and {Hayes}, Jeffrey and {Jordan}, Wendell and {Kinney}, Ellyne and {Klaene}, Mark and {Kleinman}, S.~J. and {Kron}, Richard G. and {Kresinski}, Jurek and {Lee}, Glenn and {Limmongkol}, Siriluk and {Lindenmeyer}, Carl W. and {Long}, Daniel C. and {Loomis}, Craig L. and {McGehee}, Peregrine M. and {Mantsch}, Paul M. and {Neilsen}, Jr., Eric H. and {Neswold}, Richard M. and {Newman}, Peter R. and {Nitta}, Atsuko and {Peoples}, Jr., John and {Pier}, Jeffrey R. and {Prieto}, Peter S. and {Prosapio}, Angela and {Rivetta}, Claudio and {Schneider}, Donald P. and {Snedden}, Stephanie and {Wang}, Shu-i.},
        title = "{The 2.5 m Telescope of the Sloan Digital Sky Survey}",
      journal = {\aj},
         year = 2006,
        month = apr,
       volume = {131},
       number = {4},
        pages = {2332-2359},
          doi = {10.1086/500975},
archivePrefix = {arXiv},
       eprint = {astro-ph/0602326},
 primaryClass = {astro-ph},
       adsurl = {https://ui.adsabs.harvard.edu/abs/2006AJ....131.2332G}
}

@ARTICLE{Tempel+15,
       author = {{Tempel}, E. and {Guo}, Q. and {Kipper}, R. and {Libeskind}, N.~I.},
        title = "{The alignment of satellite galaxies and cosmic filaments: observations and simulations}",
      journal = {\mnras},
         year = 2015,
        month = jul,
       volume = {450},
       number = {3},
        pages = {2727-2738},
          doi = {10.1093/mnras/stv919},
archivePrefix = {arXiv},
       eprint = {1502.02046},
 primaryClass = {astro-ph.CO},
       adsurl = {https://ui.adsabs.harvard.edu/abs/2015MNRAS.450.2727T}
}

@ARTICLE{Behroozi+13,
       author = {{Behroozi}, Peter S. and {Wechsler}, Risa H. and {Conroy}, Charlie},
        title = "{The Average Star Formation Histories of Galaxies in Dark Matter Halos from z = 0-8}",
      journal = {\apj},
         year = 2013,
        month = jun,
       volume = {770},
       number = {1},
          eid = {57},
        pages = {57},
          doi = {10.1088/0004-637X/770/1/57},
archivePrefix = {arXiv},
       eprint = {1207.6105},
 primaryClass = {astro-ph.CO},
       adsurl = {https://ui.adsabs.harvard.edu/abs/2013ApJ...770...57B}
}

@ARTICLE{Behroozi+10,
       author = {{Behroozi}, Peter S. and {Conroy}, Charlie and {Wechsler}, Risa H.},
        title = "{A Comprehensive Analysis of Uncertainties Affecting the Stellar Mass-Halo Mass Relation for 0 < z < 4}",
      journal = {\apj},
         year = 2010,
        month = jul,
       volume = {717},
       number = {1},
        pages = {379-403},
          doi = {10.1088/0004-637X/717/1/379},
archivePrefix = {arXiv},
       eprint = {1001.0015},
 primaryClass = {astro-ph.CO},
       adsurl = {https://ui.adsabs.harvard.edu/abs/2010ApJ...717..379B}
}

@ARTICLE{deIsidio+24,
       author = {{de Is{\'\i}dio}, Natanael G. and {Men{\'e}ndez-Delmestre}, K. and {Gon{\c{c}}alves}, T.~S. and {Grossi}, M. and {Rodrigues}, D.~C. and {Garavito-Camargo}, N. and {Araujo-Carvalho}, A. and {Beaklini}, P.~P.~B. and {Cavalcante-Coelho}, Y. and {Cortesi}, A. and {Quiroga-Nu{\~n}ez}, L.~H. and {Randriamampandry}, T.},
        title = "{Dark Matter Distribution in Milky Way analog Galaxies}",
      journal = {\apj},
         year = 2024,
        month = aug,
       volume = {971},
       number = {1},
          eid = {69},
        pages = {69},
          doi = {10.3847/1538-4357/ad53c8},
archivePrefix = {arXiv},
       eprint = {2310.13839},
 primaryClass = {astro-ph.GA},
       adsurl = {https://ui.adsabs.harvard.edu/abs/2024ApJ...971...69D}
}

@ARTICLE{MaNGA+15,
       author = {{Bundy}, Kevin and {Bershady}, Matthew A. and {Law}, David R. and {Yan}, Renbin and {Drory}, Niv and {MacDonald}, Nicholas and {Wake}, David A. and {Cherinka}, Brian and {S{\'a}nchez-Gallego}, Jos{\'e} R. and {Weijmans}, Anne-Marie and {Thomas}, Daniel and {Tremonti}, Christy and {Masters}, Karen and {Coccato}, Lodovico and {Diamond-Stanic}, Aleksandar M. and {Arag{\'o}n-Salamanca}, Alfonso and {Avila-Reese}, Vladimir and {Badenes}, Carles and {Falc{\'o}n-Barroso}, J{\'e}sus and {Belfiore}, Francesco and {Bizyaev}, Dmitry and {Blanc}, Guillermo A. and {Bland-Hawthorn}, Joss and {Blanton}, Michael R. and {Brownstein}, Joel R. and {Byler}, Nell and {Cappellari}, Michele and {Conroy}, Charlie and {Dutton}, Aaron A. and {Emsellem}, Eric and {Etherington}, James and {Frinchaboy}, Peter M. and {Fu}, Hai and {Gunn}, James E. and {Harding}, Paul and {Johnston}, Evelyn J. and {Kauffmann}, Guinevere and {Kinemuchi}, Karen and {Klaene}, Mark A. and {Knapen}, Johan H. and {Leauthaud}, Alexie and {Li}, Cheng and {Lin}, Lihwai and {Maiolino}, Roberto and {Malanushenko}, Viktor and {Malanushenko}, Elena and {Mao}, Shude and {Maraston}, Claudia and {McDermid}, Richard M. and {Merrifield}, Michael R. and {Nichol}, Robert C. and {Oravetz}, Daniel and {Pan}, Kaike and {Parejko}, John K. and {Sanchez}, Sebastian F. and {Schlegel}, David and {Simmons}, Audrey and {Steele}, Oliver and {Steinmetz}, Matthias and {Thanjavur}, Karun and {Thompson}, Benjamin A. and {Tinker}, Jeremy L. and {van den Bosch}, Remco C.~E. and {Westfall}, Kyle B. and {Wilkinson}, David and {Wright}, Shelley and {Xiao}, Ting and {Zhang}, Kai},
        title = "{Overview of the SDSS-IV MaNGA Survey: Mapping nearby Galaxies at Apache Point Observatory}",
      journal = {\apj},
         year = 2015,
        month = jan,
       volume = {798},
       number = {1},
          eid = {7},
        pages = {7},
          doi = {10.1088/0004-637X/798/1/7},
archivePrefix = {arXiv},
       eprint = {1412.1482},
 primaryClass = {astro-ph.GA},
       adsurl = {https://ui.adsabs.harvard.edu/abs/2015ApJ...798....7B}
}

@ARTICLE{Bloom+17b,
       author = {{Bloom}, J.~V. and {Fogarty}, L.~M.~R. and {Croom}, S.~M. and {Schaefer}, A. and {Bryant}, J.~J. and {Cortese}, L. and {Richards}, S. and {Bland-Hawthorn}, J. and {Ho}, I. -T. and {Scott}, N. and {Goldstein}, G. and {Medling}, A. and {Brough}, S. and {Sweet}, S.~M. and {Cecil}, G. and {L{\'o}pez-S{\'a}nchez}, A. and {Glazebrook}, K. and {Parker}, Q. and {Allen}, J.~T. and {Goodwin}, M. and {Green}, A.~W. and {Konstantopoulos}, I.~S. and {Lawrence}, J.~S. and {Lorente}, N. and {Owers}, M.~S. and {Sharp}, R.},
        title = "{The SAMI Galaxy Survey: asymmetry in gas kinematics and its links to stellar mass and star formation}",
      journal = {\mnras},
         year = 2017,
        month = feb,
       volume = {465},
       number = {1},
        pages = {123-148},
          doi = {10.1093/mnras/stw2605},
archivePrefix = {arXiv},
       eprint = {1610.02773},
 primaryClass = {astro-ph.GA},
       adsurl = {https://ui.adsabs.harvard.edu/abs/2017MNRAS.465..123B}
}

@ARTICLE{Loubser+22,
       author = {{Loubser}, S.~I. and {Lagos}, P. and {Babul}, A. and {O'Sullivan}, E. and {Jung}, S.~L. and {Olivares}, V. and {Kolokythas}, K.},
        title = "{Merger histories of brightest group galaxies from MUSE stellar kinematics}",
      journal = {\mnras},
         year = 2022,
        month = sep,
       volume = {515},
       number = {1},
        pages = {1104-1121},
          doi = {10.1093/mnras/stac1781},
archivePrefix = {arXiv},
       eprint = {2206.13215},
 primaryClass = {astro-ph.GA},
       adsurl = {https://ui.adsabs.harvard.edu/abs/2022MNRAS.515.1104L}
}

@ARTICLE{Krajnovic+20,
       author = {{Krajnovi{\'c}}, Davor and {Ural}, Ugur and {Kuntschner}, Harald and {Goudfrooij}, Paul and {Wolfe}, Michael and {Cappellari}, Michele and {Davies}, Roger and {de Zeeuw}, Tim P. and {Duc}, Pierre-Alain and {Emsellem}, Eric and {Karick}, Arna and {McDermid}, Richard M. and {Mei}, Simona and {Naab}, Thorsten},
        title = "{Formation channels of slowly rotating early-type galaxies}",
      journal = {\aap},
         year = 2020,
        month = mar,
       volume = {635},
          eid = {A129},
        pages = {A129},
          doi = {10.1051/0004-6361/201937040},
archivePrefix = {arXiv},
       eprint = {2001.11277},
 primaryClass = {astro-ph.GA},
       adsurl = {https://ui.adsabs.harvard.edu/abs/2020A&A...635A.129K}
}

@ARTICLE{Ene+18,
       author = {{Ene}, Irina and {Ma}, Chung-Pei and {Veale}, Melanie and {Greene}, Jenny E. and {Thomas}, Jens and {Blakeslee}, John P. and {Foster}, Caroline and {Walsh}, Jonelle L. and {Ito}, Jennifer and {Goulding}, Andy D.},
        title = "{The MASSIVE Survey - X. Misalignment between kinematic and photometric axes and intrinsic shapes of massive early-type galaxies}",
      journal = {\mnras},
         year = 2018,
        month = sep,
       volume = {479},
       number = {2},
        pages = {2810-2826},
          doi = {10.1093/mnras/sty1649},
archivePrefix = {arXiv},
       eprint = {1802.00014},
 primaryClass = {astro-ph.GA},
       adsurl = {https://ui.adsabs.harvard.edu/abs/2018MNRAS.479.2810E}
}

@ARTICLE{Lagos+22,
       author = {{Lagos}, Claudia del P. and {Emsellem}, Eric and {van de Sande}, Jesse and {Harborne}, Katherine E. and {Cortese}, Luca and {Davison}, Thomas and {Foster}, Caroline and {Wright}, Ruby J.},
        title = "{The diverse nature and formation paths of slow rotator galaxies in the EAGLE simulations}",
      journal = {\mnras},
         year = 2022,
        month = jan,
       volume = {509},
       number = {3},
        pages = {4372-4391},
          doi = {10.1093/mnras/stab3128},
archivePrefix = {arXiv},
       eprint = {2012.08060},
 primaryClass = {astro-ph.GA},
       adsurl = {https://ui.adsabs.harvard.edu/abs/2022MNRAS.509.4372L}
}

@ARTICLE{Naab+14,
       author = {{Naab}, Thorsten and {Oser}, L. and {Emsellem}, E. and {Cappellari}, Michele and {Krajnovi{\'c}}, D. and {McDermid}, R.~M. and {Alatalo}, K. and {Bayet}, E. and {Blitz}, L. and {Bois}, M. and {Bournaud}, F. and {Bureau}, M. and {Crocker}, A. and {Davies}, R.~L. and {Davis}, T.~A. and {de Zeeuw}, P.~T. and {Duc}, P.-A. and {Hirschmann}, M. and {Johansson}, P.~H. and {Khochfar}, S. and {Kuntschner}, H. and {Morganti}, R. and {Oosterloo}, T. and {Sarzi}, M. and {Scott}, N. and {Serra}, P. and {van de Ven}, G. and {Weijmans}, A. and {Young}, L.~M.},
        title = "{The ATLAS$^{3D}$ project - XXV. Two-dimensional kinematic analysis of simulated galaxies and the cosmological origin of fast and slow rotators}",
      journal = {\mnras},
         year = 2014,
        month = nov,
       volume = {444},
       number = {4},
        pages = {3357-3387},
          doi = {10.1093/mnras/stt1919},
archivePrefix = {arXiv},
       eprint = {1311.0284},
 primaryClass = {astro-ph.CO},
       adsurl = {https://ui.adsabs.harvard.edu/abs/2014MNRAS.444.3357N}
}

@ARTICLE{Bois+11,
       author = {{Bois}, Maxime and {Emsellem}, Eric and {Bournaud}, Fr{\'e}d{\'e}ric and {Alatalo}, Katherine and {Blitz}, Leo and {Bureau}, Martin and {Cappellari}, Michele and {Davies}, Roger L. and {Davis}, Timothy A. and {de Zeeuw}, P.~T. and {Duc}, Pierre-Alain and {Khochfar}, Sadegh and {Krajnovi{\'c}}, Davor and {Kuntschner}, Harald and {Lablanche}, Pierre-Yves and {McDermid}, Richard M. and {Morganti}, Raffaella and {Naab}, Thorsten and {Oosterloo}, Tom and {Sarzi}, Marc and {Scott}, Nicholas and {Serra}, Paolo and {Weijmans}, Anne-Marie and {Young}, Lisa M.},
        title = "{The ATLAS$^{3D}$ project - VI. Simulations of binary galaxy mergers and the link with fast rotators, slow rotators and kinematically distinct cores}",
      journal = {\mnras},
         year = 2011,
        month = sep,
       volume = {416},
       number = {3},
        pages = {1654-1679},
          doi = {10.1111/j.1365-2966.2011.19113.x},
archivePrefix = {arXiv},
       eprint = {1105.4076},
 primaryClass = {astro-ph.CO},
       adsurl = {https://ui.adsabs.harvard.edu/abs/2011MNRAS.416.1654B}
}

@ARTICLE{Hoffman+10,
       author = {{Hoffman}, Loren and {Cox}, Thomas J. and {Dutta}, Suvendra and {Hernquist}, Lars},
        title = "{Orbital Structure of Merger Remnants. I. Effect of Gas Fraction in Pure Disk Mergers}",
      journal = {\apj},
         year = 2010,
        month = nov,
       volume = {723},
       number = {1},
        pages = {818-844},
          doi = {10.1088/0004-637X/723/1/818},
archivePrefix = {arXiv},
       eprint = {1001.0799},
 primaryClass = {astro-ph.CO},
       adsurl = {https://ui.adsabs.harvard.edu/abs/2010ApJ...723..818H}
}

@ARTICLE{MaNGA+22,
       author = {{Abdurro'uf} and {Accetta}, Katherine and {Aerts}, Conny and {Silva Aguirre}, V{\'\i}ctor and {Ahumada}, Romina and {Ajgaonkar}, Nikhil and {Filiz Ak}, N. and {Alam}, Shadab and {Allende Prieto}, Carlos and {Almeida}, Andr{\'e}s and {Anders}, Friedrich and {Anderson}, Scott F. and {Andrews}, Brett H. and {Anguiano}, Borja and {Aquino-Ort{\'\i}z}, Erik and {Arag{\'o}n-Salamanca}, Alfonso and {Argudo-Fern{\'a}ndez}, Maria and {Ata}, Metin and {Aubert}, Marie and {Avila-Reese}, Vladimir and {Badenes}, Carles and {Barb{\'a}}, Rodolfo H. and {Barger}, Kat and {Barrera-Ballesteros}, Jorge K. and {Beaton}, Rachael L. and {Beers}, Timothy C. and {Belfiore}, Francesco and {Bender}, Chad F. and {Bernardi}, Mariangela and {Bershady}, Matthew A. and {Beutler}, Florian and {Bidin}, Christian Moni and {Bird}, Jonathan C. and {Bizyaev}, Dmitry and {Blanc}, Guillermo A. and {Blanton}, Michael R. and {Boardman}, Nicholas Fraser and {Bolton}, Adam S. and {Boquien}, M{\'e}d{\'e}ric and {Borissova}, Jura and {Bovy}, Jo and {Brandt}, W.~N. and {Brown}, Jordan and {Brownstein}, Joel R. and {Brusa}, Marcella and {Buchner}, Johannes and {Bundy}, Kevin and {Burchett}, Joseph N. and {Bureau}, Martin and {Burgasser}, Adam and {Cabang}, Tuesday K. and {Campbell}, Stephanie and {Cappellari}, Michele and {Carlberg}, Joleen K. and {Wanderley}, F{\'a}bio Carneiro and {Carrera}, Ricardo and {Cash}, Jennifer and {Chen}, Yan-Ping and {Chen}, Wei-Huai and {Cherinka}, Brian and {Chiappini}, Cristina and {Choi}, Peter Doohyun and {Chojnowski}, S. Drew and {Chung}, Haeun and {Clerc}, Nicolas and {Cohen}, Roger E. and {Comerford}, Julia M. and {Comparat}, Johan and {da Costa}, Luiz and {Covey}, Kevin and {Crane}, Jeffrey D. and {Cruz-Gonzalez}, Irene and {Culhane}, Connor and {Cunha}, Katia and {Dai}, Y. Sophia and {Damke}, Guillermo and {Darling}, Jeremy and {Davidson}, Jr., James W. and {Davies}, Roger and {Dawson}, Kyle and {De Lee}, Nathan and {Diamond-Stanic}, Aleksandar M. and {Cano-D{\'\i}az}, Mariana and {S{\'a}nchez}, Helena Dom{\'\i}nguez and {Donor}, John and {Duckworth}, Chris and {Dwelly}, Tom and {Eisenstein}, Daniel J. and {Elsworth}, Yvonne P. and {Emsellem}, Eric and {Eracleous}, Mike and {Escoffier}, Stephanie and {Fan}, Xiaohui and {Farr}, Emily and {Feng}, Shuai and {Fern{\'a}ndez-Trincado}, Jos{\'e} G. and {Feuillet}, Diane and {Filipp}, Andreas and {Fillingham}, Sean P. and {Frinchaboy}, Peter M. and {Fromenteau}, Sebastien and {Galbany}, Llu{\'\i}s and {Garc{\'\i}a}, Rafael A. and {Garc{\'\i}a-Hern{\'a}ndez}, D.~A. and {Ge}, Junqiang and {Geisler}, Doug and {Gelfand}, Joseph and {G{\'e}ron}, Tobias and {Gibson}, Benjamin J. and {Goddy}, Julian and {Godoy-Rivera}, Diego and {Grabowski}, Kathleen and {Green}, Paul J. and {Greener}, Michael and {Grier}, Catherine J. and {Griffith}, Emily and {Guo}, Hong and {Guy}, Julien and {Hadjara}, Massinissa and {Harding}, Paul and {Hasselquist}, Sten and {Hayes}, Christian R. and {Hearty}, Fred and {Hern{\'a}ndez}, Jes{\'u}s and {Hill}, Lewis and {Hogg}, David W. and {Holtzman}, Jon A. and {Horta}, Danny and {Hsieh}, Bau-Ching and {Hsu}, Chin-Hao and {Hsu}, Yun-Hsin and {Huber}, Daniel and {Huertas-Company}, Marc and {Hutchinson}, Brian and {Hwang}, Ho Seong and {Ibarra-Medel}, H{\'e}ctor J. and {Chitham}, Jacob Ider and {Ilha}, Gabriele S. and {Imig}, Julie and {Jaekle}, Will and {Jayasinghe}, Tharindu and {Ji}, Xihan and {Johnson}, Jennifer A. and {Jones}, Amy and {J{\"o}nsson}, Henrik and {Katkov}, Ivan and {Khalatyan}, Dr., Arman and {Kinemuchi}, Karen and {Kisku}, Shobhit and {Knapen}, Johan H. and {Kneib}, Jean-Paul and {Kollmeier}, Juna A. and {Kong}, Miranda and {Kounkel}, Marina and {Kreckel}, Kathryn and {Krishnarao}, Dhanesh and {Lacerna}, Ivan and {Lane}, Richard R. and {Langgin}, Rachel and {Lavender}, Ramon and {Law}, David R. and {Lazarz}, Daniel and {Leung}, Henry W. and {Leung}, Ho-Hin and {Lewis}, Hannah M. and {Li}, Cheng and {Li}, Ran and {Lian}, Jianhui and {Liang}, Fu-Heng and {Lin}, Lihwai and {Lin}, Yen-Ting and {Lin}, Sicheng and {Lintott}, Chris and {Long}, Dan and {Longa-Pe{\~n}a}, Pen{\'e}lope and {L{\'o}pez-Cob{\'a}}, Carlos and {Lu}, Shengdong and {Lundgren}, Britt F. and {Luo}, Yuanze and {Mackereth}, J. Ted and {de la Macorra}, Axel and {Mahadevan}, Suvrath and {Majewski}, Steven R. and {Manchado}, Arturo and {Mandeville}, Travis and {Maraston}, Claudia and {Margalef-Bentabol}, Berta and {Masseron}, Thomas and {Masters}, Karen L. and {Mathur}, Savita and {McDermid}, Richard M. and {Mckay}, Myles and {Merloni}, Andrea and {Merrifield}, Michael and {Meszaros}, Szabolcs and {Miglio}, Andrea and {Di Mille}, Francesco and {Minniti}, Dante and {Minsley}, Rebecca and {Monachesi}, Antonela},
        title = "{The Seventeenth Data Release of the Sloan Digital Sky Surveys: Complete Release of MaNGA, MaStar, and APOGEE-2 Data}",
      journal = {\apjs},
         year = 2022,
        month = apr,
       volume = {259},
       number = {2},
          eid = {35},
        pages = {35},
          doi = {10.3847/1538-4365/ac4414},
archivePrefix = {arXiv},
       eprint = {2112.02026},
 primaryClass = {astro-ph.GA},
       adsurl = {https://ui.adsabs.harvard.edu/abs/2022ApJS..259...35A}
}

@ARTICLE{Barolo,
       author = {{Di Teodoro}, E.~M. and {Fraternali}, F.},
        title = "{$^{3D}$ BAROLO: a new 3D algorithm to derive rotation curves of galaxies}",
      journal = {\mnras},
         year = 2015,
        month = aug,
       volume = {451},
       number = {3},
        pages = {3021-3033},
          doi = {10.1093/mnras/stv1213},
archivePrefix = {arXiv},
       eprint = {1505.07834},
 primaryClass = {astro-ph.GA},
       adsurl = {https://ui.adsabs.harvard.edu/abs/2015MNRAS.451.3021D}
}

@ARTICLE{DiTeodoro+21,
       author = {{Di Teodoro}, Enrico M. and {Peek}, J.~E.~G.},
        title = "{Radial Motions and Radial Gas Flows in Local Spiral Galaxies}",
      journal = {\apj},
         year = 2021,
        month = dec,
       volume = {923},
       number = {2},
          eid = {220},
        pages = {220},
          doi = {10.3847/1538-4357/ac2cbd},
archivePrefix = {arXiv},
       eprint = {2110.01618},
 primaryClass = {astro-ph.GA},
       adsurl = {https://ui.adsabs.harvard.edu/abs/2021ApJ...923..220D}
}

@ARTICLE{SDSS-DRIV+17,
       author = {{Blanton}, Michael R. and {Bershady}, Matthew A. and {Abolfathi}, Bela and {Albareti}, Franco D. and {Allende Prieto}, Carlos and {Almeida}, Andres and {Alonso-Garc{\'\i}a}, Javier and {Anders}, Friedrich and {Anderson}, Scott F. and {Andrews}, Brett and {Aquino-Ort{\'\i}z}, Erik and {Arag{\'o}n-Salamanca}, Alfonso and {Argudo-Fern{\'a}ndez}, Maria and {Armengaud}, Eric and {Aubourg}, Eric and {Avila-Reese}, Vladimir and {Badenes}, Carles and {Bailey}, Stephen and {Barger}, Kathleen A. and {Barrera-Ballesteros}, Jorge and {Bartosz}, Curtis and {Bates}, Dominic and {Baumgarten}, Falk and {Bautista}, Julian and {Beaton}, Rachael and {Beers}, Timothy C. and {Belfiore}, Francesco and {Bender}, Chad F. and {Berlind}, Andreas A. and {Bernardi}, Mariangela and {Beutler}, Florian and {Bird}, Jonathan C. and {Bizyaev}, Dmitry and {Blanc}, Guillermo A. and {Blomqvist}, Michael and {Bolton}, Adam S. and {Boquien}, M{\'e}d{\'e}ric and {Borissova}, Jura and {van den Bosch}, Remco and {Bovy}, Jo and {Brandt}, William N. and {Brinkmann}, Jonathan and {Brownstein}, Joel R. and {Bundy}, Kevin and {Burgasser}, Adam J. and {Burtin}, Etienne and {Busca}, Nicol{\'a}s G. and {Cappellari}, Michele and {Delgado Carigi}, Maria Leticia and {Carlberg}, Joleen K. and {Carnero Rosell}, Aurelio and {Carrera}, Ricardo and {Chanover}, Nancy J. and {Cherinka}, Brian and {Cheung}, Edmond and {G{\'o}mez Maqueo Chew}, Yilen and {Chiappini}, Cristina and {Choi}, Peter Doohyun and {Chojnowski}, Drew and {Chuang}, Chia-Hsun and {Chung}, Haeun and {Cirolini}, Rafael Fernando and {Clerc}, Nicolas and {Cohen}, Roger E. and {Comparat}, Johan and {da Costa}, Luiz and {Cousinou}, Marie-Claude and {Covey}, Kevin and {Crane}, Jeffrey D. and {Croft}, Rupert A.~C. and {Cruz-Gonzalez}, Irene and {Garrido Cuadra}, Daniel and {Cunha}, Katia and {Damke}, Guillermo J. and {Darling}, Jeremy and {Davies}, Roger and {Dawson}, Kyle and {de la Macorra}, Axel and {Dell'Agli}, Flavia and {De Lee}, Nathan and {Delubac}, Timoth{\'e}e and {Di Mille}, Francesco and {Diamond-Stanic}, Aleks and {Cano-D{\'\i}az}, Mariana and {Donor}, John and {Downes}, Juan Jos{\'e} and {Drory}, Niv and {du Mas des Bourboux}, H{\'e}lion and {Duckworth}, Christopher J. and {Dwelly}, Tom and {Dyer}, Jamie and {Ebelke}, Garrett and {Eigenbrot}, Arthur D. and {Eisenstein}, Daniel J. and {Emsellem}, Eric and {Eracleous}, Mike and {Escoffier}, Stephanie and {Evans}, Michael L. and {Fan}, Xiaohui and {Fern{\'a}ndez-Alvar}, Emma and {Fernandez-Trincado}, J.~G. and {Feuillet}, Diane K. and {Finoguenov}, Alexis and {Fleming}, Scott W. and {Font-Ribera}, Andreu and {Fredrickson}, Alexander and {Freischlad}, Gordon and {Frinchaboy}, Peter M. and {Fuentes}, Carla E. and {Galbany}, Llu{\'\i}s and {Garcia-Dias}, R. and {Garc{\'\i}a-Hern{\'a}ndez}, D.~A. and {Gaulme}, Patrick and {Geisler}, Doug and {Gelfand}, Joseph D. and {Gil-Mar{\'\i}n}, H{\'e}ctor and {Gillespie}, Bruce A. and {Goddard}, Daniel and {Gonzalez-Perez}, Violeta and {Grabowski}, Kathleen and {Green}, Paul J. and {Grier}, Catherine J. and {Gunn}, James E. and {Guo}, Hong and {Guy}, Julien and {Hagen}, Alex and {Hahn}, ChangHoon and {Hall}, Matthew and {Harding}, Paul and {Hasselquist}, Sten and {Hawley}, Suzanne L. and {Hearty}, Fred and {Gonzalez Hern{\'a}ndez}, Jonay I. and {Ho}, Shirley and {Hogg}, David W. and {Holley-Bockelmann}, Kelly and {Holtzman}, Jon A. and {Holzer}, Parker H. and {Huehnerhoff}, Joseph and {Hutchinson}, Timothy A. and {Hwang}, Ho Seong and {Ibarra-Medel}, H{\'e}ctor J. and {da Silva Ilha}, Gabriele and {Ivans}, Inese I. and {Ivory}, KeShawn and {Jackson}, Kelly and {Jensen}, Trey W. and {Johnson}, Jennifer A. and {Jones}, Amy and {J{\"o}nsson}, Henrik and {Jullo}, Eric and {Kamble}, Vikrant and {Kinemuchi}, Karen and {Kirkby}, David and {Kitaura}, Francisco-Shu and {Klaene}, Mark and {Knapp}, Gillian R. and {Kneib}, Jean-Paul and {Kollmeier}, Juna A. and {Lacerna}, Ivan and {Lane}, Richard R. and {Lang}, Dustin and {Law}, David R. and {Lazarz}, Daniel and {Lee}, Youngbae and {Le Goff}, Jean-Marc and {Liang}, Fu-Heng and {Li}, Cheng and {Li}, Hongyu and {Lian}, Jianhui and {Lima}, Marcos and {Lin}, Lihwai and {Lin}, Yen-Ting and {Bertran de Lis}, Sara and {Liu}, Chao and {de Icaza Lizaola}, Miguel Angel C. and {Long}, Dan and {Lucatello}, Sara and {Lundgren}, Britt and {MacDonald}, Nicholas K. and {Deconto Machado}, Alice and {MacLeod}, Chelsea L. and {Mahadevan}, Suvrath and {Geimba Maia}, Marcio Antonio and {Maiolino}, Roberto and {Majewski}, Steven R. and {Malanushenko}, Elena and {Malanushenko}, Viktor and {Manchado}, Arturo and {Mao}, Shude and {Maraston}, Claudia and {Marques-Chaves}, Rui and {Masseron}, Thomas and {Masters}, Karen L. and {McBride}, Cameron K. and {McDermid}, Richard M. and {McGrath}, Brianne and {McGreer}, Ian D. and {Medina Pe{\~n}a}, Nicol{\'a}s and {Melendez}, Matthew},
        title = "{Sloan Digital Sky Survey IV: Mapping the Milky Way, Nearby Galaxies, and the Distant Universe}",
      journal = {\aj},
         year = 2017,
        month = jul,
       volume = {154},
       number = {1},
          eid = {28},
        pages = {28},
          doi = {10.3847/1538-3881/aa7567},
archivePrefix = {arXiv},
       eprint = {1703.00052},
 primaryClass = {astro-ph.GA},
       adsurl = {https://ui.adsabs.harvard.edu/abs/2017AJ....154...28B}
}

@ARTICLE{Yang+07,
       author = {{Yang}, Xiaohu and {Mo}, H.~J. and {van den Bosch}, Frank C. and {Pasquali}, Anna and {Li}, Cheng and {Barden}, Marco},
        title = "{Galaxy Groups in the SDSS DR4. I. The Catalog and Basic Properties}",
      journal = {\apj},
         year = 2007,
        month = dec,
       volume = {671},
       number = {1},
        pages = {153-170},
          doi = {10.1086/522027},
archivePrefix = {arXiv},
       eprint = {0707.4640},
 primaryClass = {astro-ph},
       adsurl = {https://ui.adsabs.harvard.edu/abs/2007ApJ...671..153Y}
}

@ARTICLE{Yang+05,
       author = {{Yang}, Xiaohu and {Mo}, H.~J. and {van den Bosch}, Frank C. and {Jing}, Y.~P.},
        title = "{A halo-based galaxy group finder: calibration and application to the 2dFGRS}",
      journal = {\mnras},
         year = 2005,
        month = feb,
       volume = {356},
       number = {4},
        pages = {1293-1307},
          doi = {10.1111/j.1365-2966.2005.08560.x},
archivePrefix = {arXiv},
       eprint = {astro-ph/0405234},
 primaryClass = {astro-ph},
       adsurl = {https://ui.adsabs.harvard.edu/abs/2005MNRAS.356.1293Y}
}

@ARTICLE{Pasquali+10,
       author = {{Pasquali}, Anna and {Gallazzi}, Anna and {Fontanot}, Fabio and {van den Bosch}, Frank C. and {De Lucia}, Gabriella and {Mo}, H.~J. and {Yang}, Xiaohu},
        title = "{Ages and metallicities of central and satellite galaxies: implications for galaxy formation and evolution}",
      journal = {\mnras},
         year = 2010,
        month = sep,
       volume = {407},
       number = {2},
        pages = {937-954},
          doi = {10.1111/j.1365-2966.2010.17074.x},
archivePrefix = {arXiv},
       eprint = {0912.1853},
 primaryClass = {astro-ph.CO},
       adsurl = {https://ui.adsabs.harvard.edu/abs/2010MNRAS.407..937P}
}

@ARTICLE{Slater+19,
       author = {{Slater}, R. and {Nagar}, N.~M. and {Schnorr-M{\"u}ller}, A. and {Storchi-Bergmann}, T. and {Finlez}, C. and {Lena}, D. and {Ramakrishnan}, V. and {Mundell}, C.~G. and {Riffel}, R.~A. and {Peterson}, B. and {Robinson}, A. and {Orellana}, G.},
        title = "{Outflows in the inner kiloparsec of NGC 1566 as revealed by molecular (ALMA) and ionized gas (Gemini-GMOS/IFU) kinematics}",
      journal = {\aap},
         year = 2019,
        month = jan,
       volume = {621},
          eid = {A83},
        pages = {A83},
          doi = {10.1051/0004-6361/201730634},
archivePrefix = {arXiv},
       eprint = {1804.02054},
 primaryClass = {astro-ph.GA},
       adsurl = {https://ui.adsabs.harvard.edu/abs/2019A&A...621A..83S}
}

@ARTICLE{Holmes+15,
       author = {{Holmes}, L. and {Spekkens}, K. and {S{\'a}nchez}, S.~F. and {Walcher}, C.~J. and {Garc{\'\i}a-Benito}, R. and {Mast}, D. and {Cortijo-Ferrero}, C. and {Kalinova}, V. and {Marino}, R.~A. and {Mendez-Abreu}, J. and {Barrera-Ballesteros}, J.~K.},
        title = "{The incidence of bar-like kinematic flows in CALIFA galaxies}",
      journal = {\mnras},
         year = 2015,
        month = aug,
       volume = {451},
       number = {4},
        pages = {4397-4411},
          doi = {10.1093/mnras/stv1254},
archivePrefix = {arXiv},
       eprint = {1506.01378},
 primaryClass = {astro-ph.GA},
       adsurl = {https://ui.adsabs.harvard.edu/abs/2015MNRAS.451.4397H}
}

@ARTICLE{Liu+13,
       author = {{Liu}, Guilin and {Zakamska}, Nadia L. and {Greene}, Jenny E. and {Nesvadba}, Nicole P.~H. and {Liu}, Xin},
        title = "{Observations of feedback from radio-quiet quasars - II. Kinematics of ionized gas nebulae}",
      journal = {\mnras},
         year = 2013,
        month = dec,
       volume = {436},
       number = {3},
        pages = {2576-2597},
          doi = {10.1093/mnras/stt1755},
archivePrefix = {arXiv},
       eprint = {1305.6922},
 primaryClass = {astro-ph.CO},
       adsurl = {https://ui.adsabs.harvard.edu/abs/2013MNRAS.436.2576L}
}

@ARTICLE{Simons+19,
       author = {{Simons}, Raymond C. and {Kassin}, Susan A. and {Snyder}, Gregory F. and {Primack}, Joel R. and {Ceverino}, Daniel and {Dekel}, Avishai and {Hayward}, Christopher C. and {Mandelker}, Nir and {Mantha}, Kameswara Bharadwaj and {Pacifici}, Camilla and {de la Vega}, Alexander and {Wang}, Weichen},
        title = "{Distinguishing Mergers and Disks in High-redshift Observations of Galaxy Kinematics}",
      journal = {\apj},
         year = 2019,
        month = mar,
       volume = {874},
       number = {1},
          eid = {59},
        pages = {59},
          doi = {10.3847/1538-4357/ab07c9},
archivePrefix = {arXiv},
       eprint = {1902.06762},
 primaryClass = {astro-ph.GA},
       adsurl = {https://ui.adsabs.harvard.edu/abs/2019ApJ...874...59S}
}

@ARTICLE{Kinemetry,
       author = {{Krajnovi{\'c}}, Davor and {Cappellari}, Michele and {de Zeeuw}, P. Tim and {Copin}, Yannick},
        title = "{Kinemetry: a generalization of photometry to the higher moments of the line-of-sight velocity distribution}",
      journal = {\mnras},
         year = 2006,
        month = mar,
       volume = {366},
       number = {3},
        pages = {787-802},
          doi = {10.1111/j.1365-2966.2005.09902.x},
archivePrefix = {arXiv},
       eprint = {astro-ph/0512200},
 primaryClass = {astro-ph},
       adsurl = {https://ui.adsabs.harvard.edu/abs/2006MNRAS.366..787K}
}

@ARTICLE{Feng+22,
       author = {{Feng}, Shuai and {Shen}, Shi-Yin and {Yuan}, Fang-Ting and {Dai}, Y. Sophia and {Masters}, Karen L.},
        title = "{The Velocity Map Asymmetry of Ionized Gas in MaNGA. I. The Catalog and General Properties}",
      journal = {\apjs},
         year = 2022,
        month = sep,
       volume = {262},
       number = {1},
          eid = {6},
        pages = {6},
          doi = {10.3847/1538-4365/ac80f2},
archivePrefix = {arXiv},
       eprint = {2207.06050},
 primaryClass = {astro-ph.GA},
       adsurl = {https://ui.adsabs.harvard.edu/abs/2022ApJS..262....6F}
}

@ARTICLE{Trussler+20,
       author = {{Trussler}, James and {Maiolino}, Roberto and {Maraston}, Claudia and {Peng}, Yingjie and {Thomas}, Daniel and {Goddard}, Daniel and {Lian}, Jianhui},
        title = "{Both starvation and outflows drive galaxy quenching}",
      journal = {\mnras},
         year = 2020,
        month = feb,
       volume = {491},
       number = {4},
        pages = {5406-5434},
          doi = {10.1093/mnras/stz3286},
archivePrefix = {arXiv},
       eprint = {1811.09283},
 primaryClass = {astro-ph.GA},
       adsurl = {https://ui.adsabs.harvard.edu/abs/2020MNRAS.491.5406T}
}

@ARTICLE{Bluck+20,
       author = {{Bluck}, Asa F.~L. and {Maiolino}, Roberto and {Piotrowska}, Joanna M. and {Trussler}, James and {Ellison}, Sara L. and {S{\'a}nchez}, Sebastian F. and {Thorp}, Mallory D. and {Teimoorinia}, Hossen and {Moreno}, Jorge and {Conselice}, Christopher J.},
        title = "{How do central and satellite galaxies quench? - Insights from spatially resolved spectroscopy in the MaNGA survey}",
      journal = {\mnras},
         year = 2020,
        month = nov,
       volume = {499},
       number = {1},
        pages = {230-268},
          doi = {10.1093/mnras/staa2806},
archivePrefix = {arXiv},
       eprint = {2009.05341},
 primaryClass = {astro-ph.GA},
       adsurl = {https://ui.adsabs.harvard.edu/abs/2020MNRAS.499..230B}
}

@ARTICLE{Lackner+12,
       author = {{Lackner}, C.~N. and {Cen}, R. and {Ostriker}, J.~P. and {Joung}, M.~R.},
        title = "{Building galaxies by accretion and in situ star formation}",
      journal = {\mnras},
         year = 2012,
        month = sep,
       volume = {425},
       number = {1},
        pages = {641-656},
          doi = {10.1111/j.1365-2966.2012.21525.x},
archivePrefix = {arXiv},
       eprint = {1206.0295},
 primaryClass = {astro-ph.CO},
       adsurl = {https://ui.adsabs.harvard.edu/abs/2012MNRAS.425..641L}
}

@ARTICLE{Peng+15,
       author = {{Peng}, Y. and {Maiolino}, R. and {Cochrane}, R.},
        title = "{Strangulation as the primary mechanism for shutting down star formation in galaxies}",
      journal = {\nat},
         year = 2015,
        month = may,
       volume = {521},
       number = {7551},
        pages = {192-195},
          doi = {10.1038/nature14439},
archivePrefix = {arXiv},
       eprint = {1505.03143},
 primaryClass = {astro-ph.GA},
       adsurl = {https://ui.adsabs.harvard.edu/abs/2015Natur.521..192P}
}

@ARTICLE{Belfiore+19,
       author = {{Belfiore}, Francesco and {Westfall}, Kyle B. and {Schaefer}, Adam and {Cappellari}, Michele and {Ji}, Xihan and {Bershady}, Matthew A. and {Tremonti}, Christy and {Law}, David R. and {Yan}, Renbin and {Bundy}, Kevin and {Shetty}, Shravan and {Drory}, Niv and {Thomas}, Daniel and {Emsellem}, Eric and {S{\'a}nchez}, Sebasti{\'a}n F.},
        title = "{The Data Analysis Pipeline for the SDSS-IV MaNGA IFU Galaxy Survey: Emission-line Modeling}",
      journal = {\aj},
         year = 2019,
        month = oct,
       volume = {158},
       number = {4},
          eid = {160},
        pages = {160},
          doi = {10.3847/1538-3881/ab3e4e},
archivePrefix = {arXiv},
       eprint = {1901.00866},
 primaryClass = {astro-ph.GA},
       adsurl = {https://ui.adsabs.harvard.edu/abs/2019AJ....158..160B}
}

@ARTICLE{pyPipe3D,
       author = {{S{\'a}nchez}, S.~F. and {Barrera-Ballesteros}, J.~K. and {Lacerda}, E. and {Mej{\'\i}a-Narvaez}, A. and {Camps-Fari{\~n}a}, A. and {Bruzual}, Gustavo and {Espinosa-Ponce}, C. and {Rodr{\'\i}guez-Puebla}, A. and {Calette}, A.~R. and {Ibarra-Medel}, H. and {Avila-Reese}, V. and {Hernandez-Toledo}, H. and {Bershady}, M.~A. and {Cano-Diaz}, M. and {Munguia-Cordova}, A.~M.},
        title = "{SDSS-IV MaNGA: pyPipe3D Analysis Release for 10,000 Galaxies}",
      journal = {\apjs},
         year = 2022,
        month = oct,
       volume = {262},
       number = {2},
          eid = {36},
        pages = {36},
          doi = {10.3847/1538-4365/ac7b8f},
archivePrefix = {arXiv},
       eprint = {2206.07062},
 primaryClass = {astro-ph.GA},
       adsurl = {https://ui.adsabs.harvard.edu/abs/2022ApJS..262...36S}
}

@ARTICLE{Stark+21,
       author = {{Stark}, David V. and {Masters}, Karen L. and {Avila-Reese}, Vladimir and {Riffel}, Rogemar and {Riffel}, Rogerio and {Boardman}, Nicholas Fraser and {Zheng}, Zheng and {Weijmans}, Anne-Marie and {Dillon}, Sean and {Fielder}, Catherine and {Finnegan}, Daniel and {Fofie}, Patricia and {Goddy}, Julian and {Harrington}, Emily and {Pace}, Zachary and {Rujopakarn}, Wiphu and {Samanso}, Nattida and {Shamsi}, Shoaib and {Sharma}, Anubhav and {Warrick}, Elizabeth and {Witherspoon}, Catherine and {Wolthuis}, Nathan},
        title = "{H I-MaNGA: tracing the physics of the neutral and ionized ISM with the second data release}",
      journal = {\mnras},
         year = 2021,
        month = may,
       volume = {503},
       number = {1},
        pages = {1345-1366},
          doi = {10.1093/mnras/stab566},
archivePrefix = {arXiv},
       eprint = {2101.12680},
 primaryClass = {astro-ph.GA},
       adsurl = {https://ui.adsabs.harvard.edu/abs/2021MNRAS.503.1345S}
}

@ARTICLE{Masters+19,
       author = {{Masters}, Karen L. and {Stark}, David V. and {Pace}, Zachary J. and {Phipps}, Frederika and {Rujopakarn}, Wiphu and {Samanso}, Nattida and {Harrington}, Emily and {S{\'a}nchez-Gallego}, Jos{\'e} R. and {Avila-Reese}, Vladimir and {Bershady}, Matthew and {Cherinka}, Brian and {Fielder}, Catherine E. and {Finnegan}, Daniel and {Riffel}, Rogemar A. and {Rowlands}, Kate and {Shamsi}, Shoaib and {Newnham}, Lucy and {Weijmans}, Anne-Marie and {Witherspoon}, Catherine A.},
        title = "{H I-MaNGA: H I follow-up for the MaNGA survey}",
      journal = {\mnras},
         year = 2019,
        month = sep,
       volume = {488},
       number = {3},
        pages = {3396-3405},
          doi = {10.1093/mnras/stz1889},
archivePrefix = {arXiv},
       eprint = {1901.05579},
 primaryClass = {astro-ph.GA},
       adsurl = {https://ui.adsabs.harvard.edu/abs/2019MNRAS.488.3396M}
}

@ARTICLE{Cortese+21,
       author = {{Cortese}, L. and {Catinella}, B. and {Smith}, R.},
        title = "{The Dawes Review 9: The role of cold gas stripping on the star formation quenching of satellite galaxies}",
      journal = {\pasa},
         year = 2021,
        month = aug,
       volume = {38},
          eid = {e035},
        pages = {e035},
          doi = {10.1017/pasa.2021.18},
archivePrefix = {arXiv},
       eprint = {2104.02193},
 primaryClass = {astro-ph.GA},
       adsurl = {https://ui.adsabs.harvard.edu/abs/2021PASA...38...35C}
}

@ARTICLE{Dekel+06,
       author = {{Dekel}, Avishai and {Birnboim}, Yuval},
        title = "{Galaxy bimodality due to cold flows and shock heating}",
      journal = {\mnras},
         year = 2006,
        month = may,
       volume = {368},
       number = {1},
        pages = {2-20},
          doi = {10.1111/j.1365-2966.2006.10145.x},
archivePrefix = {arXiv},
       eprint = {astro-ph/0412300},
 primaryClass = {astro-ph},
       adsurl = {https://ui.adsabs.harvard.edu/abs/2006MNRAS.368....2D}
}

@ARTICLE{Kang+08,
       author = {{Kang}, X. and {van den Bosch}, Frank C.},
        title = "{New Constraints on the Efficiencies of Ram Pressure Stripping and the Tidal Disruption of Satellite Galaxies}",
      journal = {\apjl},
         year = 2008,
        month = apr,
       volume = {676},
       number = {2},
        pages = {L101},
          doi = {10.1086/587620},
archivePrefix = {arXiv},
       eprint = {0801.1843},
 primaryClass = {astro-ph},
       adsurl = {https://ui.adsabs.harvard.edu/abs/2008ApJ...676L.101K}
}

@ARTICLE{Poggianti+19,
       author = {{Poggianti}, Bianca M. and {Ignesti}, Alessandro and {Gitti}, Myriam and {Wolter}, Anna and {Brighenti}, Fabrizio and {Biviano}, Andrea and {George}, Koshy and {Vulcani}, Benedetta and {Gullieuszik}, Marco and {Moretti}, Alessia and {Paladino}, Rosita and {Bettoni}, Daniela and {Franchetto}, Andrea and {Jaff{\'e}}, Yara L. and {Radovich}, Mario and {Roediger}, Elke and {Tomi{\v{c}}i{\'c}}, Neven and {Tonnesen}, Stephanie and {Bellhouse}, Callum and {Fritz}, Jacopo and {Omizzolo}, Alessandro},
        title = "{GASP XXIII: A Jellyfish Galaxy as an Astrophysical Laboratory of the Baryonic Cycle}",
      journal = {\apj},
         year = 2019,
        month = dec,
       volume = {887},
       number = {2},
          eid = {155},
        pages = {155},
          doi = {10.3847/1538-4357/ab5224},
archivePrefix = {arXiv},
       eprint = {1910.11622},
 primaryClass = {astro-ph.GA},
       adsurl = {https://ui.adsabs.harvard.edu/abs/2019ApJ...887..155P}
}

@ARTICLE{Cortese+19,
       author = {{Cortese}, L. and {van de Sande}, J. and {Lagos}, C.~P. and {Catinella}, B. and {Davies}, L.~J.~M. and {Croom}, S.~M. and {Brough}, S. and {Bryant}, J.~J. and {Lawrence}, J.~S. and {Owers}, M.~S. and {Richards}, S.~N. and {Sweet}, S.~M. and {Bland-Hawthorn}, J.},
        title = "{The SAMI Galaxy Survey: satellite galaxies undergo little structural change during their quenching phase}",
      journal = {\mnras},
         year = 2019,
        month = may,
       volume = {485},
       number = {2},
        pages = {2656-2665},
          doi = {10.1093/mnras/stz485},
archivePrefix = {arXiv},
       eprint = {1902.05652},
 primaryClass = {astro-ph.GA},
       adsurl = {https://ui.adsabs.harvard.edu/abs/2019MNRAS.485.2656C}
}

@ARTICLE{Naab+09,
       author = {{Naab}, Thorsten and {Johansson}, Peter H. and {Ostriker}, Jeremiah P.},
        title = "{Minor Mergers and the Size Evolution of Elliptical Galaxies}",
      journal = {\apjl},
         year = 2009,
        month = jul,
       volume = {699},
       number = {2},
        pages = {L178-L182},
          doi = {10.1088/0004-637X/699/2/L178},
archivePrefix = {arXiv},
       eprint = {0903.1636},
 primaryClass = {astro-ph.CO},
       adsurl = {https://ui.adsabs.harvard.edu/abs/2009ApJ...699L.178N}
}

@ARTICLE{Krajnovic+11,
       author = {{Krajnovi{\'c}}, Davor and {Emsellem}, Eric and {Cappellari}, Michele and {Alatalo}, Katherine and {Blitz}, Leo and {Bois}, Maxime and {Bournaud}, Fr{\'e}d{\'e}ric and {Bureau}, Martin and {Davies}, Roger L. and {Davis}, Timothy A. and {de Zeeuw}, P.~T. and {Khochfar}, Sadegh and {Kuntschner}, Harald and {Lablanche}, Pierre-Yves and {McDermid}, Richard M. and {Morganti}, Raffaella and {Naab}, Thorsten and {Oosterloo}, Tom and {Sarzi}, Marc and {Scott}, Nicholas and {Serra}, Paolo and {Weijmans}, Anne-Marie and {Young}, Lisa M.},
        title = "{The ATLAS$^{3D}$ project - II. Morphologies, kinemetric features and alignment between photometric and kinematic axes of early-type galaxies}",
      journal = {\mnras},
         year = 2011,
        month = jul,
       volume = {414},
       number = {4},
        pages = {2923-2949},
          doi = {10.1111/j.1365-2966.2011.18560.x},
archivePrefix = {arXiv},
       eprint = {1102.3801},
 primaryClass = {astro-ph.CO},
       adsurl = {https://ui.adsabs.harvard.edu/abs/2011MNRAS.414.2923K}
}

@ARTICLE{Sanchez-Garcia+23,
       author = {{S{\'a}nchez-Garc{\'\i}a}, Osbaldo and {Cervantes Sodi}, Bernardo and {Fritz}, Jacopo and {Moretti}, Alessia and {Poggianti}, Bianca M. and {George}, Koshy and {Gullieuszik}, Marco and {Vulcani}, Benedetta and {Fasano}, Giovanni and {Tawfeek}, Amira A.},
        title = "{GASP. XLV. Stellar Bars in Jellyfish Galaxies: Analysis of Ionized Gas and Stellar Populations}",
      journal = {\apj},
         year = 2023,
        month = mar,
       volume = {945},
       number = {2},
          eid = {99},
        pages = {99},
          doi = {10.3847/1538-4357/acb269},
archivePrefix = {arXiv},
       eprint = {2301.06612},
 primaryClass = {astro-ph.GA},
       adsurl = {https://ui.adsabs.harvard.edu/abs/2023ApJ...945...99S}
}

@ARTICLE{Vulcani+21,
       author = {{Vulcani}, Benedetta and {Poggianti}, Bianca M. and {Moretti}, Alessia and {Franchetto}, Andrea and {Bacchini}, Cecilia and {McGee}, Sean and {Jaff{\'e}}, Yara L. and {Mingozzi}, Matilde and {Werle}, Ariel and {Tomi{\v{c}}i{\'c}}, Neven and {Fritz}, Jacopo and {Bettoni}, Daniela and {Wolter}, Anna and {Gullieuszik}, Marco},
        title = "{GASP. XXXIII. The Ability of Spatially Resolved Data to Distinguish among the Different Physical Mechanisms Affecting Galaxies in Low-density Environments}",
      journal = {\apj},
         year = 2021,
        month = jun,
       volume = {914},
       number = {1},
          eid = {27},
        pages = {27},
          doi = {10.3847/1538-4357/abf655},
archivePrefix = {arXiv},
       eprint = {2104.02089},
 primaryClass = {astro-ph.GA},
       adsurl = {https://ui.adsabs.harvard.edu/abs/2021ApJ...914...27V}
}

@ARTICLE{Planck+18,
       author = {{Aghanim}, N. and {Akrami}, Y. and {Ashdown}, M. and {Aumont}, J. and {Baccigalupi}, C. and {Ballardini}, M. and {Banday}, A.~J. and {Barreiro}, R.~B. and {Bartolo}, N. and {Basak}, S. and {Battye}, R. and {Benabed}, K. and {Bernard}, J.-P. and {Bersanelli}, M. and {Bielewicz}, P. and {Bock}, J.~J. and {Bond}, J.~R. and {Borrill}, J. and {Bouchet}, F.~R. and {Boulanger}, F. and {Bucher}, M. and {Burigana}, C. and {Butler}, R.~C. and {Calabrese}, E. and {Cardoso}, J.-F. and {Carron}, J. and {Challinor}, A. and {Chiang}, H.~C. and {Chluba}, J. and {Colombo}, L.~P.~L. and {Combet}, C. and {Contreras}, D. and {Crill}, B.~P. and {Cuttaia}, F. and {de Bernardis}, P. and {de Zotti}, G. and {Delabrouille}, J. and {Delouis}, J.-M. and {Di Valentino}, E. and {Diego}, J.~M. and {Dor{\'e}}, O. and {Douspis}, M. and {Ducout}, A. and {Dupac}, X. and {Dusini}, S. and {Efstathiou}, G. and {Elsner}, F. and {En{\ss}lin}, T.~A. and {Eriksen}, H.~K. and {Fantaye}, Y. and {Farhang}, M. and {Fergusson}, J. and {Fernandez-Cobos}, R. and {Finelli}, F. and {Forastieri}, F. and {Frailis}, M. and {Fraisse}, A.~A. and {Franceschi}, E. and {Frolov}, A. and {Galeotta}, S. and {Galli}, S. and {Ganga}, K. and {G{\'e}nova-Santos}, R.~T. and {Gerbino}, M. and {Ghosh}, T. and {Gonz{\'a}lez-Nuevo}, J. and {G{\'o}rski}, K.~M. and {Gratton}, S. and {Gruppuso}, A. and {Gudmundsson}, J.~E. and {Hamann}, J. and {Handley}, W. and {Hansen}, F.~K. and {Herranz}, D. and {Hildebrandt}, S.~R. and {Hivon}, E. and {Huang}, Z. and {Jaffe}, A.~H. and {Jones}, W.~C. and {Karakci}, A. and {Keih{\"a}nen}, E. and {Keskitalo}, R. and {Kiiveri}, K. and {Kim}, J. and {Kisner}, T.~S. and {Knox}, L. and {Krachmalnicoff}, N. and {Kunz}, M. and {Kurki-Suonio}, H. and {Lagache}, G. and {Lamarre}, J.-M. and {Lasenby}, A. and {Lattanzi}, M. and {Lawrence}, C.~R. and {Le Jeune}, M. and {Lemos}, P. and {Lesgourgues}, J. and {Levrier}, F. and {Lewis}, A. and {Liguori}, M. and {Lilje}, P.~B. and {Lilley}, M. and {Lindholm}, V. and {L{\'o}pez-Caniego}, M. and {Lubin}, P.~M. and {Ma}, Y.-Z. and {Mac{\'\i}as-P{\'e}rez}, J.~F. and {Maggio}, G. and {Maino}, D. and {Mandolesi}, N. and {Mangilli}, A. and {Marcos-Caballero}, A. and {Maris}, M. and {Martin}, P.~G. and {Martinelli}, M. and {Mart{\'\i}nez-Gonz{\'a}lez}, E. and {Matarrese}, S. and {Mauri}, N. and {McEwen}, J.~D. and {Meinhold}, P.~R. and {Melchiorri}, A. and {Mennella}, A. and {Migliaccio}, M. and {Millea}, M. and {Mitra}, S. and {Miville-Desch{\^e}nes}, M.-A. and {Molinari}, D. and {Montier}, L. and {Morgante}, G. and {Moss}, A. and {Natoli}, P. and {N{\o}rgaard-Nielsen}, H.~U. and {Pagano}, L. and {Paoletti}, D. and {Partridge}, B. and {Patanchon}, G. and {Peiris}, H.~V. and {Perrotta}, F. and {Pettorino}, V. and {Piacentini}, F. and {Polastri}, L. and {Polenta}, G. and {Puget}, J.-L. and {Rachen}, J.~P. and {Reinecke}, M. and {Remazeilles}, M. and {Renzi}, A. and {Rocha}, G. and {Rosset}, C. and {Roudier}, G. and {Rubi{\~n}o-Mart{\'\i}n}, J.~A. and {Ruiz-Granados}, B. and {Salvati}, L. and {Sandri}, M. and {Savelainen}, M. and {Scott}, D. and {Shellard}, E.~P.~S. and {Sirignano}, C. and {Sirri}, G. and {Spencer}, L.~D. and {Sunyaev}, R. and {Suur-Uski}, A.-S. and {Tauber}, J.~A. and {Tavagnacco}, D. and {Tenti}, M. and {Toffolatti}, L. and {Tomasi}, M. and {Trombetti}, T. and {Valenziano}, L. and {Valiviita}, J. and {Van Tent}, B. and {Vibert}, L. and {Vielva}, P. and {Villa}, F. and {Vittorio}, N. and {Wandelt}, B.~D. and {Wehus}, I.~K. and {White}, M. and {White}, S.~D.~M. and {Zacchei}, A. and {Zonca}, A.},
        title = "{Planck 2018 results. VI. Cosmological parameters}",
      journal = {\aap},
         year = 2020,
        month = sep,
       volume = {641},
          eid = {A6},
        pages = {A6},
          doi = {10.1051/0004-6361/201833910},
archivePrefix = {arXiv},
       eprint = {1807.06209},
 primaryClass = {astro-ph.CO},
       adsurl = {https://ui.adsabs.harvard.edu/abs/2020A&A...641A...6P}
}

@ARTICLE{GASP33,
       author = {{Vulcani}, Benedetta and {Poggianti}, Bianca M. and {Moretti}, Alessia and {Franchetto}, Andrea and {Bacchini}, Cecilia and {McGee}, Sean and {Jaff{\'e}}, Yara L. and {Mingozzi}, Matilde and {Werle}, Ariel and {Tomi{\v{c}}i{\'c}}, Neven and {Fritz}, Jacopo and {Bettoni}, Daniela and {Wolter}, Anna and {Gullieuszik}, Marco},
        title = "{GASP. XXXIII. The Ability of Spatially Resolved Data to Distinguish among the Different Physical Mechanisms Affecting Galaxies in Low-density Environments}",
      journal = {\apj},
         year = 2021,
        month = jun,
       volume = {914},
       number = {1},
          eid = {27},
        pages = {27},
          doi = {10.3847/1538-4357/abf655},
archivePrefix = {arXiv},
       eprint = {2104.02089},
 primaryClass = {astro-ph.GA},
       adsurl = {https://ui.adsabs.harvard.edu/abs/2021ApJ...914...27V}
}

@ARTICLE{GASP12,
       author = {{Vulcani}, Benedetta and {Poggianti}, Bianca M. and {Jaff{\'e}}, Yara L. and {Moretti}, Alessia and {Fritz}, Jacopo and {Gullieuszik}, Marco and {Bettoni}, Daniela and {Fasano}, Giovanni and {Tonnesen}, Stephanie and {McGee}, Sean},
        title = "{GASP - XII. The variety of physical processes occurring in a single galaxy group in formation}",
      journal = {\mnras},
         year = 2018,
        month = nov,
       volume = {480},
       number = {3},
        pages = {3152-3169},
          doi = {10.1093/mnras/sty2095},
archivePrefix = {arXiv},
       eprint = {1809.02668},
 primaryClass = {astro-ph.GA},
       adsurl = {https://ui.adsabs.harvard.edu/abs/2018MNRAS.480.3152V}
}

@ARTICLE{Vulcani+18,
       author = {{Vulcani}, Benedetta and {Poggianti}, Bianca M. and {Gullieuszik}, Marco and {Moretti}, Alessia and {Tonnesen}, Stephanie and {Jaff{\'e}}, Yara L. and {Fritz}, Jacopo and {Fasano}, Giovanni and {Bettoni}, Daniela},
        title = "{Enhanced Star Formation in Both Disks and Ram-pressure-stripped Tails of GASP Jellyfish Galaxies}",
      journal = {\apjl},
         year = 2018,
        month = oct,
       volume = {866},
       number = {2},
          eid = {L25},
        pages = {L25},
          doi = {10.3847/2041-8213/aae68b},
archivePrefix = {arXiv},
       eprint = {1810.05164},
 primaryClass = {astro-ph.GA},
       adsurl = {https://ui.adsabs.harvard.edu/abs/2018ApJ...866L..25V}
}

@ARTICLE{Jaffe+18,
       author = {{Jaff{\'e}}, Yara L. and {Poggianti}, Bianca M. and {Moretti}, Alessia and {Gullieuszik}, Marco and {Smith}, Rory and {Vulcani}, Benedetta and {Fasano}, Giovanni and {Fritz}, Jacopo and {Tonnesen}, Stephanie and {Bettoni}, Daniela and {Hau}, George and {Biviano}, Andrea and {Bellhouse}, Callum and {McGee}, Sean},
        title = "{GASP. IX. Jellyfish galaxies in phase-space: an orbital study of intense ram-pressure stripping in clusters}",
      journal = {\mnras},
         year = 2018,
        month = jun,
       volume = {476},
       number = {4},
        pages = {4753-4764},
          doi = {10.1093/mnras/sty500},
archivePrefix = {arXiv},
       eprint = {1802.07297},
 primaryClass = {astro-ph.GA},
       adsurl = {https://ui.adsabs.harvard.edu/abs/2018MNRAS.476.4753J}
}

@ARTICLE{Gunn+72,
       author = {{Gunn}, James E. and {Gott}, III, J. Richard},
        title = "{On the Infall of Matter Into Clusters of Galaxies and Some Effects on Their Evolution}",
      journal = {\apj},
         year = 1972,
        month = aug,
       volume = {176},
        pages = {1},
          doi = {10.1086/151605},
       adsurl = {https://ui.adsabs.harvard.edu/abs/1972ApJ...176....1G}
}

@ARTICLE{Behroozi+19,
       author = {{Behroozi}, Peter and {Wechsler}, Risa H. and {Hearin}, Andrew P. and {Conroy}, Charlie},
        title = "{UNIVERSEMACHINE: The correlation between galaxy growth and dark matter halo assembly from z = 0-10}",
      journal = {\mnras},
         year = 2019,
        month = sep,
       volume = {488},
       number = {3},
        pages = {3143-3194},
          doi = {10.1093/mnras/stz1182},
archivePrefix = {arXiv},
       eprint = {1806.07893},
 primaryClass = {astro-ph.GA},
       adsurl = {https://ui.adsabs.harvard.edu/abs/2019MNRAS.488.3143B}
}

@ARTICLE{Croton+06,
       author = {{Croton}, Darren J. and {Springel}, Volker and {White}, Simon D.~M. and {De Lucia}, G. and {Frenk}, C.~S. and {Gao}, L. and {Jenkins}, A. and {Kauffmann}, G. and {Navarro}, J.~F. and {Yoshida}, N.},
        title = "{The many lives of active galactic nuclei: cooling flows, black holes and the luminosities and colours of galaxies}",
      journal = {\mnras},
         year = 2006,
        month = jan,
       volume = {365},
       number = {1},
        pages = {11-28},
          doi = {10.1111/j.1365-2966.2005.09675.x},
archivePrefix = {arXiv},
       eprint = {astro-ph/0508046},
 primaryClass = {astro-ph},
       adsurl = {https://ui.adsabs.harvard.edu/abs/2006MNRAS.365...11C}
}

@ARTICLE{Baxter+25,
       author = {{Baxter}, Devontae C. and {Fillingham}, Sean P. and {Coil}, Alison L. and {Cooper}, Michael C.},
        title = "{The Importance of Gas Starvation in Driving Satellite Quenching in Galaxy Groups at z \raisebox{-0.5ex}\textasciitilde 0.8}",
      journal = {\apj},
         year = 2025,
        month = jan,
       volume = {979},
       number = {1},
          eid = {41},
        pages = {41},
          doi = {10.3847/1538-4357/ad9aa4},
archivePrefix = {arXiv},
       eprint = {2412.02766},
 primaryClass = {astro-ph.GA},
       adsurl = {https://ui.adsabs.harvard.edu/abs/2025ApJ...979...41B}
}

@ARTICLE{Lim+25,
       author = {{Lim}, Seunghwan and {Tacchella}, Sandro and {Maiolino}, Roberto and {Schaye}, Joop and {Schaller}, Matthieu},
        title = "{In-situ vs. ex-situ drivers of galaxy quenching: ubiquity of main sequence and critical black hole mass from the FLAMINGO simulation}",
      journal = {arXiv e-prints},
         year = 2025,
        month = apr,
          eid = {arXiv:2504.02027},
        pages = {arXiv:2504.02027},
          doi = {10.48550/arXiv.2504.02027},
archivePrefix = {arXiv},
       eprint = {2504.02027},
 primaryClass = {astro-ph.GA},
       adsurl = {https://ui.adsabs.harvard.edu/abs/2025arXiv250402027L}
}

@ARTICLE{Brown+17,
       author = {{Brown}, Toby and {Catinella}, Barbara and {Cortese}, Luca and {Lagos}, Claudia del P. and {Dav{\'e}}, Romeel and {Kilborn}, Virginia and {Haynes}, Martha P. and {Giovanelli}, Riccardo and {Rafieferantsoa}, Mika},
        title = "{Cold gas stripping in satellite galaxies: from pairs to clusters}",
      journal = {\mnras},
         year = 2017,
        month = apr,
       volume = {466},
       number = {2},
        pages = {1275-1289},
          doi = {10.1093/mnras/stw2991},
archivePrefix = {arXiv},
       eprint = {1611.00896},
 primaryClass = {astro-ph.GA},
       adsurl = {https://ui.adsabs.harvard.edu/abs/2017MNRAS.466.1275B}
}

@ARTICLE{Wetzel+12,
       author = {{Wetzel}, Andrew R. and {Tinker}, Jeremy L. and {Conroy}, Charlie},
        title = "{Galaxy evolution in groups and clusters: star formation rates, red sequence fractions and the persistent bimodality}",
      journal = {\mnras},
         year = 2012,
        month = jul,
       volume = {424},
       number = {1},
        pages = {232-243},
          doi = {10.1111/j.1365-2966.2012.21188.x},
archivePrefix = {arXiv},
       eprint = {1107.5311},
 primaryClass = {astro-ph.CO},
       adsurl = {https://ui.adsabs.harvard.edu/abs/2012MNRAS.424..232W}
}

@ARTICLE{Popesso+19a,
       author = {{Popesso}, P. and {Concas}, A. and {Morselli}, L. and {Schreiber}, C. and {Rodighiero}, G. and {Cresci}, G. and {Belli}, S. and {Erfanianfar}, G. and {Mancini}, C. and {Inami}, H. and {Dickinson}, M. and {Ilbert}, O. and {Pannella}, M. and {Elbaz}, D.},
        title = "{The main sequence of star-forming galaxies - I. The local relation and its bending}",
      journal = {\mnras},
         year = 2019,
        month = mar,
       volume = {483},
       number = {3},
        pages = {3213-3226},
          doi = {10.1093/mnras/sty3210},
archivePrefix = {arXiv},
       eprint = {1812.07057},
 primaryClass = {astro-ph.GA},
       adsurl = {https://ui.adsabs.harvard.edu/abs/2019MNRAS.483.3213P}
}

@ARTICLE{Gallazzi+05,
       author = {{Gallazzi}, Anna and {Charlot}, St{\'e}phane and {Brinchmann}, Jarle and {White}, Simon D.~M. and {Tremonti}, Christy A.},
        title = "{The ages and metallicities of galaxies in the local universe}",
      journal = {\mnras},
         year = 2005,
        month = sep,
       volume = {362},
       number = {1},
        pages = {41-58},
          doi = {10.1111/j.1365-2966.2005.09321.x},
archivePrefix = {arXiv},
       eprint = {astro-ph/0506539},
 primaryClass = {astro-ph},
       adsurl = {https://ui.adsabs.harvard.edu/abs/2005MNRAS.362...41G}
}

@ARTICLE{Westfall+19,
       author = {{Westfall}, Kyle B. and {Cappellari}, Michele and {Bershady}, Matthew A. and {Bundy}, Kevin and {Belfiore}, Francesco and {Ji}, Xihan and {Law}, David R. and {Schaefer}, Adam and {Shetty}, Shravan and {Tremonti}, Christy A. and {Yan}, Renbin and {Andrews}, Brett H. and {Brownstein}, Joel R. and {Cherinka}, Brian and {Coccato}, Lodovico and {Drory}, Niv and {Maraston}, Claudia and {Parikh}, Taniya and {S{\'a}nchez-Gallego}, Jos{\'e} R. and {Thomas}, Daniel and {Weijmans}, Anne-Marie and {Barrera-Ballesteros}, Jorge and {Du}, Cheng and {Goddard}, Daniel and {Li}, Niu and {Masters}, Karen and {Ibarra Medel}, H{\'e}ctor Javier and {S{\'a}nchez}, Sebasti{\'a}n F. and {Yang}, Meng and {Zheng}, Zheng and {Zhou}, Shuang},
        title = "{The Data Analysis Pipeline for the SDSS-IV MaNGA IFU Galaxy Survey: Overview}",
      journal = {\aj},
         year = 2019,
        month = dec,
       volume = {158},
       number = {6},
          eid = {231},
        pages = {231},
          doi = {10.3847/1538-3881/ab44a2},
archivePrefix = {arXiv},
       eprint = {1901.00856},
 primaryClass = {astro-ph.GA},
       adsurl = {https://ui.adsabs.harvard.edu/abs/2019AJ....158..231W}
}

@ARTICLE{Bagge+23,
       author = {{Bagge}, R.~S. and {Foster}, C. and {Battisti}, A. and {Bellstedt}, S. and {Mun}, M. and {Harborne}, K. and {Barsanti}, S. and {Mendel}, T. and {Brough}, S. and {Croom}, S.~M. and {Lagos}, C.~D.~P. and {Mukherjee}, T. and {Peng}, Y. and {Remus}, R. -S. and {Santucci}, G. and {Sharda}, P. and {Thater}, S. and {van de Sande}, J. and {Valenzuela}, L.~M. and {Wisnioski}, E. and {Zafar}, T. and {Ziegler}, B.},
        title = "{The MAGPI survey: Drivers of kinematic asymmetries in the ionised gas of z {\ensuremath{\sim}} 0.3 star-forming galaxies}",
      journal = {\pasa},
         year = 2023,
        month = dec,
       volume = {40},
          eid = {e060},
        pages = {e060},
          doi = {10.1017/pasa.2023.58},
archivePrefix = {arXiv},
       eprint = {2311.10268},
 primaryClass = {astro-ph.GA},
       adsurl = {https://ui.adsabs.harvard.edu/abs/2023PASA...40...60B}
}

@ARTICLE{Eke+04,
       author = {{Eke}, V.~R. and {Baugh}, Carlton M. and {Cole}, Shaun and {Frenk}, Carlos S. and {Norberg}, Peder and {Peacock}, John A. and {Baldry}, Ivan K. and {Bland-Hawthorn}, Joss and {Bridges}, Terry and {Cannon}, Russell and {Colless}, Matthew and {Collins}, Chris and {Couch}, Warrick and {Dalton}, Gavin and {de Propris}, Roberto and {Driver}, Simon P. and {Efstathiou}, George and {Ellis}, Richard S. and {Glazebrook}, Karl and {Jackson}, Carole and {Lahav}, Ofer and {Lewis}, Ian and {Lumsden}, Stuart and {Maddox}, Steve and {Madgwick}, Darren and {Peterson}, Bruce A. and {Sutherland}, Will and {Taylor}, Keith},
        title = "{Galaxy groups in the 2dFGRS: the group-finding algorithm and the 2PIGG catalogue}",
      journal = {\mnras},
         year = 2004,
        month = mar,
       volume = {348},
       number = {3},
        pages = {866-878},
          doi = {10.1111/j.1365-2966.2004.07408.x},
archivePrefix = {arXiv},
       eprint = {astro-ph/0402567},
 primaryClass = {astro-ph},
       adsurl = {https://ui.adsabs.harvard.edu/abs/2004MNRAS.348..866E}
}

@ARTICLE{Argudo-Fernández+15,
       author = {{Argudo-Fern{\'a}ndez}, M. and {Verley}, S. and {Bergond}, G. and {Duarte Puertas}, S. and {Ramos Carmona}, E. and {Sabater}, J. and {Fern{\'a}ndez Lorenzo}, M. and {Espada}, D. and {Sulentic}, J. and {Ruiz}, J.~E. and {Leon}, S.},
        title = "{Catalogues of isolated galaxies, isolated pairs, and isolated triplets in the local Universe}",
      journal = {\aap},
         year = 2015,
        month = jun,
       volume = {578},
          eid = {A110},
        pages = {A110},
          doi = {10.1051/0004-6361/201526016},
archivePrefix = {arXiv},
       eprint = {1504.00117},
 primaryClass = {astro-ph.GA},
       adsurl = {https://ui.adsabs.harvard.edu/abs/2015A&A...578A.110A}
}

@ARTICLE{Larson+80,
       author = {{Larson}, R.~B. and {Tinsley}, B.~M. and {Caldwell}, C.~N.},
        title = "{The evolution of disk galaxies and the origin of S0 galaxies}",
      journal = {\apj},
         year = 1980,
        month = may,
       volume = {237},
        pages = {692-707},
          doi = {10.1086/157917},
       adsurl = {https://ui.adsabs.harvard.edu/abs/1980ApJ...237..692L}
}

@ARTICLE{Moore+98,
       author = {{Moore}, Ben and {Lake}, George and {Katz}, Neal},
        title = "{Morphological Transformation from Galaxy Harassment}",
      journal = {\apj},
         year = 1998,
        month = mar,
       volume = {495},
       number = {1},
        pages = {139-151},
          doi = {10.1086/305264},
archivePrefix = {arXiv},
       eprint = {astro-ph/9701211},
 primaryClass = {astro-ph},
       adsurl = {https://ui.adsabs.harvard.edu/abs/1998ApJ...495..139M}
}

@ARTICLE{Tinker+21,
       author = {{Tinker}, Jeremy L.},
        title = "{A Self-Calibrating Halo-Based Group Finder: Application to SDSS}",
      journal = {\apj},
         year = 2021,
        month = dec,
       volume = {923},
       number = {2},
          eid = {154},
        pages = {154},
          doi = {10.3847/1538-4357/ac2aaa},
archivePrefix = {arXiv},
       eprint = {2010.02946},
 primaryClass = {astro-ph.CO},
       adsurl = {https://ui.adsabs.harvard.edu/abs/2021ApJ...923..154T}
}

@ARTICLE{Wang+20,
       author = {{Wang}, Bitao and {Cappellari}, Michele and {Peng}, Yingjie and {Graham}, Mark},
        title = "{SDSS-IV MaNGA: The kinematic-morphology of galaxies on the mass versus star-formation relation in different environments}",
      journal = {\mnras},
         year = 2020,
        month = jun,
       volume = {495},
       number = {2},
        pages = {1958-1977},
          doi = {10.1093/mnras/staa1325},
archivePrefix = {arXiv},
       eprint = {2002.09011},
 primaryClass = {astro-ph.GA},
       adsurl = {https://ui.adsabs.harvard.edu/abs/2020MNRAS.495.1958W}
}

@ARTICLE{Wang+24,
       author = {{Wang}, Bitao and {Peng}, Yingjie and {Cappellari}, Michele and {Gao}, Hua and {Mo}, Houjun},
        title = "{On the Kinematic Nature of Apparent Disks at High Redshifts: Local Counterparts are Not Dominated by Ordered Rotation but by Tangentially Anisotropic Random Motion}",
      journal = {\apjl},
         year = 2024,
        month = sep,
       volume = {973},
       number = {1},
          eid = {L29},
        pages = {L29},
          doi = {10.3847/2041-8213/ad772d},
archivePrefix = {arXiv},
       eprint = {2408.10735},
 primaryClass = {astro-ph.GA},
       adsurl = {https://ui.adsabs.harvard.edu/abs/2024ApJ...973L..29W}
}

@ARTICLE{Yang+21,
       author = {{Yang}, Xiaohu and {Xu}, Haojie and {He}, Min and {Gu}, Yizhou and {Katsianis}, Antonios and {Meng}, Jiacheng and {Shi}, Feng and {Zou}, Hu and {Zhang}, Youcai and {Liu}, Chengze and {Wang}, Zhaoyu and {Dong}, Fuyu and {Lu}, Yi and {Li}, Qingyang and {Chen}, Yangyao and {Wang}, Huiyuan and {Mo}, Houjun and {Fu}, Jian and {Guo}, Hong and {Leauthaud}, Alexie and {Luo}, Yu and {Zhang}, Jun and {Zu}, Ying},
        title = "{An Extended Halo-based Group/Cluster Finder: Application to the DESI Legacy Imaging Surveys DR8}",
      journal = {\apj},
         year = 2021,
        month = mar,
       volume = {909},
       number = {2},
          eid = {143},
        pages = {143},
          doi = {10.3847/1538-4357/abddb2},
archivePrefix = {arXiv},
       eprint = {2012.14998},
 primaryClass = {astro-ph.GA},
       adsurl = {https://ui.adsabs.harvard.edu/abs/2021ApJ...909..143Y}
}

@ARTICLE{Marini24a,
       author = {{Marini}, I. and {Popesso}, P. and {Dolag}, K. and {Bravo}, M. and {Robotham}, A. and {Tempel}, E. and {Li}, Q. and {Yang}, X. and {Csizi}, B. and {Behroozi}, P. and {Biffi}, V. and {Biviano}, A. and {Lamer}, G. and {Malavasi}, N. and {Mazengo}, D. and {Toptun}, V.},
        title = "{Detecting clusters and groups of galaxies populating the local Universe in large optical spectroscopic surveys}",
      journal = {\aap},
         year = 2025,
        month = feb,
       volume = {694},
          eid = {A207},
        pages = {A207},
          doi = {10.1051/0004-6361/202452028},
archivePrefix = {arXiv},
       eprint = {2411.16455},
 primaryClass = {astro-ph.GA},
       adsurl = {https://ui.adsabs.harvard.edu/abs/2025A&A...694A.207M}
}

\begin{appendix}

\onecolumn
\section{Signal-to-noise ratio in MaNGA data}
\label{Appendix_A}
\begin{figure*}[htbp!]
  \centering
\includegraphics[width=1\textwidth, page=1]
{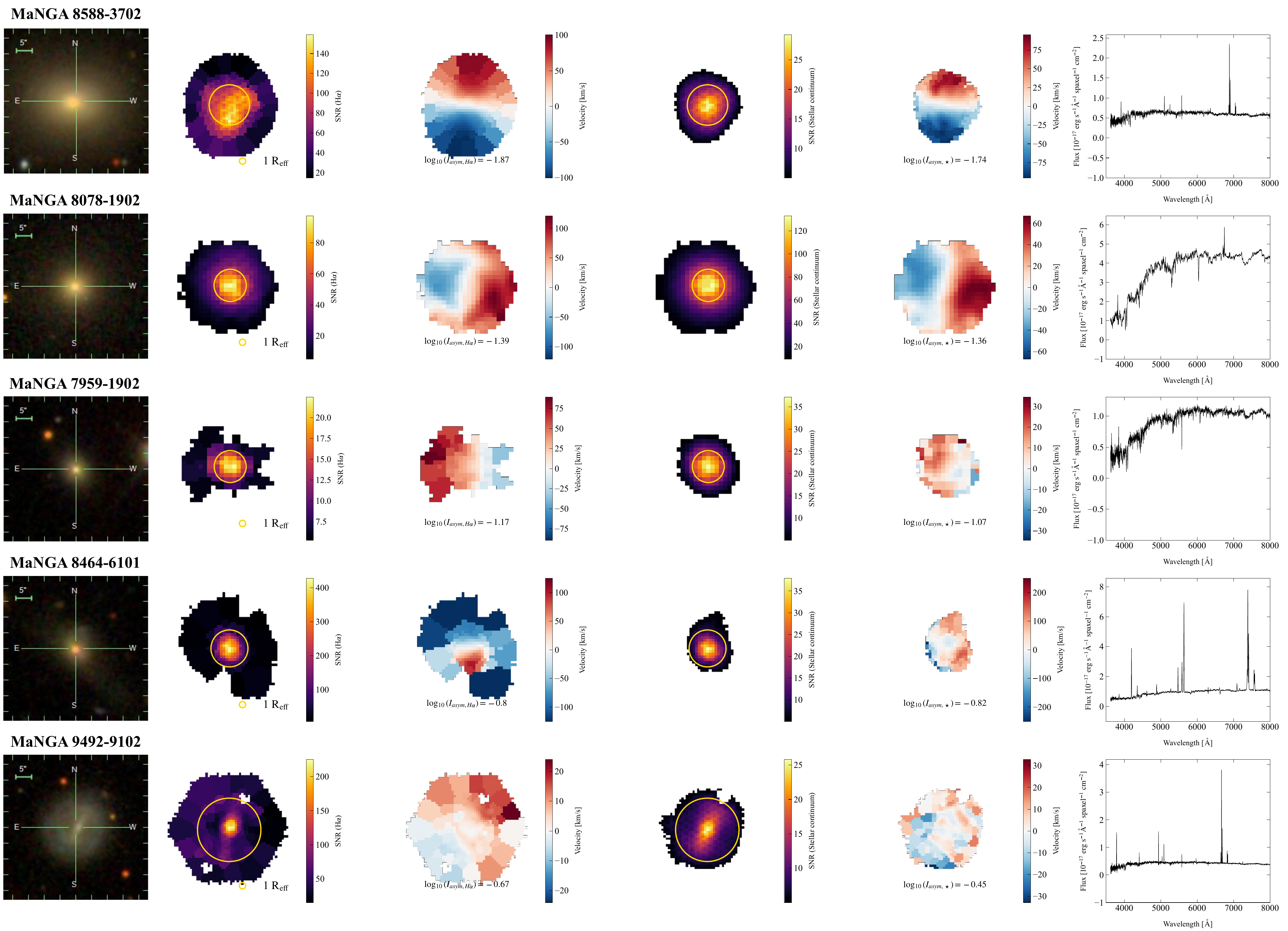}
    \caption{Examples of galaxies with different levels of kinematic asymmetry from the most kinematically symmetric (top panels), to the most asymmetric (bottom panels).
    \textit{First~column:} Stellar emission maps of the galaxy from SDSS;
    \textit{Second~column:} Average signal-to-noise map of the H$\alpha$ emission; 
    \textit{Third~column:} H$\alpha$ velocity map of each galaxy along with its corresponding asymmetry parameter;
    \textit{Fourth~column:} Average signal-to-noise map of the stellar component;
    \textit{Fifth~column:} Stellar velocity map of each galaxy along with its corresponding asymmetry parameter;
    \textit{Sixth~column:} Spectrum of each galaxy showing emission lines (particularly prominent in the last two rows) and stellar absorption features.
    The yellow circle displayed in the second and fourth columns corresponds to 1~R$_{\rm eff}$ of the galaxy. The white background corresponds to spaxels with S/N < 5.}
    \label{fig:first_fig}
\end{figure*}

As we mention in $\S$\ref{selection_criteria}, all spaxels taken into account in both gaseous and stellar velocity maps analyzed in this work must have at least a minumum signal-to-noise ratio (SNR) of 5. For each MaNGA galaxy, we calculate the SNR of the H$\alpha$ line following Equation (1) from \cite{Belfiore+19}:
      \begin{equation}\label{eq3}
      \left(\frac{S}{N}\right)_{\rm{H\alpha}}=\frac{\rm{Flux_{\rm{H\alpha}}}}{\rm{Err}_{\rm{Flux,~H\alpha}}},
      \end{equation}
where $\rm Err_{Flux,H\alpha}$ is the error on the flux of the H$\alpha$ line.
Similarly, we calculate the SNR per spaxel for as follows:
      \begin{equation}\label{eq4}
            \left(\frac{S}{N}\right)_{\rm{spaxel}} = \rm Flux_{spaxel}\times\sqrt{\rm Var_{\rm spaxel}^{-1}},
      \end{equation}
where $\rm Flux_{spaxel}$ is the flux intensity (in our case, either $\rm H\alpha$ or stellar continuum), and $\rm Var_{\rm spaxel}^{-1}$ is the inverse variance of the flux as provided by MaNGA Data Analysis Pipeline.
Spaxels with SNR $\geq$ 5 were considered valid for subsequent \textsc{kinemetry} analysis. Spaxels below this threshold were masked (set to NaN) in the velocity maps before performing the asymmetry calculations.
This SNR-based masking ensures that only reliable measurements contribute to the kinematic asymmetry analysis, minimizing noise contamination in the results.
Examples of SNR maps are displayed in the second and fourth panels of Fig.~\ref{fig:first_fig} for H$\alpha$ and stellar velocity maps, respectively.
All galaxies analyzed in this work have at least 90\% of valid spaxels within 1~$R_{\rm eff}$.

\twocolumn
\section{\textsc{Kinemetry} on GASP data}\label{Appendix_B}

\begin{figure}[htbp!]
  \centering
\includegraphics[width=0.5\textwidth]
{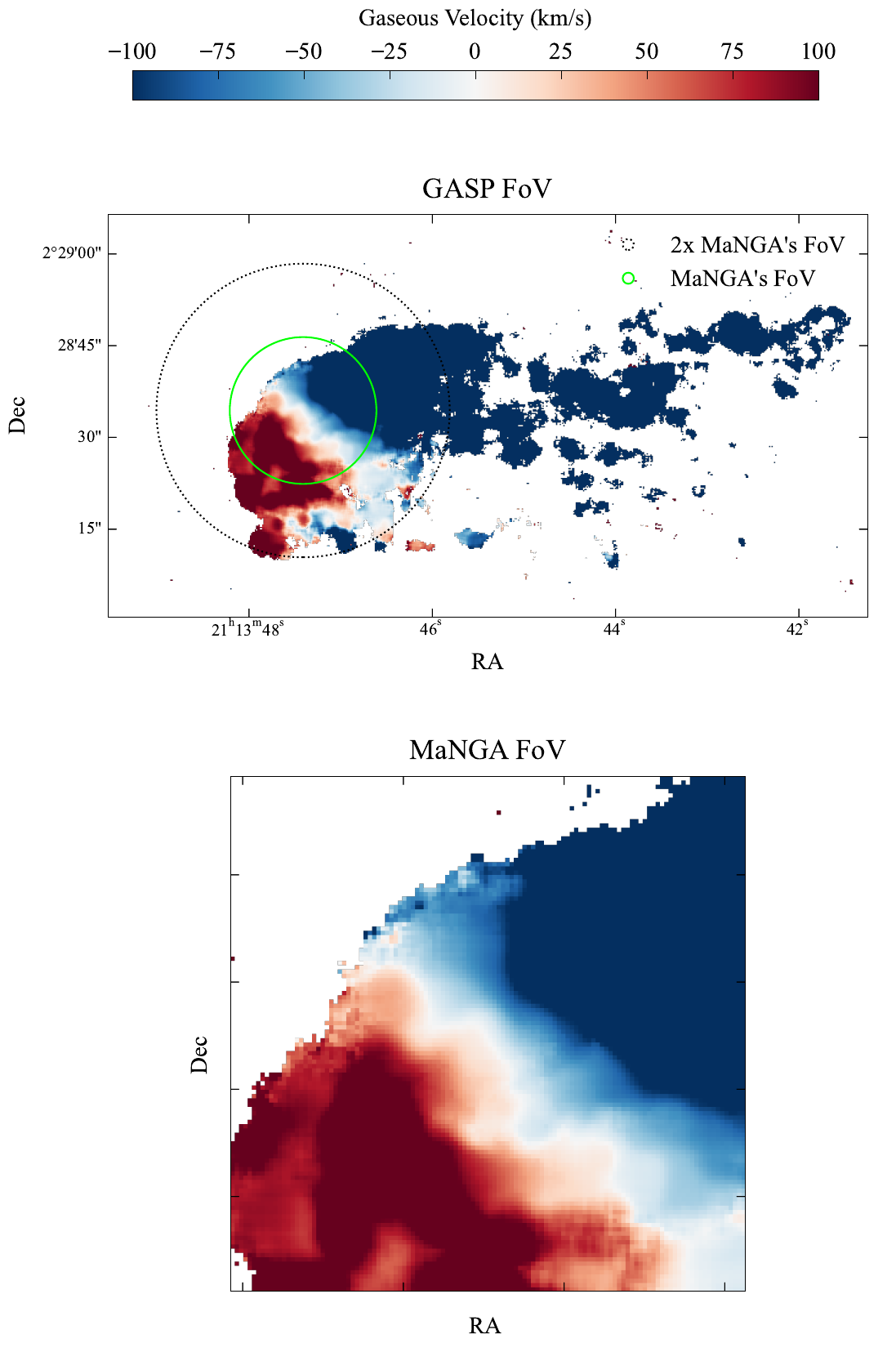}
    \caption{We show three different apertures matching the GASP FoV (original data; \textit{top~panel}), and MaNGA's average FoV (\textit{bottom~panel}) for the jellyfish galaxy JO206.
    As shown in the \textit{bottom~panel}, using only MaNGA's average FoV would miss the tail-like substructure, resulting in the expected lower asymmetry parameter highlighted in Fig.~\ref{fig:MaNGA_FoV}.
    }
    \label{fig:JO206_FoVs}
\end{figure}

As discussed in Section~\ref{jellyfish}, we analyzed 23 randomly selected galaxies from the GAs Stripping Phenomena in galaxies with MUSE (GASP) survey to enable a direct comparison of the detectability of ram pressure features in limited-field-of-view (FoV) surveys such as MaNGA. We present in Fig.~\ref{fig:JO206_FoVs} the kinematic map of JO206, showing the characteristic jellyfish tails as observed by the GASP FoV (\textit{top~panel}).

\begin{table}[htbp!]
\begin{center}
\caption{GASP galaxies analyzed in this work.}
\label{tab:GASP}
    \begin{tabular}{cccc}

    \hline
    Galaxy ID & $I_{asym}$ & $I_{asym}$ & RPS \\
     & \footnotesize{(GASP FoV)} & \footnotesize{(MaNGA FoV)} & confirmed? \\
    \midrule
JO17   & $0.040_{-0.002}^{+0.002}$ & $0.038_{-0.001}^{+0.001}$ & Mild (1)\\ [0.10cm]
JO171  & $0.222_{-0.002}^{+0.002}$ & $0.181_{-0.002}^{+0.002}$ & Extreme (1)\\[0.10cm]
JO175  & $0.054_{-0.001}^{+0.001}$ & $0.031_{-0.001}^{+0.001}$ & Extreme (1)\\[0.10cm]
JO190  & $0.250_{-0.003}^{+0.003}$ & $0.251_{-0.003}^{+0.003}$ & Merger + RPS (2)\\[0.10cm]
JO204  & $0.073_{-0.001}^{+0.001}$ & $0.054_{-0.001}^{+0.001}$ & Extreme (1)\\[0.10cm]
JO206  & $0.075_{-0.001}^{+0.001}$ & $0.041_{-0.001}^{+0.001}$ & Extreme (1)\\[0.10cm]
JO69   & $0.043_{-0.005}^{+0.005}$ & $0.049_{-0.004}^{+0.004}$ & Strong (1)\\[0.10cm]
JO95   & $0.128_{-0.008}^{+0.008}$ & $0.110_{-0.005}^{+0.005}$ & Strong (1)\\[0.10cm]
JW100  & $0.240_{-0.001}^{+0.001}$ & $0.190_{-0.001}^{+0.001}$ & Extreme (1)\\[0.10cm]
P96244 & $0.027_{-0.001}^{+0.001}$ & $0.016_{-0.001}^{+0.001}$ & Yes (2)\\[0.10cm]
P96949 & $0.166_{-0.004}^{+0.004}$ & $0.144_{-0.004}^{+0.004}$ & Merger\\[0.10cm]
JO20   & $0.102_{-0.002}^{+0.002}$ & $0.104_{-0.001}^{+0.001}$ & Merger\\[0.10cm]
JO24   & $0.084_{-0.003}^{+0.003}$ & $0.079_{-0.002}^{+0.002}$ & Strong (1)\\[0.10cm]
JO45   & $0.067_{-0.004}^{+0.004}$ & $0.093_{-0.003}^{+0.003}$ & Mild (1)\\[0.10cm]
P13384 & $0.044_{-0.006}^{+0.006}$ & $0.045_{-0.004}^{+0.004}$ & No (3)\\[0.10cm]
P16762 & $0.138_{-0.011}^{+0.011}$ & $0.128_{-0.007}^{+0.007}$ & No (3)\\[0.10cm]
P18060 & $0.066_{-0.004}^{+0.004}$ & $0.068_{-0.003}^{+0.003}$ & No (3)\\[0.10cm]
P20159 & $0.080_{-0.003}^{+0.003}$ & $0.102_{-0.002}^{+0.002}$ & Yes (2)\\[0.10cm]
P20769 & $0.053_{-0.003}^{+0.003}$ & $0.053_{-0.002}^{+0.002}$ & No (3)\\[0.10cm]
P20883 & $0.024_{-0.002}^{+0.002}$ & $0.019_{-0.001}^{+0.001}$ & No (3)\\[0.10cm]
P21734 & $0.024_{-0.001}^{+0.001}$ & $0.011_{-0.001}^{+0.001}$ & No (3)\\[0.10cm]
P5055  & $0.053_{-0.001}^{+0.001}$ & $0.053_{-0.001}^{+0.001}$ & Yes (2)\\[0.10cm]
P954   & $0.028_{-0.002}^{+0.002}$ & $0.034_{-0.002}^{+0.002}$ & No (3)\\[0.10cm]

\hline
\end{tabular}\\
\vspace{0.1cm}
\end{center}
    \footnotesize \bf{Notes:} \normalfont References in column 4: (1)~\citet{Poggianti+25}; (2)~\citet{Vulcani+21}; (3)~\citet{Sanchez-Garcia+23}.
\end{table}

\begin{figure}[htbp!]
  \centering
\includegraphics[width=0.48\textwidth]
{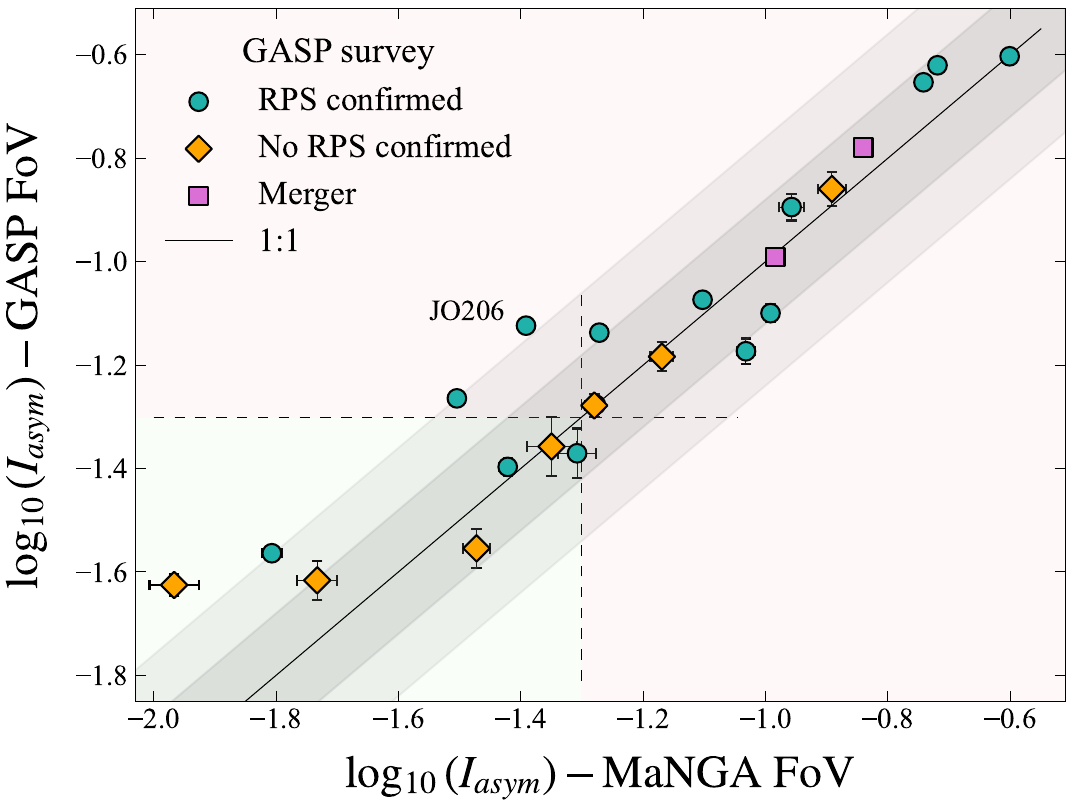}
    \caption{Comparison between the asymmetry indices of GASP galaxies calculated applying a limited aperture with both the average ($\sim$24~arcsec) FoV of the MaNGA survey (\textit{x axis}) and the FoV of GASP survey (\textit{y axis}). 
    We highlight in red the GASP galaxies with strong jellyfish features in the kinematic maps.
    Except for JO206 (Fig.~\ref{fig:JO206_FoVs}), JO171, and JW100, we verify that the asymmetry estimates show good agreement within 2$\sigma$ of confidence level.
    }
    \label{fig:MaNGA_FoV}
\end{figure}

\onecolumn
\section{AGN contribution}\label{Appendix_C}
\begin{figure}[htbp!]
  \centering
\includegraphics[width=0.78\textwidth]
{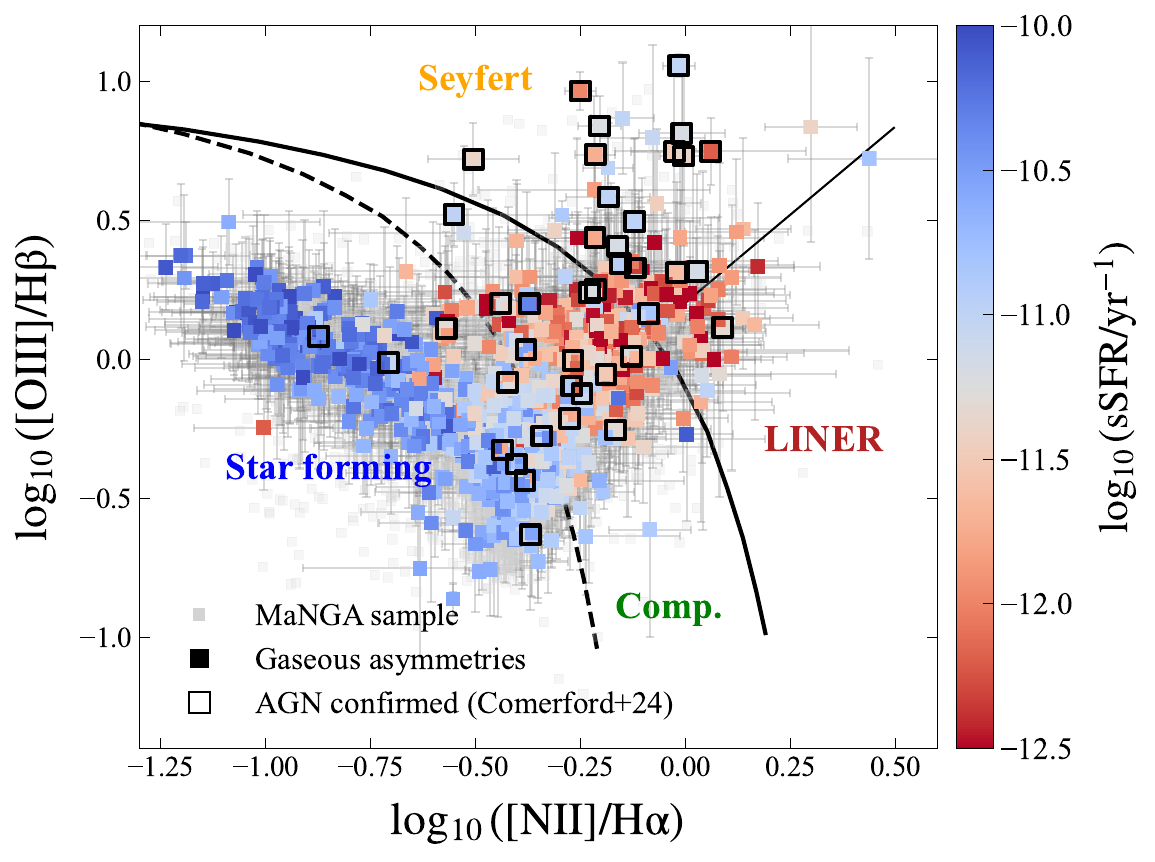}
    \caption{BPT diagram for galaxies exhibiting asymmetries in both gaseous and stellar components. Galaxies confirmed as AGN hosts from \cite{Comerford+24} are highlighted with black square contours overlaid on the data points. The \textit{dashed} and \textit{solid} lines demarcate the AGN from star-forming galaxies based on the BPT classification from \cite{Kauffmann+03} and \cite{Kewley+06}, respectively.}
    \label{fig:BPT}
\end{figure}

\begin{figure*}[htbp!]
  \centering
\includegraphics[width=0.93\textwidth, page=1]
{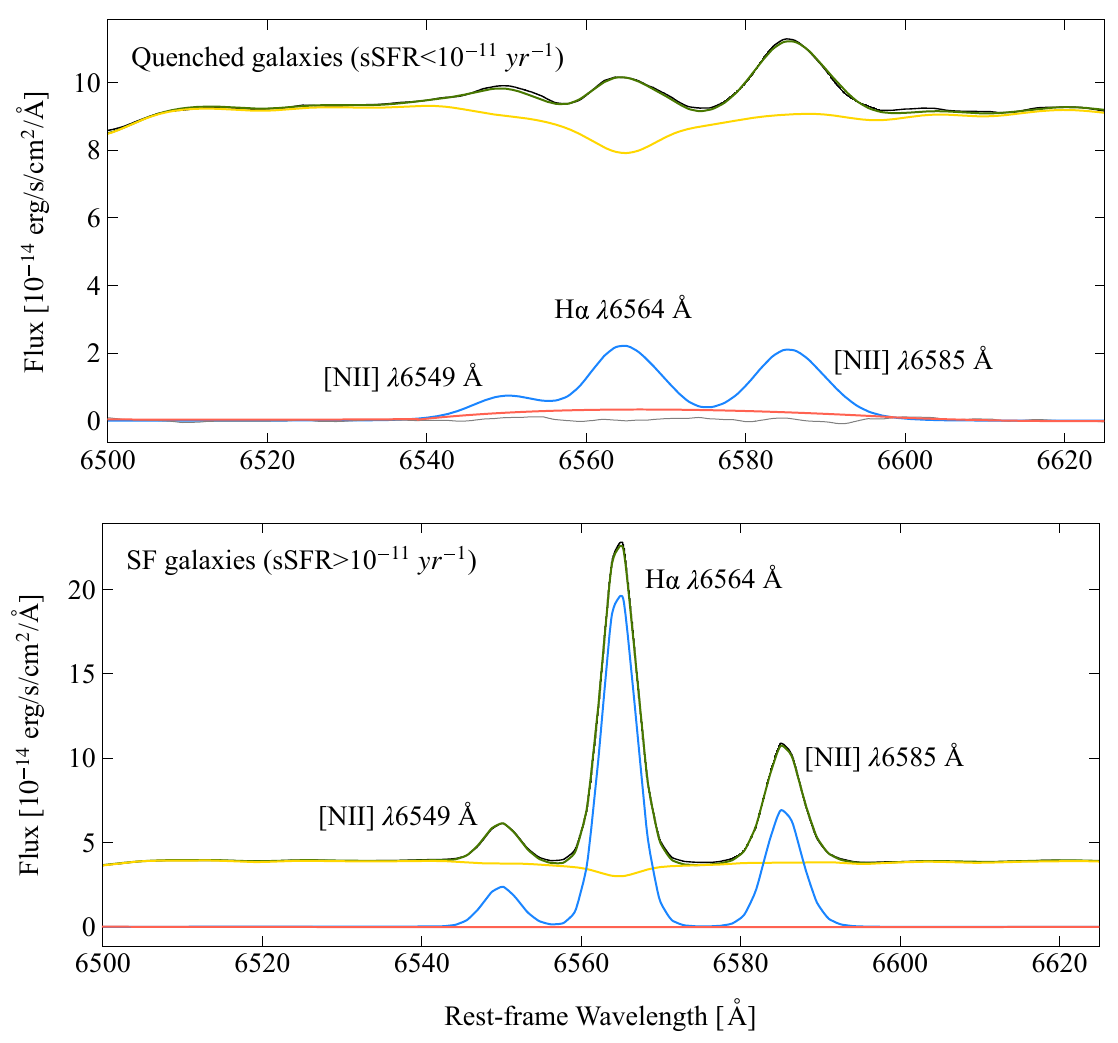}
    \caption{The stacked spectrum of 20 randomly selected AGN host galaxies, as confirmed by \cite[][]{Comerford+24} (\textit{top panel}), reveals a broad H$\alpha$ component that is likely linked with outflows, which might be causing the observed kinematic gaseous asymmetries. The \textit{middle panel} presents the stacked spectrum of quenched galaxies, showing no visible broad component, while the \textit{bottom panel} illustrates a perfect single Gaussian fitting for emission lines without any potential outflow contribution.}
    \label{fig:stacked}
\end{figure*}

Outflows from AGN or star formation could also disturb gas while leaving stars unaffected (i.e., Case 3 in $\S$\ref{obs_results}). To test this, we cross-matched our subsample of gaseous kinematically disturbed galaxies with the MaNGA AGN catalog \citep{Comerford+24}.
We highlight the cross-matched galaxies (i.e., those with gaseous asymmetries and a confirmed AGN) with black contours in Fig.~\ref{fig:BPT}. In the top panel of Fig.~\ref{fig:stacked}, we show the stacked spectrum of 20 randomly selected AGN host galaxies. We centered the emission on the H$\alpha$ line to search for a potential broad component that could arise from AGN outflows.

For galaxies with gaseous kinematic asymmetries but no confirmed AGN, we also analyzed their stacked spectrum and found that it is well fitted by a single Gaussian, suggesting no broadening in the H$\alpha$ line. This is displayed for both non-AGN confirmed, quenched galaxies (middle panel of Fig.~\ref{fig:stacked}) and non-AGN confirmed, star-forming galaxies (bottom panel of Fig.~\ref{fig:stacked}).
\begin{figure*}[htbp!]
  \centering
\includegraphics[width=0.96\textwidth]{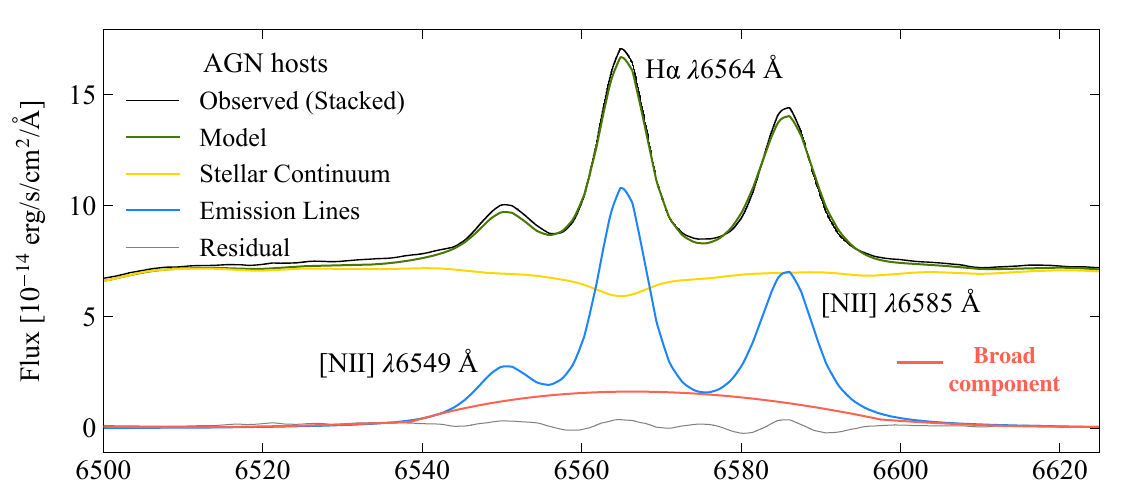}
\end{figure*}

\end{appendix}
\end{document}